\documentclass[11pt]{article}
\usepackage[utf8]{inputenc}
\usepackage[T1]{fontenc}
\usepackage[margin=1in]{geometry}
\usepackage{amsmath,amssymb}
\usepackage{subcaption}
\usepackage{graphicx}
\DeclareGraphicsExtensions{.pdf,.png}
\usepackage{booktabs}
\usepackage{array}
\usepackage{cleveref}
\usepackage{url}
 \usepackage{apacite}
 \usepackage{natbib}
\usepackage{authblk}
\usepackage{lineno}

\usepackage{iftex}
\ifXeTeX
\usepackage{xeCJK}
\fi

\title{Earthquake Aftershock Forecasting using Conditional Generative Models}

\author{Weiqiang Zhu\\
  \small Department of Earth and Planetary Science, University of California, Berkeley\\
  \small Berkeley Seismological Laboratory, University of California, Berkeley\\
  \small \texttt{zhuwq@berkeley.edu}}

\date{}

\begin{document}
\maketitle

\begin{abstract}
Forecasting how aftershocks evolve in space and time after a large earthquake is a central problem in statistical seismology and underpins operational earthquake forecasting.
Existing forecasting methods rest on statistical point-process models such as the epidemic-type aftershock sequence (ETAS) and Reasenberg--Jones models, which prescribe a fixed decay in time and an isotropic kernel in space. They match the average Omori--Utsu and Gutenberg--Richter statistics well but do not capture the fault-controlled spatial patterns of real sequences or the productivity that varies among sequences.
Neural point-process models relax these fixed forms but keep the event-by-event view and have not consistently surpassed ETAS on common benchmarks.
Rather than modeling individual events as a point process, we recast aftershock forecasting as conditional generation of a spatiotemporal field.
We develop QuakeGen, a diffusion model that generates the evolving fields of aftershock rate and maximum magnitude conditioned on recent seismicity and the forecasting horizon.
The same conditional generative framework can be trained on rich seismic catalogs to forecast global aftershock sequences and regional daily seismicity, and could further condition on physical fields such as fault geometry or geodetic deformation.
On global sequences, the data-driven approach outperforms the operational USGS Reasenberg--Jones forecast, recovering 
the fault-controlled, anisotropic spatial structure that fixed kernels cannot express. On daily forecasting, QuakeGen also matches the well-tuned ETAS baselines, which neural point-process models have yet to surpass on the regional benchmark.
Conditional generative modeling, which has transformed prediction in fields as diverse as weather forecasting and protein structure prediction, holds the potential to forecast more accurately how earthquake sequences unfold in space and time.
\end{abstract}

\section{Introduction}

A large earthquake is followed by thousands of aftershocks over the following seconds to years, so modeling their number and spatial distribution is a key component of earthquake forecasting that is operational today \citep{jordan2011operational, hardebeck2024aftershock, mizrahi2024developing}. Aftershock forecasting relies on empirical statistical laws, with event numbers decaying in time along the Omori--Utsu law \citep{omori1895after, utsu1961statistical, utsu1995centenary} and their magnitudes following the Gutenberg--Richter distribution \citep{gutenbergrichter1944frequency}. 
These laws, however, describe only the average behavior of a sequence. A real aftershock sequence departs from these statistical representations in both space and time. Its productivity varies widely from one sequence to the next \citep[e.g.,][]{dascher2020controls, trugman2023coherent}, and its rate rebounds as each large aftershock triggers its own secondary aftershocks. Its events cluster on the ruptured fault and in the lobes of increased Coulomb stress \citep{king1994static, stein1999role} rather than spreading symmetrically around the mainshock. Dynamic triggering produces further aftershocks even in the static stress shadows \citep{hardebeck2022shadows, brodsky2014uses}. An accurate forecast must therefore capture this sequence-specific structure, which the average laws do not describe.

Operational forecasting systems build on these empirical statistical laws, fitting fixed functional forms to past seismicity with a set of adjustable parameters. The Reasenberg--Jones model \citep{reasenberg1989earthquake, reasenberg1994update}, which underlies the USGS aftershock advisories \citep{barall2025oaf, vanderelst2025forecaster} and the Short-Term Earthquake Probability (STEP) model \citep{gerstenberger2005realtime}, describes the aftershocks of a single mainshock as an Omori--Utsu decay in time and a Gutenberg--Richter distribution in magnitude. The epidemic-type aftershock sequence (ETAS) model \citep{ogata1988statistical, ogata1998spacetime} extends this to an entire catalog by letting every earthquake, not the mainshock alone, trigger its own aftershocks through the same Omori decay and a power-law kernel in space, so the sequence becomes a self-exciting cascade above a constant background rate, and itself underlies operational forecasts such as UCERF3-ETAS \citep{field2017ucerf3}.
In practice the Reasenberg--Jones parameters are generic, set by tectonic region \citep{page2016three}, so they capture a region's average productivity but not the wide variation from one sequence to the next. ETAS instead must be inverted for each region or sequence, an inversion sensitive to the completeness magnitude \citep{seif2017estimating, mizrahi2021embracing}, which fluctuates through a sequence as small aftershocks are lost beneath the coda of larger ones \citep{vanderelst2021bpositive}. More fundamentally, the functional form is fixed in advance and shared across sequences, so it cannot express what sets one sequence apart from another. The spatial kernel is isotropic in its standard operational form \citep{ogata1998spacetime, felzer2006decay}, a simplification of the magnitude-dependent, anisotropic pattern that real aftershocks follow \citep{hainzl2008impact, vanderelst2015larger}. In the first hours after a mainshock, when only the mainshock and a few aftershocks have been recorded, the forecast is therefore a near-circular halo that carries limited information about the rupture's orientation. The fault-aligned pattern emerges only later, as aftershocks accumulate and their superposed kernels trace it. A finite-fault model could supply that geometry sooner and sharpen the early forecast \citep{cattania2018forecasting}, but it must be inverted separately for each mainshock and is rarely available in the first hours, when the forecast matters most.

Physics-based models instead start from mechanics rather than catalog statistics. Aftershocks concentrate where the mainshock raised the static Coulomb stress \citep{king1994static, stein1999role}, and the application of rate-and-state friction relationships converts that stress change into an evolving seismicity rate \citep{dieterich1994constitutive}. These models still prescribe the form of the stress-to-rate relation and the friction parameters that govern it, and they need the coseismic stress change as input, which must be computed from a finite-fault rupture model. That rupture model is itself inverted from seismic and geodetic recordings, is non-unique, and depends on an assumed fault geometry, and it may be poorly resolved in the first days after a mainshock \citep{goldberg2022beyond}. In prospective tests, physics-based forecasts seldom outperform ETAS \citep{cattania2018forecasting, hardebeck2021spatial}. Both the statistical and physics-based families rest on the same assumption, that a few fixed functions can link past and future earthquakes across a process as heterogeneous as seismicity, whose behavior changes with tectonic setting, fault geometry, and driving mechanism. Modern measurements now resolve that heterogeneity in ever finer detail, from dense, high-resolution catalogs built by template matching and deep learning \citep[e.g.,][]{ross2019searching, shelly2020, tan2021} to InSAR and GPS geodesy \citep[e.g.,][]{massonnet1993displacement, segall1997gps, xu2020surface} and distributed fiber-optic sensing \citep[e.g.,][]{zhan2020distributed, li2023asperities, li2025minute}. A model that learns the link between past and future seismicity directly from these data, with no kernel fixed in advance, could sharpen forecasts as the observations grow richer \citep{beroza2021machine}.

This shift from fixed rules to learned representations has already transformed earthquake monitoring \citep{bergen2019machine, mousavi2022deep}. Classical detectors such as the short-term-to-long-term amplitude ratio (STA/LTA) apply a fixed function to the waveform and register a threshold crossing \citep{allen1978automatic}; template matching sharpens their sensitivity about tenfold but recognizes only waveforms that repeat \citep{gibbons2006detection, peng2009migration, ross2019searching}. A neural network instead learns, from millions of labeled seismograms, which features separate an earthquake from noise, a representation far richer than any hand-designed function can express \citep[e.g.,][]{perol2018convolutional, ross2018generalized, zhu2018phasenet, mousavi2020eqtransformer}. Such networks now detect events and pick phases about as reliably as a seismic analyst and recover many small earthquakes that classical detectors miss \citep{munchmeyer2022which}. The resulting deep-learning catalogs, with up to an order of magnitude more earthquakes, sharpen the image of the active fault networks and the fluids that drive seismicity across tectonic, subduction, and volcanic settings \citep[e.g.,][]{suzuki2025forearc, wilding2023magmatic, tan2025clearer}. Forecasting could gain in the same way, though feeding denser catalogs to fixed statistical models has so far brought only limited gains \citep{mancini2022use}. A learned representation can express structure that fixed kernels cannot, and the enhanced catalogs now supply the spatial and temporal detail to learn it from.

Applying deep learning to forecasting has proved harder than applying it to monitoring. Early work predicted the failure of laboratory faults \citep{rouet2017machine}. The same idea, inferring the state of a fault from its continuous seismic signal, has since reached real faults, tracking slow slip in Cascadia and fault displacement at K\={\i}lauea \citep{rouetleduc2019continuous, johnson2025automatic}, though these models estimate the current fault state rather than forecast ahead. Early work also trained a neural network to forecast aftershock locations from the static stress change around a rupture \citep{devries2018deep}, improving on the classical Coulomb failure criterion. A simple two-parameter model was soon shown to match that accuracy \citep{mignan2019one}, tempering the apparent gain from deep learning. Neural point processes instead model the time evolution of a sequence, keeping the event-by-event view of ETAS but letting a network learn the triggering function rather than prescribe it \citep[e.g.,][]{du2016recurrent, mei2017neural, chen2021neural, stockman2023forecasting, dascher2023using, zhan2026flexible}. On a shared benchmark, these models have yet to surpass a well-calibrated ETAS \citep{stockman2025earthquakenpp}. A few recent models step outside the point-process view, forecasting the earthquake rate on a spatial grid and learning fault-like structure from the catalog \citep{zlydenko2023neural, zhang2024forecasting} or applying a time-series transformer to nowcast a region's seismicity \citep{rundle2024quakegpt}.
 Across these forecasting studies, neural models match the statistical baselines, but their gains remain modest, and forecasting has yet to see the step change that learned representations brought to monitoring.

Deep learning has, however, transformed forecasting in other scientific fields, most visibly in weather prediction \citep{lam2023graphcast, bi2023pangu}. These advances center on generative models, which learn the distribution of possible futures and sample from it. A deterministic network, such as those used for earthquake detection, produces a single prediction that would average every sequence consistent with the past, blurring the sharp, fault-aligned structure of a real catalog. 
A generative model instead draws samples, each a realization conditioned on the past that retains that structure, with the spread across the ensemble quantifying the forecast uncertainty.
Diffusion models are now the leading generative framework. They generate a sample by reversing a gradual noising of the data \citep{sohl2015deep, ho2020denoising, song2021scorebased} and excel at producing high-dimensional fields, from natural images \citep{rombach2022high} and video \citep{ho2022video} to protein structures \citep{abramson2024accurate} and the forecasting of evolving spatiotemporal fields \citep{wen2023diffstg, gao2023prediff}, including weather ensembles more accurate than the best operational forecasts \citep{price2025probabilistic}. 
In seismology, diffusion models already aid seismic imaging and full-waveform inversion \citep{wang2023prior}, data processing such as denoising and interpolation \citep{durall2023seismic}, and ground-motion synthesis \citep{huang2025ground}. For earthquake forecasting, however, they remain relatively unexplored, applied mainly to placing one event at a time within a point process \citep{yuan2023spatiotemporal}.

An earthquake sequence can be viewed as a spatiotemporal field, a single stochastic realization of the process set by the faults and stress conditions that produced it. Forecasting such a field is a natural fit for generative models, which learn the distribution of seismicity and the link between a sequence's past and its future directly from the catalog. In this work, we develop QuakeGen, a conditional diffusion model that forecasts future seismicity as spatiotemporal fields of earthquake frequency and maximum magnitude, conditioned on the seismicity already observed. Because the model is generative, sampling it repeatedly yields an ensemble of possible futures whose mean or median serves as the forecast and whose spread reflects the uncertainty.
Beyond the past seismicity, the same model can condition on other fields that become available after a large earthquake, such as its rupture geometry and surrounding deformation. 
We train and evaluate QuakeGen in two settings, a global one for aftershock sequences anchored to mainshocks worldwide and a regional one for the daily seismicity of a single region, and compare each against the statistical model used in that setting. 
Because it learns the mapping from past to future seismicity directly from data rather than fitting a prescribed functional form to each region, a single QuakeGen model replaces the per-sequence, per-region calibration that operational models require and would improve as catalogs grow denser.
The same framework for forecasting evolving seismicity could extend beyond mainshock--aftershock sequences to other scenarios, from fluid-driven swarms and injection-induced seismicity to sequences driven by slow fault slip, pointing toward a general approach to forecasting how seismicity evolves in space and time.

\section{Methods}

\subsection{Seismicity forecasting as conditional generative modeling}

\begin{figure}[htbp]
  \centering
  \includegraphics[width=\textwidth]{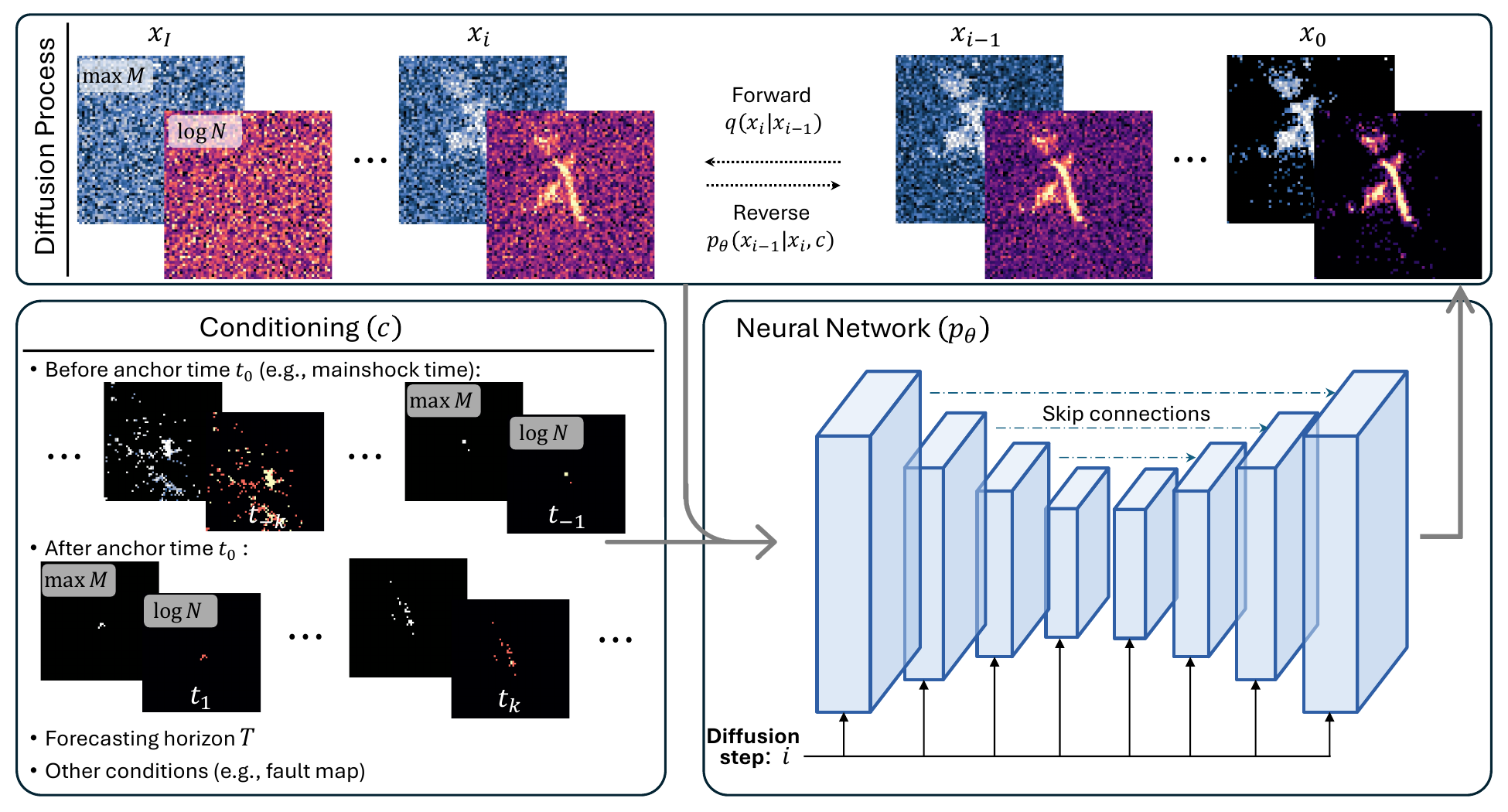}
  \caption{QuakeGen forecasts a seismicity sequence as a two-channel spatial field, the log-count $\log N$ and the maximum magnitude $\max M$ per cell, with a conditional denoising diffusion model. Top: the diffusion process links a clean forecast field $\mathbf{x}_0$ to pure noise $\mathbf{x}_I$; the fixed forward process $q(\mathbf{x}_i \mid \mathbf{x}_{i-1})$ adds a little Gaussian noise at each step, and the learned reverse process $p_\theta(\mathbf{x}_{i-1} \mid \mathbf{x}_i, \mathbf{c})$ removes it, denoising a noise sample back to a forecast field conditioned on $\mathbf{c}$. Bottom left: the conditioning $\mathbf{c}$ stacks cumulative log-count and maximum-magnitude maps over nested time windows before the forecast anchor $t_0$ (for the global model, the mainshock time) and, in the global setting, short windows after it, and further includes the scalar forecast horizon $T$ and optional physical fields such as a fault map. Bottom right: the denoiser is a U-Net with skip connections that reads the noisy field, the conditioning $\mathbf{c}$, and the diffusion step $i$, and outputs the update that forms the next, slightly denoised field $\mathbf{x}_{i-1}$.}
  \label{fig:quakegen}
\end{figure}

Statistical forecasting models describe seismicity as a point process, predicting the rate of future earthquakes from the times, locations, and magnitudes already recorded \citep{ogata2017statistics}. The Reasenberg--Jones and ETAS models prescribe this rate in closed form and issue the forecast as the expected number of earthquakes in each space--time--magnitude bin.
QuakeGen instead represents a sequence as a spatiotemporal field (Fig.~\ref{fig:quakegen}), so one model captures its full spatial pattern at once rather than one event at a time, an advantage that grows for dense sequences such as aftershocks and swarms.
We lay an $n_x \times n_y$ grid of cells $\Omega_{ij}$ over the forecast region; for a forecast anchor $t_0$ and a horizon $T$, the forecast target over the window $(t_0, t_0 + T)$ is the pair of fields
\begin{equation}
N_{ij} = \#\big\{ e : t_e \in (t_0, t_0 + T),\ \mathbf{x}_e \in \Omega_{ij} \big\},
\qquad
M^{\max}_{ij} = \max\big\{ M_e : t_e \in (t_0, t_0 + T),\ \mathbf{x}_e \in \Omega_{ij} \big\},
\label{eq:target}
\end{equation}
the number of earthquakes and the largest magnitude in each cell.
We encode the two fields of Eq.~\ref{eq:target} into the two channels the model generates. The count $N_{ij}$ ranges from zero to thousands across cells, so we compress it with a logarithm:
\begin{equation}
\mathrm{logN}_{ij} = \frac{\log_{10}(N_{ij} + \alpha) - \log_{10}\alpha}{\beta},
\qquad
\mathrm{maxM}_{ij} = \frac{M^{\max}_{ij}}{\gamma},
\label{eq:encoding}
\end{equation}
where the offset $\alpha$ fixes an empty cell at zero and the divisors $\beta$ and $\gamma$ bring both channels near $[0, 1]$, the range commonly used in diffusion models to keep training stable.
QuakeGen learns the joint conditional distribution of these two channels over the grid,
\begin{equation}
(\mathrm{logN},\, \mathrm{maxM}) \;\sim\; p_\theta\big(\,\cdot \bigm| \mathbf{c} \big),
\label{eq:task}
\end{equation}
where the conditioning $\mathbf{c}$ gathers the seismicity recorded up to the anchor $t_0$ on the same grid and the forecast horizon $T$.

The spatiotemporal forecasting approach provides a flexible framework for modeling seismic sequence evolution. We apply this formulation in two settings, a global one for mainshock--aftershock sequences worldwide and a regional one for the daily seismicity of a single region. The two share the same architecture and differ in the grid geometry, the scaling constants $(\alpha, \beta, \gamma)$ of Eq.~\ref{eq:encoding}, and the conditioning fields (Table~\ref{tab:hyperparams}).
In the global setting, we choose the forecast anchor $t_0$ to be the time of a mainshock, the forecast domain to be centered on the epicenter, and the domain size to be proportional to the rupture length $L$.
Here we adopt the empirical Wells--Coppersmith relation to set the surface rupture length $\log_{10} L = -3.22 + 0.69\,M$ \citep{wellscoppersmith1994}, and set the grid half-extent to $2L$, clamped to $[16, 400]$ km, so the $64 \times 64$ grid spans a box between $32$ and $800$ km across.
Other scaling relations could serve equally well, since the goal is to have the same grid resolve a small M~$5$ sequence and a great M~$8$ rupture at proportionate resolution. 
In the regional setting, we choose the forecast anchor $t_0$ to be any time within the study period, fix the cell size at $1.25$ km, and allow random sampling of the grid domain within the study region. Here the count map is a daily rate, the cumulative count divided by the window length in days, so the network forecasts the same horizon-invariant quantity at every horizon.

The forecast field is generated to be consistent with the information available when the forecast is made, which the network receives as a set of conditioning fields $\mathbf{c}$ (Fig.~\ref{fig:quakegen}). In this work that information is the seismicity already observed, and each conditioning field is the frequency and maximum magnitude accumulated over a past time window ending at the anchor time $t_0$, the mainshock time in the global case or any chosen time in the regional case.
We stack several conditioning windows whose lengths grow geometrically, from the last few hours to the last month, so the conditioning carries both the slowly varying long-term background and the rapidly changing recent activity.
In the global setting we add a few short windows just after the mainshock, up to a conditioning cutoff $t_\mathrm{c}$, at finer temporal resolution than the pre-mainshock windows, because the aftershock rate changes fastest immediately after a mainshock and the first recorded aftershocks are the strongest predictor of the hours that follow.
In the regional setting we instead add two static maps of the long-term rate and largest magnitude over the entire training period, which give the network the spatial background of the region.
The forecast horizon $T$ is supplied to the network as an additional scalar input, so one trained model produces forecasts at any horizon in its range rather than a separate model per horizon.
In this work we condition only on past earthquakes, as ETAS does, but the same input channels could carry any other field that bears on the sequence, such as a mapped fault network, the mainshock moment tensor, and coseismic deformation from geodetic measurements.
Additional hyperparameters, such as window lengths, channel counts, and scaling constants, are collected in Table~\ref{tab:hyperparams}.

\begin{figure}[htbp]
  \centering
  \includegraphics[width=\textwidth]{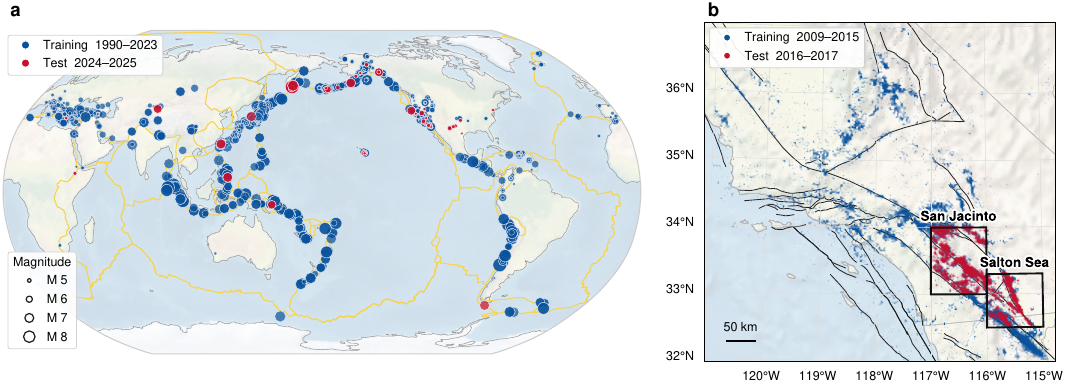}
  \caption{Earthquake catalogs for the two forecasting settings. (a) Global mainshocks ($M \ge 4.5$) from the USGS ComCat and ISC Bulletin, split into training events (blue) and held-out test events (red) and sized by magnitude. (b) Regional QTM template-matching catalog of southern California \citep{ross2019searching}, with mapped fault traces (black) and training (blue) and test (red) events. A single model is trained over the full region and evaluated on the Salton Sea and San Jacinto sub-regions (boxes) defined by the EarthquakeNPP benchmark \citep{stockman2025earthquakenpp}.}
  \label{fig:catalog}
\end{figure}

\subsection{Denoising diffusion model}
\label{sec:model}

A generative model learns the distribution of forecast fields conditioned on the observations, $p_\theta(\cdot \mid \mathbf{c})$ of Eq.~\ref{eq:task}, and produces a forecast by drawing samples from it.
Seismicity is inherently stochastic, so the same observed past can evolve into many different futures. A deterministic network trained to output a single field would return their average and blur the sharp fault-aligned structure of a real sequence, while a generative model allows each sample to be one realization of the coming seismicity, and many samples together form an ensemble whose spread reflects the forecast uncertainty.
We use the denoising diffusion model, the class of generative models behind recent image and weather generators \citep{ho2020denoising, song2021scorebased}, for earthquake forecasting in this work.
The denoising diffusion model uses two processes that run in opposite directions (Fig.~\ref{fig:quakegen}).
The forward process takes the forecast field $\mathbf{x}_0 = (\mathrm{logN}, \mathrm{maxM})$ of Eq.~\ref{eq:encoding} and corrupts it in many small steps, each adding a little Gaussian noise, until only featureless noise remains.
This corruption is fixed and involves no learning; after $i$ steps the field is a known blend of the clean field and standard Gaussian noise $\boldsymbol{\epsilon} \sim \mathcal{N}(\mathbf{0}, \mathbf{I})$,
\begin{equation}
\mathbf{x}_i = \sqrt{\bar\alpha_i}\,\mathbf{x}_0 + \sqrt{1 - \bar\alpha_i}\,\boldsymbol{\epsilon},
\label{eq:forward}
\end{equation}
where the weight $\bar\alpha_i$ follows a sigmoid schedule that falls from near one (almost clean) to near zero (almost pure noise).
The reverse process is what the network learns: given a noisy field and its noise level, it removes a small amount of noise to return a slightly cleaner field,
\begin{equation}
p_\theta(\mathbf{x}_{i-1}\mid \mathbf{x}_i, \mathbf{c}) = \mathcal{N}\!\big(\mathbf{x}_{i-1};\ \boldsymbol{\mu}_\theta(\mathbf{x}_i, i, \mathbf{c}),\ \sigma_i^2\mathbf{I}\big),
\label{eq:reverse}
\end{equation}
where $\mathbf{c}$ collects all the conditioning information, such as the observed-seismicity windows, the forecast horizon, and other fields (e.g., a fault map). The mean $\boldsymbol{\mu}_\theta$ is computed by a neural network that reads the noisy field $\mathbf{x}_i$, the diffusion step $i$, and the conditioning $\mathbf{c}$, while the variance $\sigma_i^2$ follows the noise schedule of the forward process.
For the denoising diffusion model, training reduces to a straightforward regression. We take a real observed field, pick a random noise level, add the corresponding noise, and train the network to recover what was added so that the step can be undone, minimizing the mean squared error of its prediction.
We use the $\mathbf{v}$-prediction target \citep{salimans2022progressive}, a fixed combination of the clean field and the added noise,
\begin{equation}
\mathbf{v}_i = \sqrt{\bar\alpha_i}\,\boldsymbol{\epsilon} - \sqrt{1 - \bar\alpha_i}\,\mathbf{x}_0,
\label{eq:vpred}
\end{equation}
which keeps the regression equally well-scaled at every noise level, from nearly clean to nearly pure noise.

The denoiser is a U-Net \citep{ronneberger2015unet}, a convolutional network with an encoder that downsamples the field through a few resolution levels and a decoder that upsamples it back, linked by skip connections that pass fine detail from each encoder level to the matching decoder level (Fig.~\ref{fig:quakegen}).
The same architecture has proved effective across a range of seismic tasks, including phase picking \citep{zhu2018phasenet, zhu2023phasenetdas} and waveform denoising \citep{zhu2019seismic}, on both seismometer and distributed acoustic sensing data.
The conditioning maps are stacked with the noisy field as extra input channels, so the network compares the field it is denoising against the observed seismicity at the same resolution.
The noise level controlled by the diffusion step and the forecast horizon each enter through a sinusoidal embedding. The two are combined and applied inside the network's residual blocks as a feature-wise scale-and-shift (FiLM) modulation \citep{perez2018film}.

Once the model is trained, generation runs the corruption in reverse. Starting from pure random noise, the denoiser removes a little noise at each step, with the conditioning $\mathbf{c}$ supplied throughout, until a clean field $\mathbf{x}_0$ remains (Fig.~\ref{fig:quakegen}).
That field is one draw from $p_\theta(\cdot \mid \mathbf{c})$ of Eq.~\ref{eq:task}, a forecast of the sequence conditioned on the seismicity already observed; a different initial noise field yields a different realization.

\subsection{Earthquake catalogs}

We assemble two datasets, one for each forecasting setting, a global catalog of mainshock--aftershock sequences and a regional catalog of continuous southern California seismicity (Fig.~\ref{fig:catalog}).

The global dataset is assembled from the USGS ANSS Comprehensive Catalog \citep{usgs2017comcat} and the ISC Bulletin \citep{isc2026bulletin} (Fig.~\ref{fig:catalog}a). We select mainshocks of M~$4.5$ and above and shallower than $30$ km to anchor the forecasts, while keeping their aftershocks at all magnitudes.
We apply a simple filtering that keeps only the largest event within a $100$ km, $8$-day space--time window, so that most aftershocks of a large event are not themselves selected as mainshocks.
Around each mainshock, we collect every cataloged earthquake within $\pm 768$ h and within the mainshock's forecast box, and discard sequences with fewer than $100$ aftershocks.
We set no completeness threshold and bin every cataloged earthquake, so the model sees the seismicity as it was actually recorded, including the well-known loss of small events in the hours after a large one, and learns to forecast through that incompleteness rather than around an imposed completeness magnitude.
When both agencies record the same earthquake, we keep both records as separate sequences rather than choosing between them, since the two catalogs differ slightly in magnitudes, locations, and detected aftershocks.
The resulting dataset contains $2{,}126$ mainshock sequences with magnitudes from $4.5$ to $9.1$. We train on the $2{,}046$ sequences of 1990--2023 and hold out the $80$ mainshocks of the most recent two years, 2024--2025, as a strictly prospective test set. The training and test sets contain $2.2$ million and $55{,}931$ events, respectively, within the forecast grid and the $\pm768$-hour window of each mainshock.

The regional dataset is the QTM template-matching catalog of southern California \citep{ross2019searching} (Fig.~\ref{fig:catalog}b). We train a single model over the full QTM region, drawing random $80 \times 80$ km windows concentrated on the fault network with the forecast anchor at random times, so the model sees many spatiotemporal settings.
We evaluate it on the Salton Sea and San Jacinto sub-regions defined by the EarthquakeNPP benchmark \citep{stockman2025earthquakenpp} (Fig.~\ref{fig:catalog}b). 
We apply the same completeness cut ($M_c = 1.0$) and train/test split as the benchmark, so that QuakeGen and the ETAS baseline are compared on identical data.
The processed catalog contains $127{,}789$ events of M~$1.0$ and above from 2008 through 2017. We draw training anchors from 2009--2015, a window containing $100{,}847$ events, and test on the held-out years 2016--2017. The Salton Sea and San Jacinto evaluation catalogs contain $45{,}570$ and $21{,}291$ events, of which $4{,}104$ and $4{,}400$ fall in the test years and are scored.

\subsection{Model training and inference}
\label{sec:traininfer}

Each training step draws a random sequence, a random forecast horizon, and a random noise level, and updates the network with the Adam optimizer (Table~\ref{tab:hyperparams}). We keep an exponential moving average of the weights for use at inference, which stabilizes diffusion sampling.
The horizon is drawn log-uniformly over each model's range so that all horizons are learned together by the single shared network.
We augment each training sample to enlarge the effective dataset and prevent the network from memorizing individual sequences \citep{zhu2020seismic}. The augmentations include random rotations and reflections, small magnitude jitter, a small spatial shift, and light blurring of the count maps. These transformations approximate the observational uncertainty in the catalogs, such as errors in event magnitudes and locations.

To produce a forecast we draw an ensemble of $K$ fields. Each field starts from an independent draw of random noise and is carried through the reverse process to a distinct sample, so the $K$ fields differ and together represent the forecast distribution.
We accelerate generation with the DDIM sampler \citep{song2021denoising}, which produces a field in $250$ steps instead of the $1000$ steps used in training at negligible loss of quality. DDIM sampling is deterministic, so the ensemble spread comes entirely from the independent initial-noise draws rather than from injected sampling noise.
We invert the encoding of Eq.~\ref{eq:encoding} to turn each generated field back into event counts and maximum magnitudes. We summarize the ensemble by its mean count field, which gives the expected number of aftershocks and their spatial pattern over the forecast window, and by an upper percentile of the maximum-magnitude field.
The global model forecasts each mainshock on its single box. The regional domain is larger than the network's $64 \times 64$ grid, so we forecast it as a set of overlapping $80$ km tiles and average them where they overlap.

We score every model with a common set of metrics (Supplementary Information).
Two come from the standard gridded point-process log-likelihood that the Collaboratory for the Study of Earthquake Predictability (CSEP) uses for prospective forecast testing, split into a temporal and a spatial part \citep{gneiting2007strictly, iturrieta2024evaluation, savran2022pycsep}.
The temporal log-likelihood (TLL) rewards forecasting the right event rate over the window while penalizing over-forecasting. 
The spatial information gain (IG) is the spatial log-likelihood of how a model places events, referenced to a spatially uniform forecast and reported in nats per earthquake. 
We also include metrics that compare the ensemble-mean predicted count field with the observed count field, each isolating a different error.
A productivity ratio $N_{\mathrm{pred}}/N_{\mathrm{obs}}$, ideal at one, compares the predicted and observed total number of events, i.e., whether a model over- or under-forecasts.
The Wasserstein-1 distance, reported in number of grid cells, measures how far the predicted activity is displaced from the observed and so isolates spatial mislocation from total count.
The spatial RMSE measures the cell-by-cell count error.
The mean-squared-error skill score (MSESS) is the fraction of the observed field's spatial variance the forecast explains.

\begin{table}[htbp]
  \centering
  \small \caption{Hyperparameters and design choices for the global (per-event, mainshock-anchored) and regional (fixed-region, day-anchored) models. $M$ denotes the mainshock magnitude. $L$ denotes the rupture length calculated by the empirical Wells--Coppersmith relation \citep{wellscoppersmith1994}. Rows that span both columns share the same value.}
  \label{tab:hyperparams}
  \begin{tabular}{@{}>{\raggedright\arraybackslash}p{0.27\textwidth}>{\raggedright\arraybackslash}p{0.34\textwidth}>{\raggedright\arraybackslash}p{0.31\textwidth}@{}}
    \toprule
	& Global & Regional \\
    \midrule
    \multicolumn{3}{@{}l}{\emph{Field representation}} \\
    Output channels        & \multicolumn{2}{l}{$\mathrm{logN}$ (count), $\mathrm{maxM}$ (maximum magnitude)} \\
    Grid                   & \multicolumn{2}{l}{$64 \times 64$} \\
    Grid half-extent       & $2L$, clamped to $[16, 400]$ km  & $40$ km, fixed \\
    Cell size              & $0.5$--$12.5$ km, based on  $M$ & $1.25$ km, fixed \\
    Count scale $\beta$    & $0.3\,M$ & $4.0$ \\
    Count offset $\alpha$  & $1$ & $0.01$ \\
    Magnitude scale $\gamma$ & $M$ & $6.0$ \\
    \addlinespace
    \multicolumn{3}{@{}l}{\emph{Conditioning}} \\
    Pre-anchor windows (start)  & $-768, -192, -48, -12, -3$ h & $-768, -384, -192, -96, -48, -24$ h \\
    Post-anchor windows (end)   & $0.1, 0.2, 0.4, 0.8, 1.6$ h & none \\
    Static history channels        & none & 2 (long-term $\log N$, $\max M$) \\
    \addlinespace
    \multicolumn{3}{@{}l}{\emph{Training}} \\
    Optimizer              & \multicolumn{2}{l}{Adam, learning rate $10^{-4}$, EMA decay $0.995$} \\
    Batch size             & \multicolumn{2}{l}{$128$} \\
    Steps                  & $5 \times 10^{5}$ & $2.5 \times 10^{5}$ \\
    Horizon sampling       & LogUniform$(0.1, 768)$ h & LogUniform$(24, 168)$ h \\
    \addlinespace
    \multicolumn{3}{@{}l}{\emph{Inference}} \\
    Sampler                & \multicolumn{2}{l}{DDIM, $250$ steps} \\
    Ensemble size $K$      & $100$ per mainshock & $50$ per tile \\
    Evaluation horizons    & $3, 12, 48, 168, 720$ h & $24, 48, 96, 168$ h \\
    \addlinespace
    \multicolumn{3}{@{}l}{\emph{Data construction}} \\
    Catalog                & USGS ComCat / ISC Bulletin & QTM \citep{ross2019searching} \\
    Mainshock selection    & $M \ge 4.5$, depth $\le 30$ km & fixed region \\
    Completeness cut       & none (raw catalog) & catalog $M_c = 1.0$ \\
    Train / test           & 1990--2023 / 2024--2025 & 2009--2015 / 2016--2017 \\
    \bottomrule
  \end{tabular}
\end{table}

\section{Results}

We evaluate QuakeGen in two settings, global aftershock forecasting and regional daily forecasting.
An aftershock sequence supplies strong conditioning, a clear mainshock and its early aftershocks, but demands that the forecast recover the fault-controlled spatial structure of the whole sequence from a few sparse hours of early data, while regional seismicity has a stable long-term rate along the fault system but a spatial pattern that shifts from one day to the next. 

\subsection{Global aftershock forecasting}

We first apply QuakeGen to global aftershock sequences.
For each mainshock we condition on the seismicity over the 768 hours before it and on the early aftershocks over the first $t_\mathrm{c} = 1.6$ hours after. We withhold direct source information, such as the moment tensor, a finite-fault model, or geodetic measurements, so the model must infer the rupture geometry from the early aftershocks alone. The two windows are chosen ad hoc to give the model enough context while keeping the task compatible with rapid post-earthquake response.
The model takes the forecast horizon as an input, so one trained model forecasts every lead time from hours to a month, with no retraining and no separate model per horizon. 
We train on horizons up to 768 hours, matching the pre-mainshock window, and score five fixed horizons from 3 to 720 hours to compare against the operational baseline, the USGS Reasenberg--Jones model (hereafter USGS-RJ) with the generic priors of \citet{page2016three} as implemented in the USGS Operational Aftershock Forecasting system \citep{barall2025oaf} (Supplementary Information).

\begin{figure}[htbp]
  \centering
  \includegraphics[width=\textwidth]{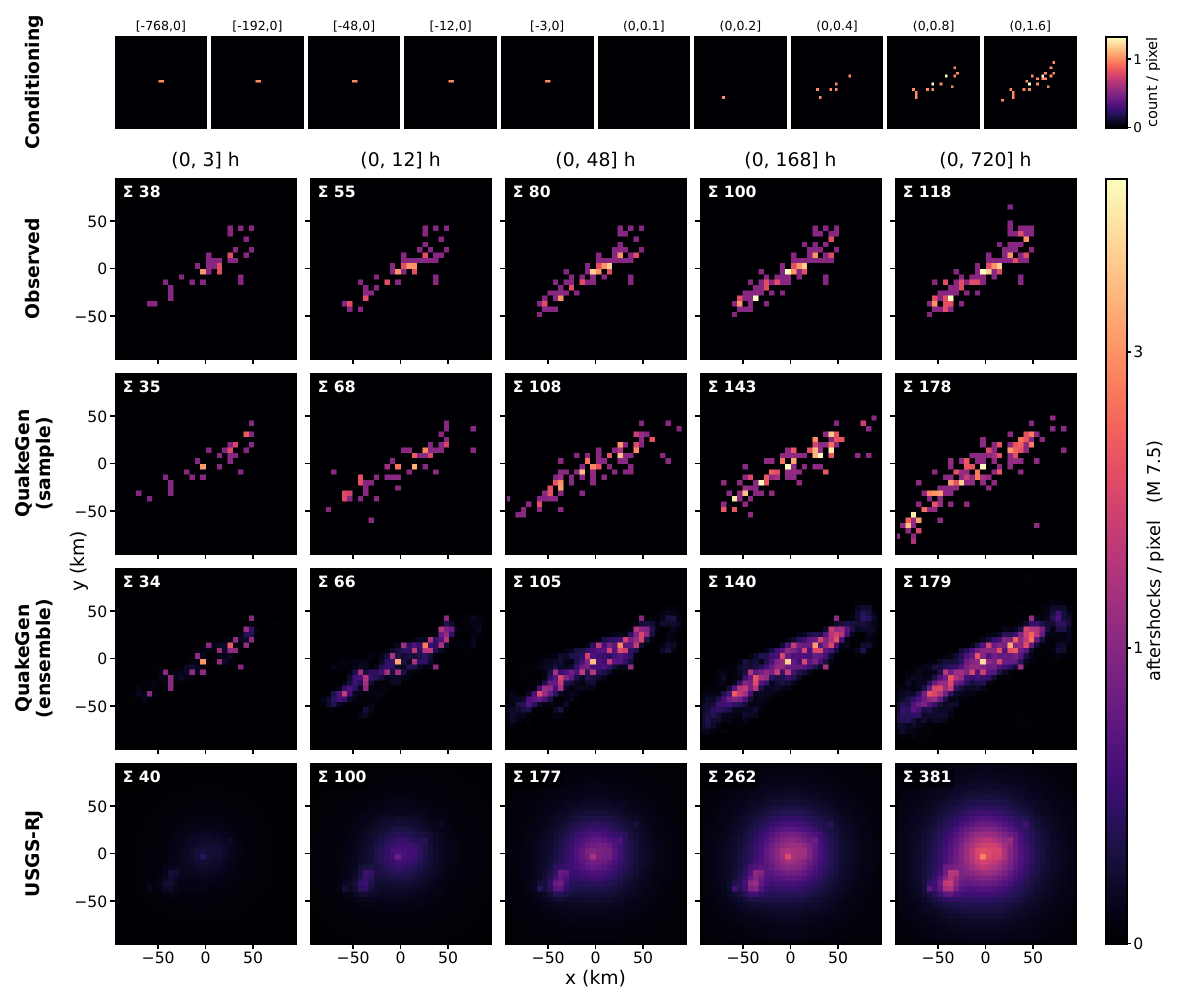}
  \caption{Global aftershock count forecast for the M~$7.5$ Noto Peninsula, Japan earthquake of 1 January 2024. Top row, the conditioning seismicity in ten nested windows, from 768 hours before the mainshock to the first 1.6 hours after. Lower rows, the observed count field, the median QuakeGen sample, the QuakeGen ensemble mean, and the operational USGS Reasenberg--Jones forecast, at forecast horizons of 3, 12, 48, 168, and 720 hours. Color gives aftershocks per pixel and $\Sigma$ the box total; axes are kilometers from the mainshock. The observed aftershocks trace a narrow northeast--southwest fault, which the median QuakeGen sample reproduces as discrete events and the ensemble mean as a smooth rate, while Reasenberg--Jones spreads a broad circular halo.}
  \label{fig:ex_noto}
\end{figure}

\begin{figure}[htbp]
  \centering
  \includegraphics[width=\textwidth]{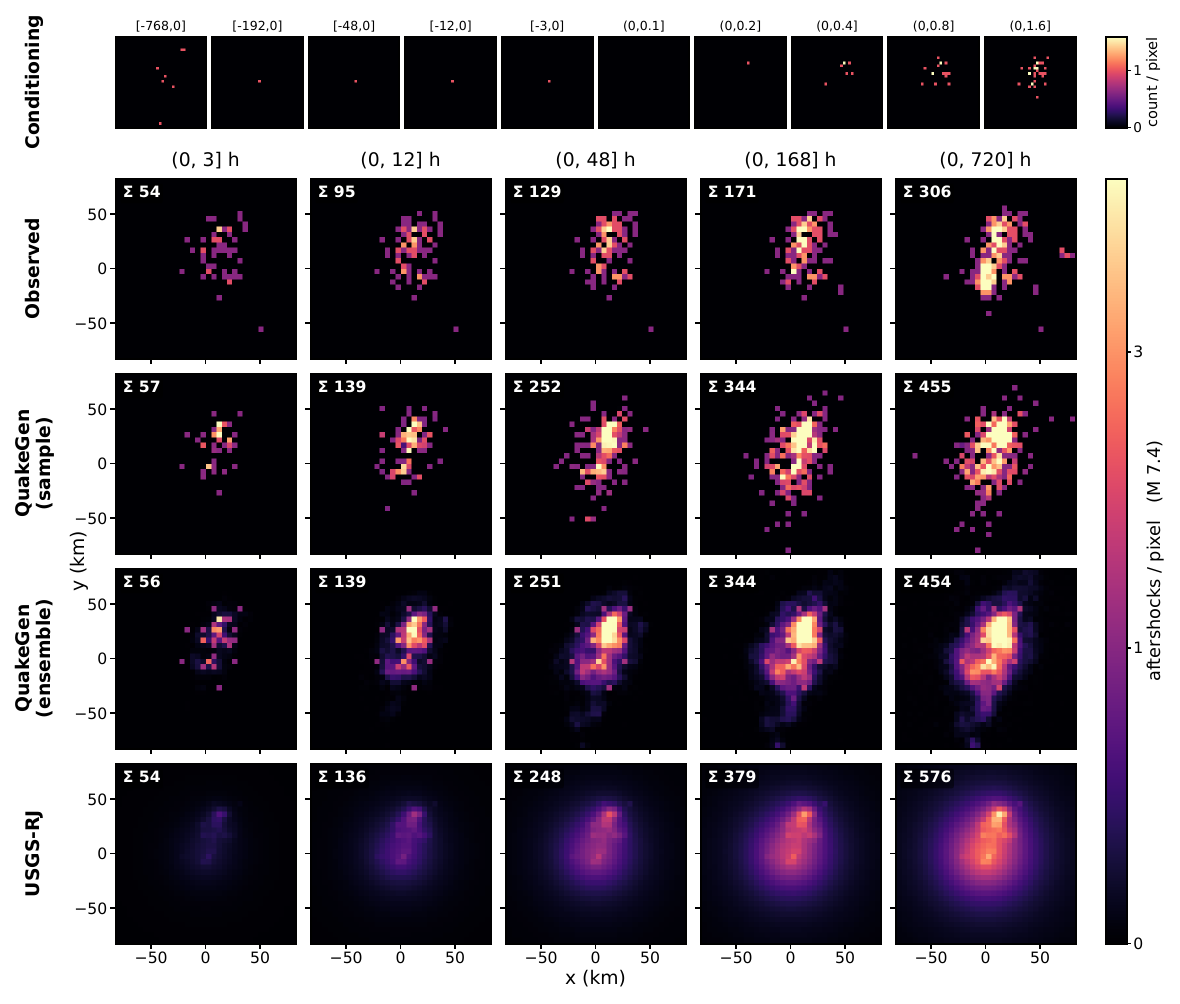}
  \caption{Global aftershock count forecast for the M~$7.4$ Hualien, Taiwan earthquake of 2 April 2024, in the layout of Fig.~\ref{fig:ex_noto}. The aftershocks form a north--south elongated cluster with two density maxima. The median QuakeGen sample and the ensemble mean reproduce both concentrations, while the Reasenberg--Jones forecast is a broad smooth maximum.}
  \label{fig:ex_taiwan}
\end{figure}

\begin{figure}[htbp]
  \centering
  \includegraphics[width=\textwidth]{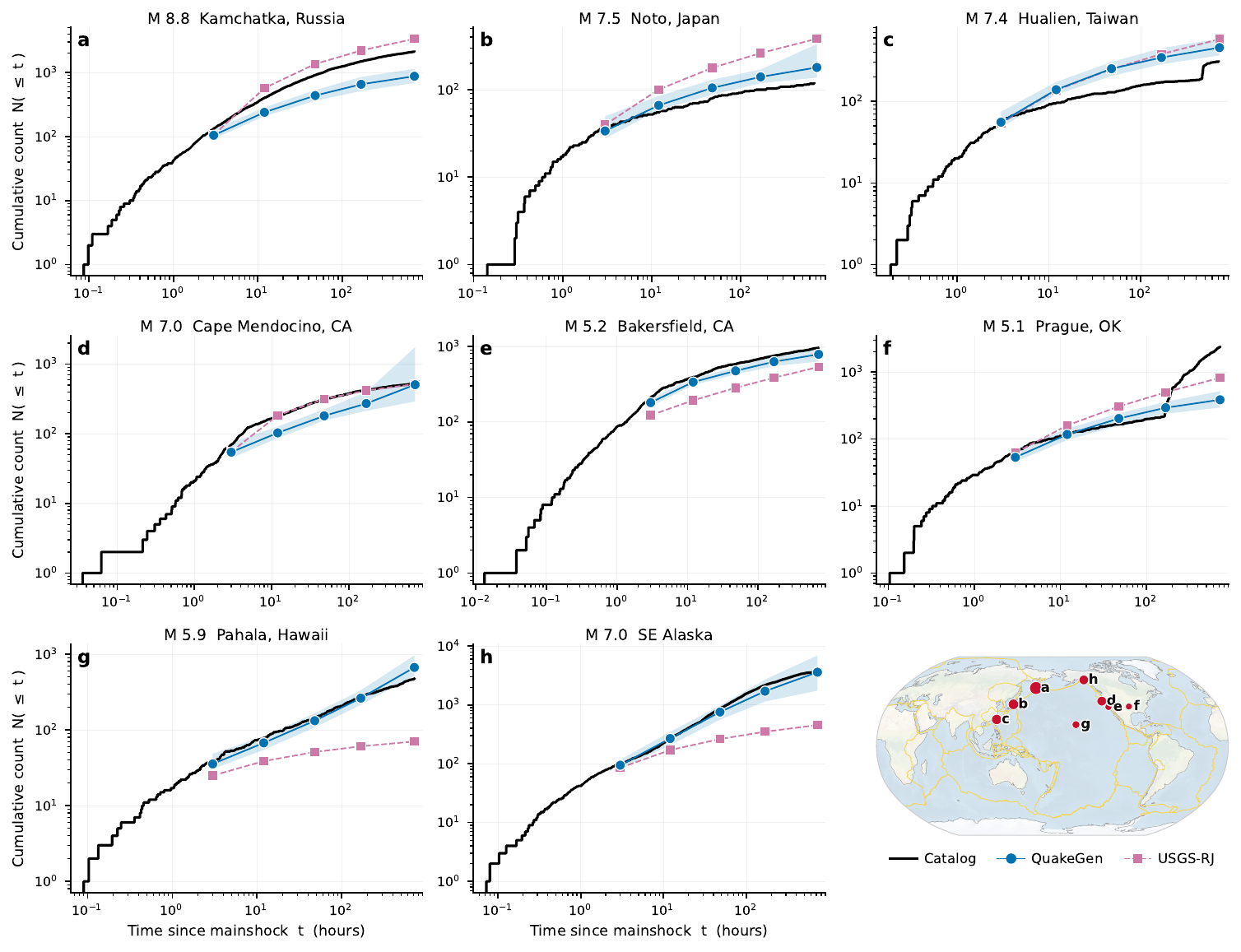}
  \caption{Cumulative aftershock count $N(\le t)$ for eight of the 2024--2025 test mainshocks (a--h), spanning M~$5.1$ to $8.8$ and a range of tectonic settings. Black, the observed catalog; blue, the QuakeGen ensemble-mean forecast at the five scored horizons, with the 95\% ensemble interval shaded; pink dashed, the USGS Reasenberg--Jones forecast. The last panel locates the events on a Pacific-centered map with plate boundaries.}
  \label{fig:temporal_global}
\end{figure}

\begin{figure}[htbp]
  \centering
  \includegraphics[width=\textwidth]{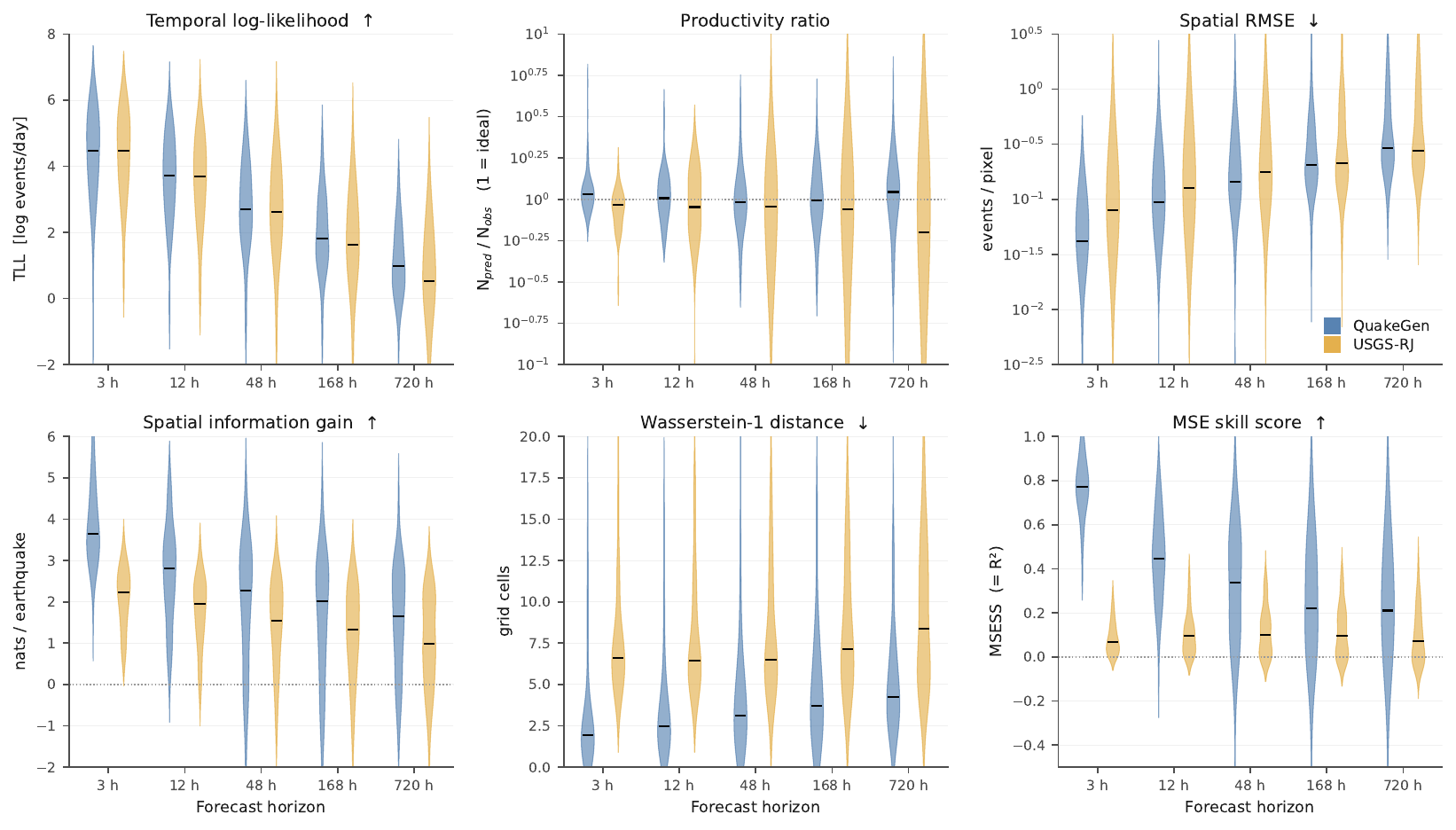}
  \caption{Global forecast skill, QuakeGen (blue) against the operational USGS Reasenberg--Jones model (orange), over the 80 mainshocks of 2024--2025. Each violin is the distribution across events of a per-event metric at a given horizon; the black tick marks the median, and the arrow in each panel title gives the preferred direction. Top row, temporal log-likelihood, productivity ratio $N_{\mathrm{pred}}/N_{\mathrm{obs}}$ (ideal at one), and spatial RMSE. Bottom row, spatial information gain, Wasserstein-1 distance in grid cells, and the mean-squared-error skill score.}
  \label{fig:global_metrics}
\end{figure}

Because a diffusion model learns a distribution rather than a single field, we draw an ensemble of $K=100$ realizations for each forecast and report both a representative median sample and the ensemble mean (Figs.~\ref{fig:ex_noto}, \ref{fig:ex_taiwan}).
USGS-RJ builds the Omori decay in time and the Gutenberg--Richter distribution in magnitude into its equations, so it satisfies these empirical laws by construction. QuakeGen instead learns from data alone, yet its forecasts reproduce the Omori-like decay of cumulative counts at every horizon from hours to a month, for mainshocks spanning several magnitude units, and across settings from tectonic to volcanic and induced sequences (Fig.~\ref{fig:temporal_global}).
QuakeGen places aftershocks more accurately in space than USGS-RJ fit to the same early data (Fig.~\ref{fig:global_metrics}; Table~\ref{tab:performance}). Across the 80 mainshocks of M~$4.5$ and above from 2024--2025 held out of training, it leads USGS-RJ on spatial information gain, Wasserstein-1 distance, and the mean-squared-error skill score at every horizon, and on RMSE at all horizons but the longest. 

This spatial advantage comes from structure that a radial kernel cannot represent. Real aftershocks do not spread symmetrically about the mainshock. They align with the ruptured fault and the adjacent lobes of increased Coulomb stress, an anisotropy QuakeGen learns directly from the early aftershocks.
Each generated sample of QuakeGen carries the granular texture of a real catalog, discrete events strung along the fault, while the ensemble mean gives the smooth underlying rate.
After the M~$7.5$ Noto Peninsula, Japan earthquake, the aftershocks extend roughly 150 km along a northeast--southwest trend, and a single QuakeGen sample reproduces that elongation while USGS-RJ spreads a broad circular halo (Fig.~\ref{fig:ex_noto}).
The M~$7.4$ Hualien, Taiwan sequence is more complex, a north--south elongated cluster with two density maxima, and QuakeGen reproduces both concentrations, whereas USGS-RJ produces one broad, smooth maximum (Fig.~\ref{fig:ex_taiwan}). 
Six additional test sequences, spanning settings from induced Oklahoma seismicity to subduction ruptures in Alaska and Kamchatka, are shown in Figs.~\ref{fig:si_freq_kamchatka}--\ref{fig:si_freq_alaska}.

\begin{figure}[htbp]
  \centering
  \includegraphics[width=\textwidth]{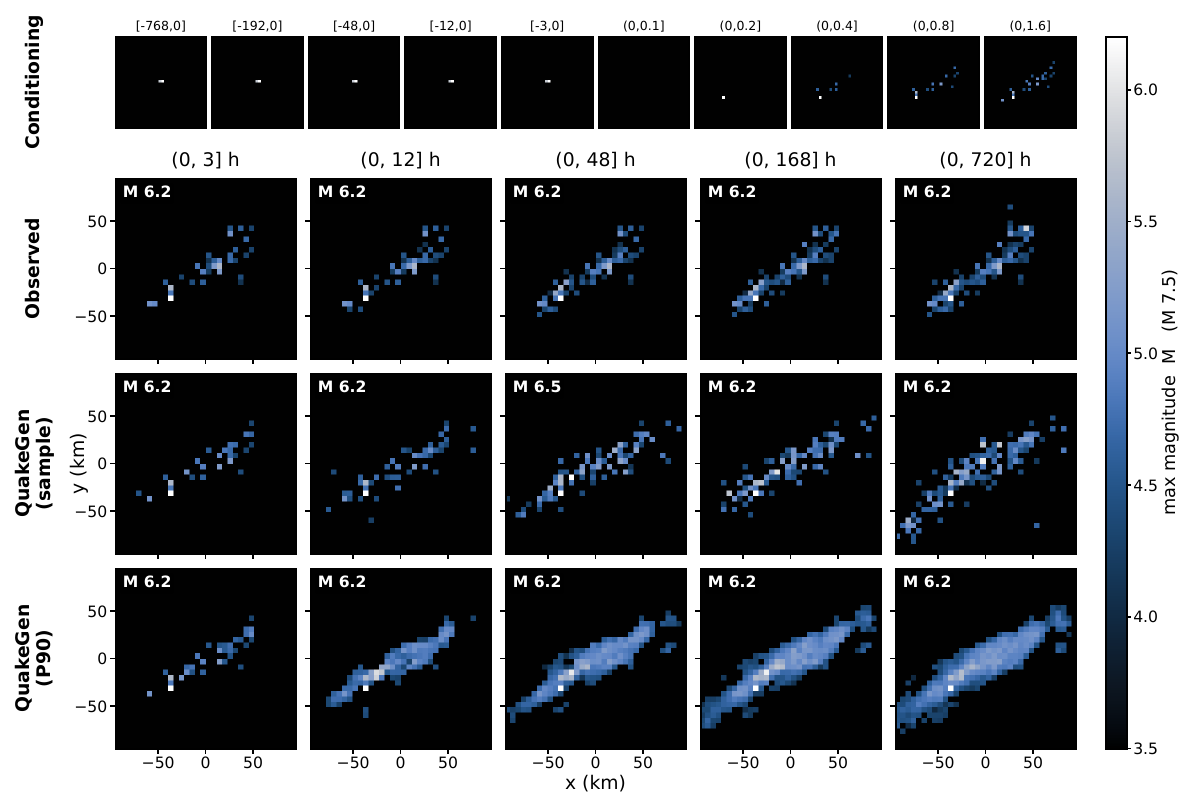}
  \caption{Maximum-magnitude field forecast for the Noto sequence, in the layout of Fig.~\ref{fig:ex_noto}. The conditioning row shows the maximum-magnitude channel; the lower rows are the observed field, the median QuakeGen sample, and the ensemble 90th percentile (P90), reported in place of the ensemble mean, which is biased low by empty cells. QuakeGen concentrates the largest expected aftershocks along the ruptured fault.}
  \label{fig:ex_mag}
\end{figure}

QuakeGen also conditions on and forecasts the largest magnitude expected in each cell. Forecasting the largest expected aftershock is usually cast as Bayesian extreme-value inference on the magnitude distribution \citep[e.g.,][]{shcherbakov2019forecasting}. QuakeGen instead produces this forecast as a spatial field, alongside the counts. Because the two channels are generated jointly, one realization states both how many aftershocks to expect in a cell and how large the largest is likely to be.
The ensemble-mean maximum magnitude is biased low, because cells with no forecast earthquake default to zero and pull the average down, so we report the 90th percentile (P90) across the ensemble instead.
The forecast concentrates the largest aftershocks on the ruptured fault, matching where it concentrates the counts (Fig.~\ref{fig:ex_mag}; Figs.~\ref{fig:si_mag_kamchatka}--\ref{fig:si_mag_alaska} for the other test sequences).
In addition to the frequency and magnitude channels, other channels could in principle carry any forecasting quantity for which training labels exist, such as expected ground-motion intensity \citep{ren2026learning}.

The largest earthquakes are the hardest case for both models. For the M~$8.8$ Kamchatka mainshock, the largest in the test period, QuakeGen recovers the central elongation of the several-hundred-kilometer rupture but not its full along-strike extent and undercounts the total, while USGS-RJ spreads a broad radial halo (Fig.~\ref{fig:si_freq_kamchatka}). Such large events are rare in the training record, and their first $t_\mathrm{c} = 1.6$ hours of aftershocks reveal little of the eventual productivity. Forecasting them well remains an open problem for data-driven and statistical models alike.

\subsection{Regional daily forecasting}

The same conditional generative framework applies beyond aftershock sequences. We next train QuakeGen on continuous regional seismicity and forecast the daily earthquake rate of southern California. This setting inverts the difficulty of the aftershock problem, with a relatively stable overall rate but a spatial pattern far harder to anticipate. We test it against locally tuned ETAS, the baseline that neural point processes have yet to surpass on the EarthquakeNPP benchmark \citep{stockman2025earthquakenpp}.

\begin{figure}[htbp]
  \centering
  \includegraphics[width=\textwidth]{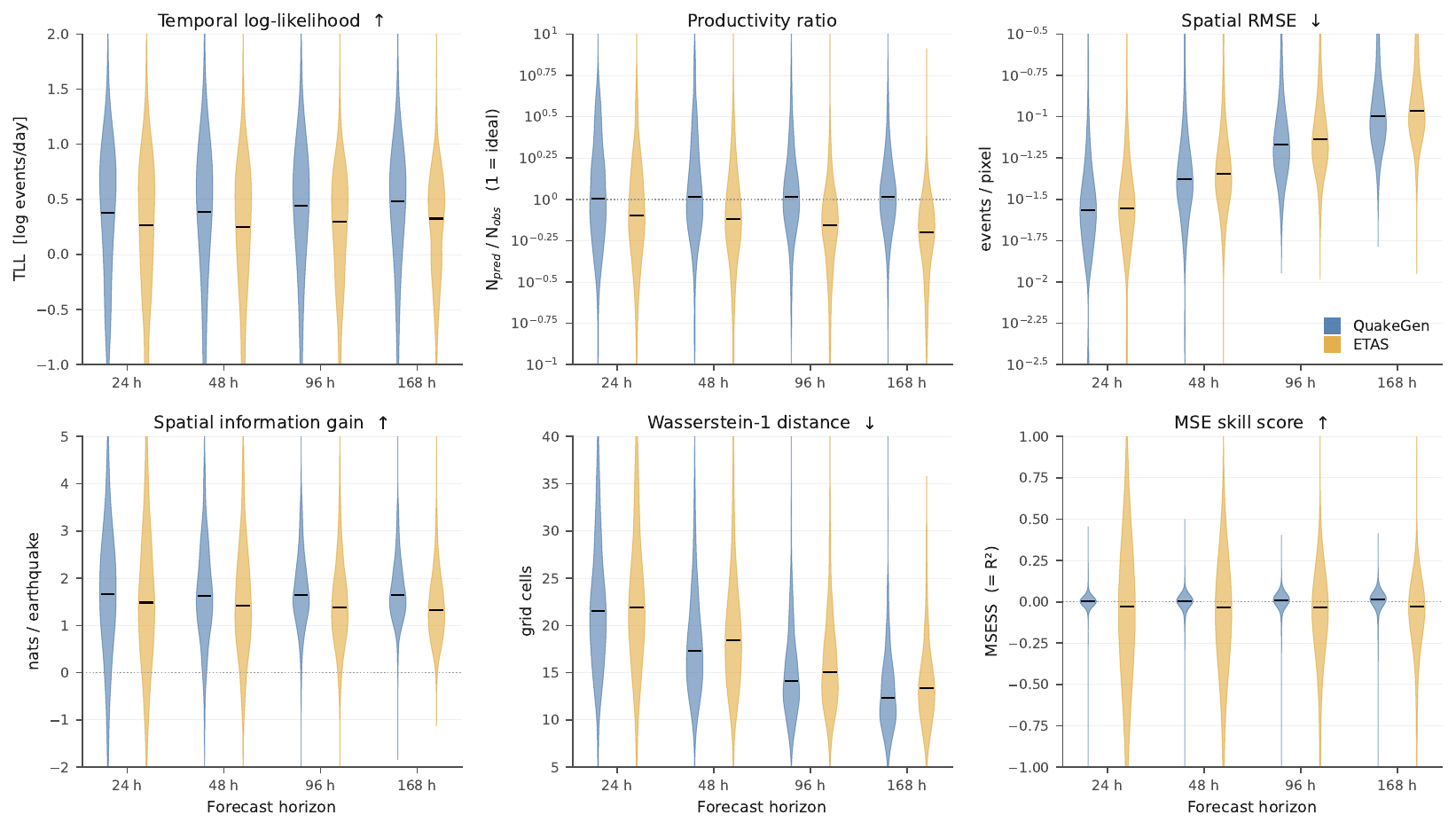}
  \caption{Regional forecast skill, QuakeGen (blue) against ETAS (orange), pooled over the Salton Sea and San Jacinto daily forecasts, 2016--2017. Panels and metrics as in Fig.~\ref{fig:global_metrics}.}
  \label{fig:regional_metrics}
\end{figure}

We train on the QTM template-matching catalog of southern California \citep{ross2019searching}, which resolves roughly ten times as many earthquakes as the standard network catalog and underlies the benchmark, and test on its 2016--2017 Salton Sea and San Jacinto sequences (Methods). ETAS is optimized separately for each region (Supplementary Information), while a single QuakeGen model is trained once over the entire QTM region and forecasts both regions, learning the regional seismicity directly from the catalog.
QuakeGen carries more spatial information per earthquake than the two tuned ETAS models at every horizon 
and a higher temporal log-likelihood, with the productivity ratio holding near one across horizons (Fig.~\ref{fig:regional_metrics}; Table~\ref{tab:performance}).

\begin{figure}[htbp]
  \centering
  \includegraphics[width=\textwidth]{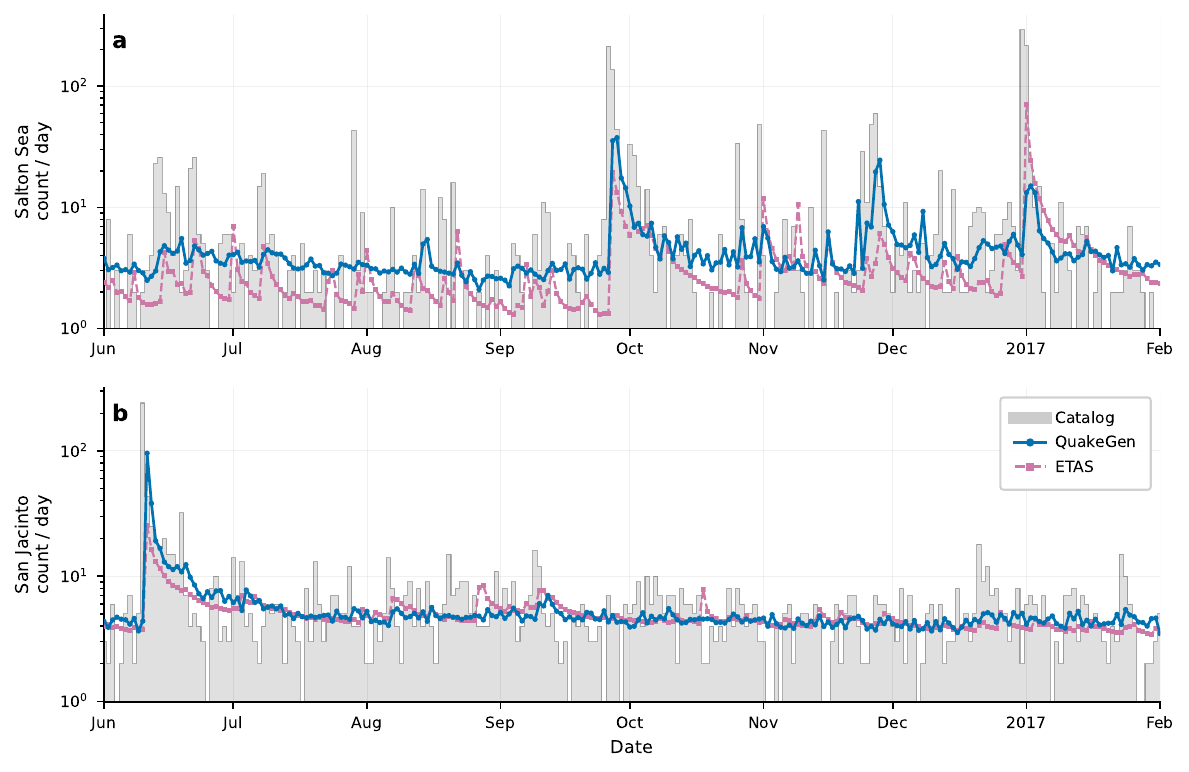}
  \caption{Regional daily forecasts for the Salton Sea (a) and San Jacinto (b) regions, June 2016 to February 2017. Gray, the observed daily counts; blue, QuakeGen; pink dashed, ETAS.}
  \label{fig:temporal_regional}
\end{figure}

The daily forecasts follow the observed counts through quiet background and sharp bursts alike (Fig.~\ref{fig:temporal_regional}). 
When activity surges, as in the September 2016 Salton Sea swarm and the June 2016 Borrego Springs sequence along the San Jacinto fault zone, both forecasts rise with the burst and decay as it subsides. 
We display the one-day forecast, but the model is trained across horizons from 24 to 168 hours (Fig.~\ref{fig:regional_metrics}).

\begin{figure}[htbp]
  \centering
  \includegraphics[width=\textwidth]{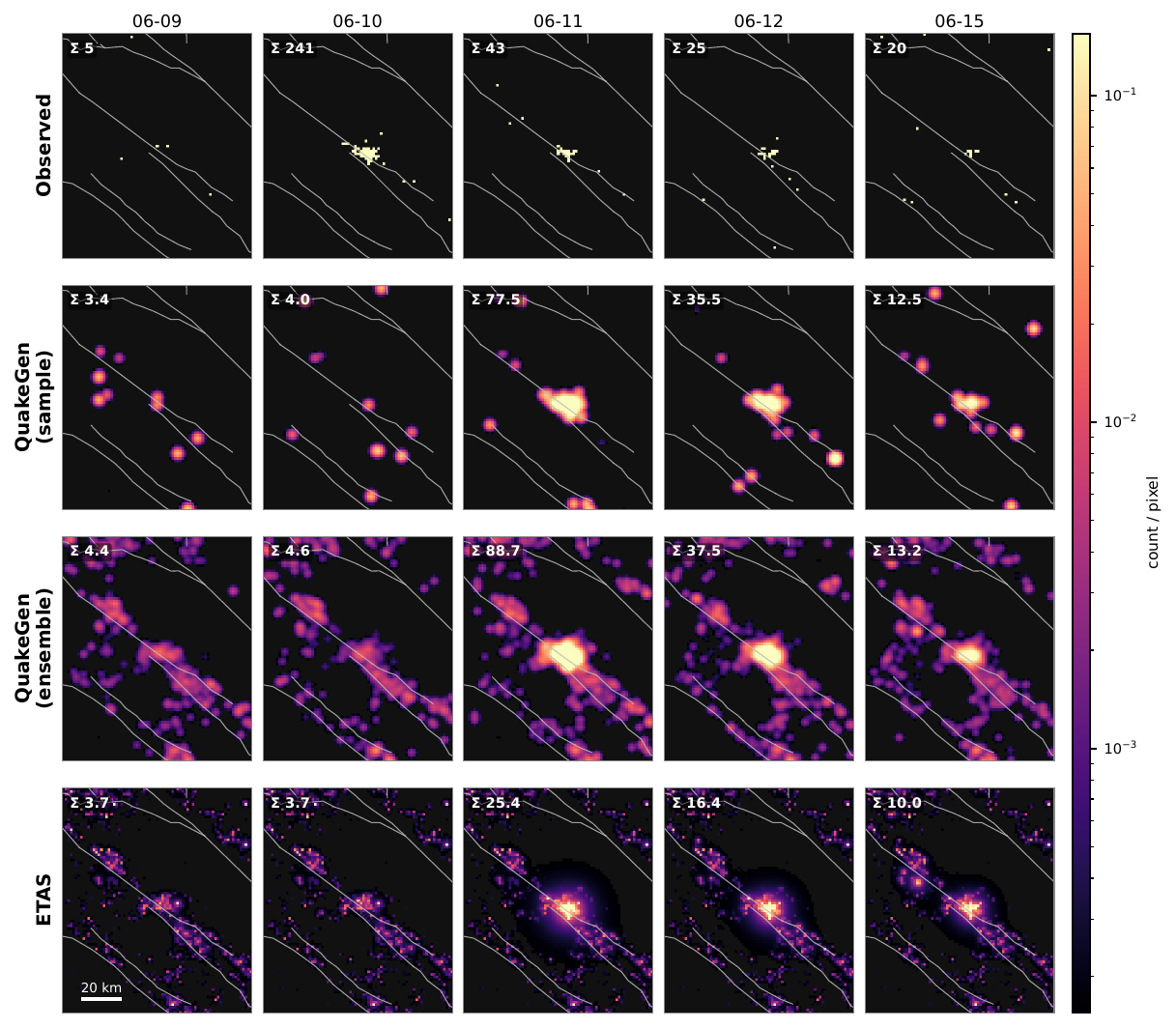}
  \caption{Daily 24-hour rate forecasts for the San Jacinto region during the June 2016 M~$5.2$ Borrego Springs sequence, on selected days spanning the sequence. Rows: observed daily count, the median QuakeGen sample, the QuakeGen ensemble mean, and ETAS; mapped fault traces are overlaid in gray. Both the QuakeGen ensemble and ETAS concentrate the forecast rate on the active fault strands.}
  \label{fig:regional_rate_sj}
\end{figure}

The spatial forecasts of the two models look broadly alike, concentrating the forecast rate on the active strands of the San Jacinto and southern San Andreas fault systems (Figs.~\ref{fig:regional_rate_sj}, \ref{fig:si_regional_rate}).
The single sample and the ensemble mean, however, diverge far more than in the aftershock setting. A single sample rarely coincides with the cells where events actually fall, as the precise location of individual events is difficult to anticipate day to day, while the ensemble mean traces the whole fault network.
QuakeGen reaches this fault-aligned pattern by learning the region's geometry from its training catalog, so each sample is one draw from the learned distribution of where activity is likely.
This wide spread reflects that the day-to-day location of background seismicity is far less predictable than that of aftershocks, which the mainshock and its early events already localize.
Even so, a single QuakeGen model, trained once over southern California, matches ETAS tuned separately for each region, pointing toward a single forecasting model for an entire fault system such as all of California.

\begin{table}[htbp]
  \centering
  \small \setlength{\tabcolsep}{5pt} \caption{Forecast skill by setting and horizon, reported as the median per-unit value of each metric (the black ticks in Figs.~\ref{fig:global_metrics} and \ref{fig:regional_metrics}). Arrows give the preferred direction, and the productivity ratio $N_{\mathrm{pred}}/N_{\mathrm{obs}}$ is ideal at one. Global forecasts are compared with the USGS Reasenberg--Jones model over the 80 test sequences of 2024--2025, and regional forecasts with ETAS over the 2016--2017 Salton Sea and San Jacinto test days.}
  \label{tab:performance}
  \begin{tabular}{llrrrrrr}
    \toprule
    Horizon & Model & TLL$\,\uparrow$ & $N_{\mathrm{pred}}/N_{\mathrm{obs}}$ & RMSE$\,\downarrow$ & IG$\,\uparrow$ & $W_1\,\downarrow$ & MSESS$\,\uparrow$ \\
    \midrule
    \multicolumn{8}{l}{\emph{Global} --- 80 mainshocks, 2024--2025} \\
    \addlinespace[2pt]
    $3$\,h   & QuakeGen & $4.48$ & $1.08$ & $0.041$ & $3.65$ & $1.93$ & $0.773$ \\
             & USGS-RJ  &$4.47$ & $0.93$ & $0.079$ & $2.23$ & $6.62$ & $0.067$ \\
    $12$\,h  & QuakeGen & $3.71$ & $1.02$ & $0.095$ & $2.81$ & $2.50$ & $0.447$ \\
             & USGS-RJ  &$3.70$ & $0.90$ & $0.126$ & $1.95$ & $6.45$ & $0.097$ \\
    $48$\,h  & QuakeGen & $2.70$ & $0.96$ & $0.145$ & $2.28$ & $3.14$ & $0.336$ \\
             & USGS-RJ  &$2.62$ & $0.91$ & $0.177$ & $1.54$ & $6.49$ & $0.099$ \\
    $168$\,h & QuakeGen & $1.82$ & $0.98$ & $0.204$ & $2.02$ & $3.69$ & $0.222$ \\
             & USGS-RJ  &$1.63$ & $0.87$ & $0.213$ & $1.33$ & $7.14$ & $0.095$ \\
    $720$\,h & QuakeGen & $0.99$ & $1.11$ & $0.294$ & $1.64$ & $4.26$ & $0.211$ \\
             & USGS-RJ  &$0.53$ & $0.63$ & $0.275$ & $0.99$ & $8.35$ & $0.070$ \\
    \midrule
    \multicolumn{8}{l}{\emph{Regional} --- Salton Sea $+$ San Jacinto, 2016--2017} \\
    \addlinespace[2pt]
    $24$\,h  & QuakeGen & $0.38$ & $1.01$ & $0.027$ & $1.66$ & $21.51$ & $0.003$ \\
             & ETAS     & $0.27$ & $0.80$ & $0.028$ & $1.49$ & $21.92$ & $-0.029$ \\
    $48$\,h  & QuakeGen & $0.38$ & $1.03$ & $0.042$ & $1.62$ & $17.35$ & $0.005$ \\
             & ETAS     & $0.25$ & $0.76$ & $0.045$ & $1.42$ & $18.44$ & $-0.036$ \\
    $96$\,h  & QuakeGen & $0.44$ & $1.03$ & $0.068$ & $1.65$ & $14.10$ & $0.010$ \\
             & ETAS     & $0.30$ & $0.70$ & $0.073$ & $1.38$ & $15.06$ & $-0.034$ \\
    $168$\,h & QuakeGen & $0.48$ & $1.04$ & $0.101$ & $1.64$ & $12.34$ & $0.015$ \\
             & ETAS     & $0.33$ & $0.63$ & $0.108$ & $1.33$ & $13.33$ & $-0.028$ \\
    \bottomrule
  \end{tabular}
\end{table}

\subsection{Generalization across catalogs}
\label{sec:generalization}

\begin{figure}[htbp]
  \centering
  \includegraphics[width=\textwidth]{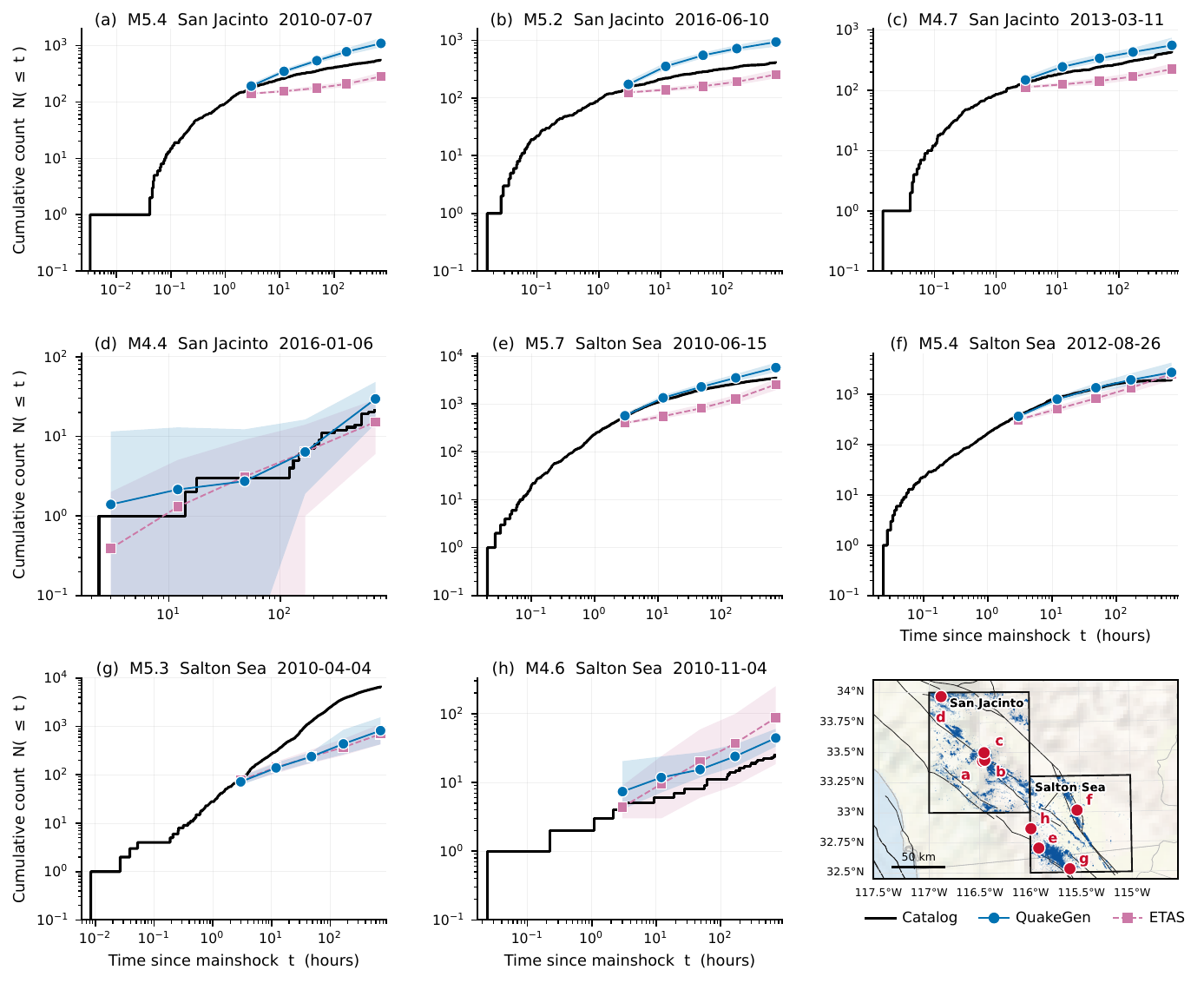}
  \caption{Cumulative aftershock count $N(\le t)$ for the four largest mainshock sequences in each of the QTM San Jacinto (a--d) and Salton Sea (e--h) regions, forecast by the global model without retraining and ordered by magnitude. Black, the observed catalog; blue, the QuakeGen ensemble mean at the five scored horizons, with its 95\% interval shaded; pink dashed, the branching ETAS simulation. The last panel locates the events within the two regional boxes.}
  \label{fig:qtm_temporal}
\end{figure}

\begin{figure}[htbp]
  \centering
  \includegraphics[width=\textwidth]{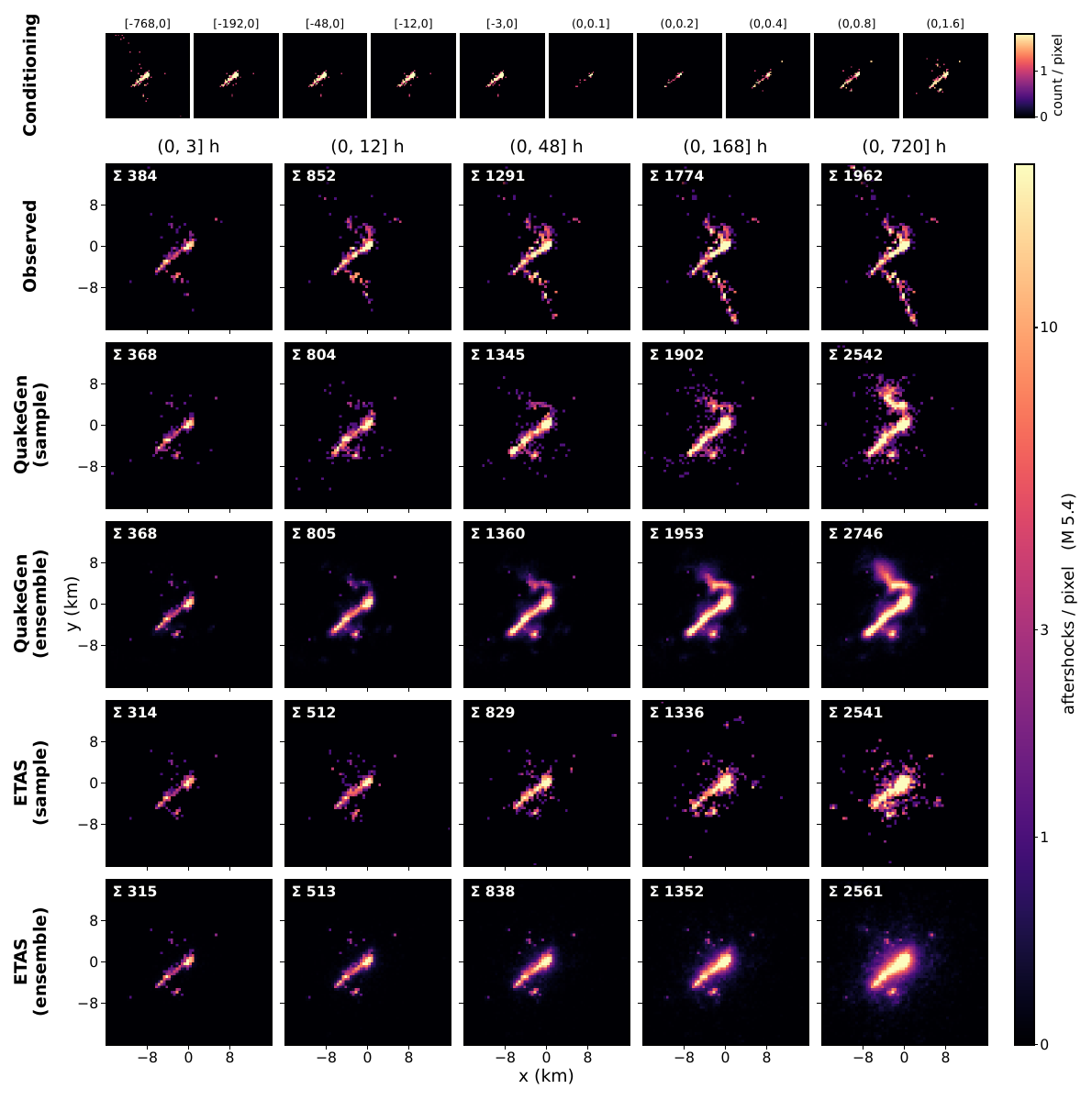}
  \caption{Global model applied to the M~$5.4$ Salton Sea (Brawley) swarm of 26 August 2012, in the layout of Fig.~\ref{fig:ex_noto}, with the region-specific branching ETAS simulation in place of Reasenberg--Jones.}
  \label{fig:qtm_example}
\end{figure}

The global and regional evaluations test generalization in time, on years held out of each training catalog.
The catalogs themselves are also improving, as deep-learning and template-matching pipelines deliver enhanced catalogs with up to ten times more small earthquakes. 
The added small events trace the faults and the early evolution of each sequence in finer detail, supplying the spatiotemporal structure for data-driven forecasting \citep{beroza2021machine}.
To explore this potential, we apply the global model, trained on the routine network catalogs, to the enhanced QTM catalog \citep{ross2019searching} without retraining, and forecast the four largest mainshock sequences in each of its Salton Sea and San Jacinto regions (M~$4.4$ to M~$5.7$).
We condition on the same windows as in the global setting and draw an ensemble of $K = 100$ realizations. 
For the baseline we simulate the region-specific ETAS as a branching process, in which every event triggers its own simulated aftershocks over successive generations \citep{mizrahi2021embracing}. We draw $K = 100$ such stochastic catalog continuations, so the two models start from the same conditioning seismicity and are compared as predictive distributions over both counts and magnitudes (Supplementary Information).
Across the eight sequences, the global model matches the ETAS models tuned to each region. Its cumulative counts track the observed totals as closely as the ETAS continuations (Fig.~\ref{fig:qtm_temporal}), and its samples reproduce the fault-aligned geometry of the sequences (Fig.~\ref{fig:qtm_example}; Figs.~\ref{fig:si_qtm_1}--\ref{fig:si_qtm_7}).
The model was never trained on an enhanced catalog, yet it performs similarly to an ETAS model fitted directly to the QTM data, suggesting that QuakeGen has learned how aftershock sequences evolve in space and time rather than the statistics of a single catalog.
Training on the enhanced catalogs now accumulating worldwide would continue to sharpen the forecasts of QuakeGen and future data-driven models.

\section{Discussion}

QuakeGen forecasts an earthquake sequence as an evolving spatiotemporal field using diffusion models. It learns the distribution of the whole field and returns a probabilistic forecast of where events fall, how fast their rate decays, and the largest magnitude expected. 
This is a different object from a point process, which prescribes how each event raises the rate of subsequent events and builds the forecast one event at a time. Neural point processes replace the prescribed triggering function with a learned one but keep that event-by-event view, and their gains over a well-tuned ETAS come mainly from better handling of catalog incompleteness rather than from sharper spatial forecasting \citep{stockman2023forecasting, stockman2025earthquakenpp}; ETAS itself has since been extended to model this incompleteness directly \citep{hainzl2024aftershock}.
Generating the spatiotemporal field allows the model to capture the fault-aligned anisotropic pattern that an isotropic kernel cannot express. It infers how a sequence will evolve by conditioning on dense fields of background seismicity and early aftershocks.
On the global sequences it outperforms the Reasenberg--Jones model that underlies public aftershock forecasting. On the densely recorded California sequences it matches the well-tuned ETAS benchmark. These results demonstrate that conditional generative modeling of spatiotemporal fields is a competitive alternative to the point-process framework that earthquake forecasting has rested on.

The advantage of the data-driven approach is that it learns the link between a sequence's past and its future from data rather than prescribing it in advance.
ETAS prescribes an Omori decay and an isotropic kernel and infers a set of controlling parameters (Tables~\ref{tab:etasparams} and \ref{tab:rj-generic}); physics-based models instead fix how the Coulomb stress change maps to the seismicity rate.
QuakeGen imposes no fixed spatiotemporal form, reaches the fault-aligned geometry directly from the catalog, and tracks a productivity that varies widely from sequence to sequence.
The advantage is evident on the global sequences, where each forecast is conditioned on a mainshock and only its first aftershocks. With so few events recorded, an isotropic kernel can draw only a near-radial halo around the mainshock, and no single ETAS calibration covers the range of tectonic settings globally. The advantage is smaller in the regional setting, where the dense catalog already lets the superposed ETAS kernels trace the fault network, so ETAS captures the fault-aligned pattern about as well as QuakeGen and the improvement is modest.

Because QuakeGen models seismicity as an evolving spatiotemporal field rather than a mainshock-triggered cascade, it is not tied to the mainshock--aftershock setting. Many sequences are driven by processes other than a single mainshock, while the seismicity pattern itself encodes the underlying forcing. 
Swarms are commonly driven by fluid migration or aseismic fault creep, and injection-induced sequences are triggered by the fluid pumped at a well. In both, the activity migrates through space and time as the driver evolves \citep[e.g.,][]{vidale2006survey, lohman2007earthquake, ross2020fault, ellsworth2013injection}. 
The same spatiotemporal learning framework can track this migration from the observed seismicity by implicitly modeling the underlying driver without an explicit physical model, while ETAS requires purpose-built, non-stationary formulations for each such case \citep[e.g.,][]{llenos2009modeling, llenos2013modeling, kumazawa2024nonstationary}.
Beyond producing forecasts, the trained model offers a way to study the driven sequences. Because the forecast depends directly on the observed seismicity, perturbing that input and measuring the response quantifies the forecast's sensitivity to the early spatiotemporal patterns.

QuakeGen can also condition on other measured fields that bear on a sequence. Aftershocks concentrate on and near the ruptured fault, and forecasts sharpen when distance is measured from a finite-fault model rather than a point source \citep{cattania2018forecasting}. The natural fields to add are therefore the mapped fault network \citep{plesch2007community, field2014uniform}, as in UCERF3-ETAS \citep{field2017ucerf3, field2021improvements}, and the rupture geometry from a finite-fault or moment-tensor inversion \citep{hartzell1983inversion, ekstrom2012global}, which set the static Coulomb stress change that drives the early aftershocks and has itself been used to forecast their spatial evolution \citep{toda2005forecasting} and to track stress changes through successive earthquake cycles \citep{freed2007implications, pollitz2008stress}. 
Surface deformation from InSAR and GPS is another field worth adding, since it records the postseismic afterslip, poroelastic rebound, and viscoelastic relaxation that redistribute the coseismic stress and reshape the aftershock sequence \citep{perfettini2004postseismic, burgmann2008rheology}. When a sequence has an external driver that is directly measured, such as the fluid volume injected at a well, that record can enter as a conditioning field in the same way \citep[e.g.,][]{ellsworth2013injection, shapiro2010seismogenic, langenbruch2016how, kim2023stress, schultz2026forecasting}. Each of these is measured routinely yet used unevenly in forecasting, and a single model that ingests them together, in the way machine learning across the geosciences increasingly fuses heterogeneous data, would make aftershock forecasting a unified problem rather than a separate method per observation.

The framework's central limitation is its reliance on large training datasets. The largest earthquakes are the ones a forecast must get right and the rarest in the record, so the model trains on many small sequences and few great ones and is least constrained where it matters most, as the Kamchatka sequence showed. Physics-based simulations offer one way forward. Multicycle simulators generate large synthetic catalogs of great earthquakes and their aftershocks \citep{dieterich2010earthquake, shaw2018physics}, and dynamic rupture models reproduce the rupture of individual great earthquakes \citep{kozdon2013rupture, jia2023complex}; both could augment the few observed examples at the top of the magnitude range. The enhanced catalogs that machine learning now produces for monitoring offer another direction, supplying far more of the fine spatiotemporal structure the model learns from, just as the dense QTM catalog already underlies the fault-aligned skill of the regional model.
Other limitations lie in the current implementation rather than the conditional generative framework. The model forecasts seismicity in map view, collapsing the depth distribution of the events. Extending the fields and the U-Net to three dimensions would recover the depth distribution at a higher computational cost. Although the model forecasts any horizon, it does not yet update its conditioning as new events arrive. Running it autoregressively, re-conditioning on the seismicity observed so far, as the regional daily model already does, would let it track a sequence continuously and better suit operational use.

Aftershock forecasting is tractable because the sequence is an abundant and evolving record whose early activity constrains what follows. 
However, predicting an individual mainshock from seismicity data is far harder, for the opposite reason, that little clear precursory signal exists to condition on \citep{geller1997critical, cicerone2009systematic, jordan2011operational}.
The seismic-gap hypothesis failed prospective tests \citep{kagan1991seismic}, and the densely instrumented Parkfield experiment recorded its anticipated earthquake a decade late and without detected precursors \citep{bakun2005implications}. The nucleation processes that might produce such a signal, whether cascading foreshocks, aseismic slip, or fluid flow, remain debated \citep{helmstetter2003foreshocks, kato2021generation, peng2024physical}. 
In this work, QuakeGen forecasts only the tractable part, the days and weeks after a damaging earthquake. As its conditioning expands, the same framework could reach from aftershocks toward the fainter, sparser signals of foreshocks, precursory slip, and other potential precursory processes \citep[e.g.,][]{kato2012propagation, bouchon2013long, trugman2019pervasive, gulia2019realtime, bletery2023precursory, karimpouli2026preparatory} and, ultimately, toward forecasting seismic sequences more generally.

\section{Conclusions}
QuakeGen recasts aftershock forecasting as conditional generation of an evolving spatiotemporal field, using a diffusion model to forecast the fields of aftershock rate and maximum magnitude from the seismicity already observed.
Unlike the empirical laws that forecasting has long relied on, it prescribes no fixed decay or kernel; the Omori-like decay, the sequence-to-sequence variation in productivity, and the fault-aligned geometry of each sequence are learned from the catalog alone.
Trained once and applied without refitting for each sequence or region, it outperforms the operational Reasenberg--Jones model on global aftershock sequences and matches the region-specific ETAS baseline that neural point-process models have yet to surpass on the regional daily benchmark.
QuakeGen could also condition on the fault-geometry, deformation, and stress fields that become available after a large earthquake and extend beyond mainshock--aftershock sequences to swarms and foreshocks.
The data-driven diffusion model underlying QuakeGen sharpens as the data grow richer, giving earthquake forecasting the same capacity for improvement that has transformed prediction in fields as diverse as weather forecasting and protein structure.

\section*{Data and code availability}

The global catalogs are public through the USGS and ISC FDSN web services. The regional QTM catalog \citep{ross2019searching} is distributed by the Southern California Earthquake Data Center and packaged by the EarthquakeNPP benchmark \citep{stockman2025earthquakenpp}.
The Reasenberg--Jones baseline uses the USGS \texttt{opensha-oaf} code \citep{barall2025oaf} (\url{https://code.usgs.gov/esc/oaf/opensha-oaf}).
The ETAS baseline uses the implementation of \citet{mizrahi2021embracing} (\url{https://github.com/lmizrahi/etas}), and forecasts are scored with the gridded log-likelihood and field metrics, following CSEP forecast-testing practice \citep{savran2022pycsep}.
QuakeGen source code and model weights are at \url{https://ai4eps.github.io/QuakeGen/}.
Forecast outputs are archived at \url{https://osf.io/q7hax}.

\section*{Acknowledgements}

This work was supported by the U.S. Department of Energy, Office of Science, Basic Energy Sciences, under Award DE-SC0026120. This research used the Lawrencium computational cluster provided by the IT Division at Lawrence Berkeley National Laboratory (supported by the Director, Office of Science, Office of Basic Energy Sciences, of the U.S. Department of Energy under Contract No. DE-AC02-05CH11231).

This work began during the days when my father, Jinquan Chen (1963.2.19--2026.5.26), was fighting cancer. Even in the late stages of his illness, he constantly encouraged me to focus on my studies, which gave me the motivation to finish this work. In his final days in the hospital, he still took care of himself as much as he could, not wanting to take up my time. He repeatedly told me not to blame myself for not making the right decisions about his treatment.
My father spent his entire life as a math teacher at a village middle school. Seeing many people come to bid him farewell, I understood that the life he chose was rewarding enough. As a teacher, a father, and a friend, he gave me everything he could. It was his example and guidance that shaped the person I am today.

\ifXeTeX
这个工作开始于我爸爸陈金泉老师(1963.2.19--2026.5.26)与癌症抗争的日子里。即使癌症晚期，他仍一再叮嘱我安心自己的工作，这成为我完成这个工作的最大动力。
在生命的最后一段时光，他仍在医院尽力料理自己的起居，不愿占用我的时间; 一遍遍安慰我不要为治疗方案的选择自责，这辈子最重要的事情都已经完成，没有遗憾。
爸爸一生都是乡村中学的一名数学老师。看到那么多人来送他最后一程，我才明白他所选择的这份平凡，已经足够。作为老师，父亲，朋友，他无条件给予我的一切，我已再没有机会报答, 是他的言传身教才让我成为今天的我。
\fi

\section*{Declaration of AI use}

During the preparation of this work, the author used Claude (Anthropic) to assist with data collection, code development (reproducing the baseline models for the benchmarks), results visualization, and language editing and proofreading. The author reviewed and edited all outputs and takes responsibility for the scientific content, experimental design, analysis, and conclusions.

\clearpage
\section*{Supplementary Information}
\setcounter{section}{0}
\setcounter{equation}{0}
\setcounter{figure}{0}
\setcounter{table}{0}
\renewcommand{\thesection}{S\arabic{section}}
\renewcommand{\thesubsection}{S\arabic{section}.\arabic{subsection}}
\renewcommand{\theequation}{S\arabic{equation}}
\renewcommand{\thetable}{S\arabic{table}}
\renewcommand{\thefigure}{S\arabic{figure}}

\section{Baseline models}
\label{si:baselines}

We compare QuakeGen against the operational models used in each setting, a locally tuned ETAS model regionally and the Bayesian Reasenberg--Jones model globally, where no standard global ETAS calibration exists. Both are gridded onto the same cells as QuakeGen and scored by the metrics of Section~\ref{si:eval}.

\subsection{Regional ETAS baseline}
\label{si:etas}

The regional baseline is the ETAS model \citep{ogata1988statistical} with the parameters fitted for each region by the EarthquakeNPP benchmark \citep{mizrahi2021embracing, stockman2025earthquakenpp}, evaluated as a rate map on QuakeGen's $1.25$ km grid. Its space-time conditional intensity is
\begin{equation}
\lambda(t, \mathbf{x}) = \mu
  + \sum_{t_e < t} k(M_e)\, g(t - t_e)\, f(|\mathbf{x} - \mathbf{x}_e|;\, M_e),
\end{equation}
with a uniform background rate $\mu$, productivity $k(M) = k_0 \exp\!\big(a(M - M_c)\big)$, a modified-Omori temporal kernel with an exponential taper $g(\tau) = (\tau + c)^{-(1+\omega)} \exp(-\tau/\tau_d)$, and a power-law spatial kernel $f(r; M) = \big(r^2 + d \exp\!\big(\gamma(M - M_c)\big)\big)^{-(1+\rho)}$. The fitted productivity $k_0$ absorbs the spatial kernel's normalization. The expected number of events in cell $(i,j)$ over a forecast window $(t_0, t_0 + T_h)$, for forecast anchor $t_0$ and horizon $T_h$, is the intensity, conditioned on the pre-anchor history, integrated over the cell and the window,
\begin{equation}
N^{\mathrm{ETAS}}_{ij} = |\Omega|\left[
  \mu\,T_h
  + \sum_{t_e < t_0} k(M_e)\, f(|\mathbf{x}_{ij} - \mathbf{x}_e|;\, M_e)
    \int_{t_0 - t_e}^{t_0 + T_h - t_e} g(\tau)\, d\tau
\right],
\end{equation}
where $|\Omega|$ is the cell area, the spatial kernel is evaluated at the cell center $\mathbf{x}_{ij}$, and only events before the anchor ($t_e < t_0$) contribute. The parameters $(\mu, k_0, a, c, \omega, \tau_d, d, \gamma, \rho)$ and the completeness $M_c = 1.0$ are the EarthquakeNPP inversion values for each region (Table~\ref{tab:etasparams}).

To generate ensemble forecasts for the cross-catalog experiment (Section~\ref{sec:generalization}), we simulate the fitted ETAS model as a branching process, drawing $K = 100$ stochastic continuations of the observed seismicity \citep{mizrahi2021embracing}. Each continuation is seeded by the events QuakeGen conditions on, those within $[-768, +1.6]$\,h of the mainshock inside the forecast box. Every event of magnitude $M_e$, observed or simulated, produces a Poisson number of direct aftershocks,
\begin{equation}
n_e \sim \mathrm{Poisson}\big(\bar n(M_e)\big),
\qquad
\bar n(M) = k(M) \int_0^{\infty} g(\tau)\, d\tau \int_{\mathbb{R}^2} f(|\mathbf{x}|;\, M)\, d\mathbf{x},
\end{equation}
its productivity integrated over the temporal and spatial kernels defined above. Each aftershock is drawn by inverse-transform sampling of the same kernels. With a uniform variate $u \in (0, 1)$, the delay after the parent, the distance from the parent, and the magnitude are
\begin{align}
\tau &= \tau_d\, \Gamma^{-1}\!\big({-\omega},\ (1-u)\, \Gamma({-\omega},\ c/\tau_d)\big) - c, \\
r &= \Big( d\, e^{\gamma (M_p - M_c)} \big[ (1-u)^{-1/\rho} - 1 \big] \Big)^{1/2}, \\
M &= M_c - \beta^{-1} \ln(1-u),
\end{align}
where $\Gamma(\cdot,\cdot)$ is the upper incomplete gamma function, inverted in its second argument, $M_p$ is the parent magnitude, the azimuth around the parent is uniform, and $\beta$ is the region's fitted Gutenberg--Richter slope (Table~\ref{tab:etasparams}). Simulated events trigger offspring of their own by the same rules until the cascade is extinct, and background events arrive uniformly over the box and window at the fitted rate $\mu$ with the same magnitude distribution. The events in the forecast window $(t_\mathrm{c},\, t_0 + T_h)$, together with the observed events of $(0, t_\mathrm{c})$, are binned onto the same $n_x \times n_y$ grid as QuakeGen, giving $K$ cumulative count and maximum-magnitude fields scored exactly as the QuakeGen ensemble.

\begin{table}[ht]
  \centering
  \small \caption{Fitted ETAS parameters for the two EarthquakeNPP evaluation regions, from the inversion of \citet{mizrahi2021embracing} on the 2009--2015 catalog ($M_c = 1.0$), together with the Gutenberg--Richter $\beta$ used to simulate magnitudes.}
  \label{tab:etasparams}
  \begin{tabular}{@{}lrr@{}}
    \toprule
    Parameter & Salton Sea & San Jacinto \\
    \midrule
    $\log_{10}\mu$    & $-4.54$ & $-4.13$ \\
    $\log_{10}k_0$    & $-3.53$ & $-3.48$ \\
    $a$               & $1.60$  & $1.26$ \\
    $\log_{10}c$      & $-2.84$ & $-4.39$ \\
    $\omega$          & $0.035$ & $-0.181$ \\
    $\log_{10}\tau_d$ & $2.81$  & $3.01$ \\
    $\log_{10}d$      & $-2.36$ & $-2.49$ \\
    $\gamma$          & $1.17$  & $0.83$ \\
    $\rho$            & $0.61$  & $0.47$ \\
    $\beta$           & $1.75$  & $2.24$ \\
    \bottomrule
  \end{tabular}
\end{table}

\subsection{Global Reasenberg--Jones baseline}
\label{si:rj}

The global baseline is the Bayesian Reasenberg--Jones (R--J) model that underlies
operational aftershock advisories \citep{reasenberg1989earthquake}, reproducing the
USGS \texttt{opensha-oaf} implementation \citep{barall2025oaf} with the
generic priors of \citet{page2016three}. The temporal model is the modified-Omori law with a
Gutenberg--Richter productivity,
\begin{equation}
\lambda(t) = k\,(t + c)^{-p},
\qquad
k = 10^{\,a + b\,\big(M_\mathrm{main} - M_\mathrm{c}(t)\big)},
\label{eq:rj}
\end{equation}
for the rate of events above the completeness magnitude $M_\mathrm{c}(t)$ (Eq.~\ref{eq:mc}). Each mainshock receives the generic values of $b$, $p$, and $c$ and a Gaussian prior on $a$ from its tectonic regime, resolved from the epicenter as in the operational system \citep{garcia2012global} (Table~\ref{tab:rj-generic}); only the productivity $a$ is updated from data, using the aftershocks recorded before QuakeGen's conditioning cutoff $t_\mathrm{c} = 1.6$ h. $M_\mathrm{c}(t)$ serves only as the reference magnitude and does not filter the catalog, so the R--J model is calibrated on and scored against the same unfiltered events as QuakeGen, and the fitted $a$ describes the full recorded stream rather than the $M \ge M_\mathrm{c}(t)$ population. QuakeGen also reads the $768$ hours of seismicity before the mainshock, which the R--J model cannot use, since its rate begins at the mainshock and carries no background term.

\begin{table}[htbp]
\centering
\caption{Generic R--J parameters by tectonic regime \citep{page2016three}, with
$b = 1$ throughout. $N$ is the number of test mainshocks assigned to each regime.}
\label{tab:rj-generic}
\small
\begin{tabular}{lrrrrrr}
\toprule
Regime & $a_\mathrm{mean}$ & $\sigma_0$ & $\sigma_1$ & $p$ & $c$ (d) & $N$ \\
\midrule
SZ-GENERIC       & $-2.47$ & 0.49 & 570 & 0.88 & 0.0180 & 15 \\
SCR-GENERIC      & $-2.85$ & 0.49 & 870 & 0.73 & 0.0180 & 14 \\
SZ-ONSHORE       & $-2.34$ & 0.49 & 500 & 0.81 & 0.0180 & 13 \\
ANSR-SHALCON     & $-2.42$ & 0.49 & 570 & 0.98 & 0.0180 & 10 \\
ANSR-HOTSPOT     & $-3.00$ & 0.49 & 680 & 1.12 & 0.0180 &  9 \\
CAL-SCSN         & $-2.30$ & 0.50 &   0 & 0.83 & 0.0033 &  8 \\
ANSR-DEEPCON     & $-2.13$ & 0.49 & 250 & 0.98 & 0.0180 &  2 \\
ANSR-ABSLSHC     & $-2.85$ & 0.49 & 800 & 1.06 & 0.0180 &  2 \\
SZ-OUTERTR       & $-2.42$ & 0.49 & 540 & 0.92 & 0.0180 &  2 \\
CAL-NCSS         & $-2.64$ & 0.48 &   0 & 0.96 & 0.0120 &  2 \\
CAL-MENDOCINO    & $-3.18$ & 0.47 &   0 & 1.15 & 0.0500 &  2 \\
SZ-INLBACK       & $-2.43$ & 0.49 & 640 & 0.86 & 0.0180 &  1 \\
ANSR-ABSLDEC     & $-2.29$ & 0.49 & 560 & 1.01 & 0.0180 &  0 \\
ANSR-ABSLOCB     & $-2.82$ & 0.49 & 890 & 0.64 & 0.0180 &  0 \\
ANSR-OCEANBD     & $-3.19$ & 0.49 & 500 & 1.08 & 0.0180 &  0 \\
SOR-GENERIC      & $-3.04$ & 0.49 & 650 & 0.97 & 0.0180 &  0 \\
SCR-ABVSLAB      & $-2.85$ & 0.49 & 870 & 0.73 & 0.0180 &  0 \\
SOR-ABVSLAB      & $-3.04$ & 0.49 & 650 & 0.97 & 0.0180 &  0 \\
CAL-HYDROTHERMAL & $-1.79$ & 0.29 &   0 & 0.94 & 0.0260 &  0 \\
\bottomrule
\end{tabular}
\end{table}

The completeness follows the time-dependent form of \citet{page2016three},
\begin{equation}
M_\mathrm{c}(t) = \max\!\big(m_\mathrm{cat},\;
  F M_\mathrm{main} - G - H \log_{10} t\big),
\label{eq:mc}
\end{equation}
elevated immediately after the mainshock, when small aftershocks are lost beneath the coda of larger ones, and relaxing to the long-term level $m_\mathrm{cat}$, with parameters set by region (Table~\ref{tab:rj-magcomp}). The fitted $a$ is the Bayesian posterior mean combining the regime prior, a Gaussian of mean $a_\mathrm{mean}$ and width $\sigma_a = \big(\sigma_0^2 + \sigma_1^2 / 10^{\max(M_\mathrm{main},\,6)}\big)^{1/2}$, with the maximum-likelihood estimate $\hat{a} = \log_{10}(n / N_0)$, where $n$ is the observed and $N_0$ the expected count of the conditioning window at $a = 0$.

\begin{table}[htbp]
\centering
\caption{Region-dependent completeness parameters for Eq.~\ref{eq:mc}. The regions
are tested in the order listed, the first match applying, with WORLD as
the fallback.}
\label{tab:rj-magcomp}
\small
\begin{tabular}{lrrrr}
\toprule
Regions & $m_\mathrm{cat}$ & $F$ & $G$ & $H$ \\
\midrule
CAL-SCSN, CAL-NCSS, CAL-MENDOCINO,          & 3.0 & 1.0 & 4.50 & 0.75 \\
\quad CAL-HYDROTHERMAL, HAWAII &     &     &      &      \\
ALASKA                                      & 3.5 & 1.0 & 4.50 & 0.75 \\
USA, PUERTO-RICO                            & 3.5 & 0.5 & 0.25 & 1.00 \\
GUAM-MARIANAS, AMERICAN-SAMOA, WORLD        & 4.6 & 0.5 & 0.25 & 1.00 \\
\bottomrule
\end{tabular}
\end{table}

The temporal rate is distributed in space by a mixture of kernels
centered on the recorded sources, the mainshock at the box center and every
aftershock before $t_\mathrm{c}$ within $\Delta M = 3$ of the mainshock magnitude,
forming the source set $\mathcal{S}$. The conditional spatial density is the
productivity- and Omori-weighted mixture of per-source kernels,
\begin{align}
f(\mathbf{x} \mid t)
  &= \frac{\sum_{e \in \mathcal{S}} w_e(t)\, K(|\mathbf{x} - \mathbf{x}_e|;\, d_e)}
          {\sum_{e \in \mathcal{S}} w_e(t)\, I(d_e)},
  \qquad
  w_e(t) = 10^{\,a + b(M_e - m_\mathrm{cat})}\big(t - t_e + c\big)^{-p}, \\
K(r; d) &= \frac{D\,d}{2\pi\,(d^2 + r^2)\,\sqrt{D^2/4 + d^2 + r^2}},
  \qquad
  I(d) = d\,\log\!\frac{s + D/2}{s - D/2},
  \quad s = \sqrt{D^2/4 + d^2},
\end{align}
where $K$ is a three-dimensional radial kernel integrated over a seismogenic depth
$D = 10$ km, $I(d)$ is its plane integral so that $f$ integrates to one, and
$d_e = \big(7 M_{0,e}/16\sigma\big)^{1/3}$ is the circular-crack source radius for seismic moment $M_{0,e} = 10^{\,1.5 M_e + 9.0}$ N$\cdot$m and stress drop $\sigma = 2$ MPa
\citep{eshelby1957determination, keilisborok1959estimation}.
Because $f$ is normalized over the whole plane rather than the forecast box, aftershocks expected beyond the box edges are excluded from the gridded forecast. The cumulative grid for horizon $T_h$ is
\begin{equation}
N^{\mathrm{RJ}}_{ij} = |\Omega| \int_0^{T_h}
  \lambda(t)\, f(\mathbf{x}_{ij} \mid t)\, dt,
\end{equation}
evaluated numerically, because $f(\mathbf{x} \mid t)$ depends on $t$ both through the
source weights $w_e(t)$ and through the activation of sources as $t$ passes each
$t_e$.

\section{Evaluation metrics}
\label{si:eval}

All models are scored on forecast grids of the same form \citep{schorlemmer2007earthquake, woessner2011retrospective, stockman2025earthquakenpp}. For each horizon $T_h$, the grid $\hat N^{(h)}_{ij}$ gives the expected number of earthquakes in each cell over the forecast window $(t_0, t_0 + T_h)$ and is compared with the observed count field $N_{ij}$ of Eq.~\ref{eq:target}. For the statistical baselines, the grid is the modeled rate integrated over the window (Section~\ref{si:baselines}); for QuakeGen and the branching ETAS simulation, the grid is the mean of the $K$ sampled fields. 
In the global setting, each test mainshock is scored on its own grid at horizons of 3, 12, 48, 168, and 720 h after the mainshock. In the regional setting, forecast anchors are placed every 24 h through the test window, and the events inside each window $(t_0, t_0 + T_h)$ are scored against that anchor's grid, so an event enters several overlapping windows at horizons beyond one day. QuakeGen and ETAS are scored on the same anchors and windows.

We report six metrics: the temporal log-likelihood, the spatial information gain, the productivity ratio, the spatial RMSE, the Wasserstein-1 distance, and the mean-squared-error skill score (Figs.~\ref{fig:global_metrics}, \ref{fig:regional_metrics}; Table~\ref{tab:performance}).
For the temporal log-likelihood and the spatial information gain, every aftershock $e$ inside the window $(t_0, t_0 + T_h)$ receives two scores against that horizon's grid \citep{gneiting2007strictly},
\begin{align}
\ell^{t}_e &= \log\!\left(\frac{\sum_{ij}\hat N^{(h)}_{ij}}{T_h}\right), \\
\ell^{s}_e &= \log\!\big(n_\mathrm{cell}\; \hat p_{i_e j_e}\big),
\qquad
\hat p_{ij} = \frac{\hat N^{(h)}_{ij}}{\sum_{ij}\hat N^{(h)}_{ij}},
\end{align}
so $\ell^{t}_e$ is the log of the forecast's mean daily event rate over the window, with $T_h$ in days, and $\ell^{s}_e$ the log probability the forecast assigns to the event's cell, relative to a spatially uniform forecast over the $n_\mathrm{cell} = n_x n_y$ cells. A spatial density floor $\varepsilon = 10^{-6}$ events/km$^2$ keeps the logarithms finite. Pooling over the scored aftershocks of one forecast unit, a mainshock in the global setting or one anchor in the regional setting, gives the two metrics,
\begin{equation}
\mathrm{TLL} = \frac{1}{N}\sum_e \ell^{t}_e - \frac{\Lambda}{N},
\qquad
\mathrm{IG} = \frac{1}{N}\sum_e \ell^{s}_e,
\end{equation}
where $N$ is the number of scored aftershocks and $\Lambda$ is the total predicted count over the scored windows; the $-\Lambda/N$ term penalizes over-forecasting. A spatially uniform forecast scores $\mathrm{IG} = 0$, so $\mathrm{IG}$ measures the information gained over the uniform forecast, in nats per earthquake \citep{harte2005entropy}. A gain of $g$ nats means the forecast concentrates $e^{g}$ times more probability where events occur. No box dimension enters the score, so $\mathrm{IG}$ is comparable across the widely different global boxes of Fig.~\ref{fig:catalog}.
The other four metrics compare $\hat N_{ij}$ with $N_{ij}$ directly. The \emph{productivity ratio}
\begin{equation}
\frac{N_\mathrm{pred}}{N_\mathrm{obs}} = \frac{\sum_{ij}\hat N_{ij}}{\sum_{ij} N_{ij}}
\end{equation}
compares the predicted and observed total number of events and is ideal at one. The \emph{spatial RMSE}
\begin{equation}
\mathrm{RMSE} = \sqrt{\frac{1}{n_\mathrm{cell}}
  \sum_{ij}\big(\hat N_{ij} - N_{ij}\big)^2}
\end{equation}
is the cell-by-cell count error. The \emph{Wasserstein-1 distance} treats the normalized forecast $\hat p$ and the normalized observed field $p_{ij} = N_{ij} / \sum_{ij} N_{ij}$ as spatial probability distributions and reports the mean distance over which probability mass must move to turn one into the other,
\begin{equation}
W_1 = \min_{\pi \in \Pi(\hat p,\, p)}
  \sum_{u, v} \pi_{uv}\,\lVert \mathbf{u} - \mathbf{v} \rVert,
\end{equation}
where the minimum is over transport plans $\pi$ with the two fields as marginals and distance is measured in grid cells, so $W_1$ is comparable across box sizes and isolates spatial displacement from total count. The \emph{mean-squared-error skill score}
\begin{equation}
\mathrm{MSESS} = 1 - \frac{\sum_{ij}(\hat N_{ij} - N_{ij})^2}
  {\sum_{ij}(N_{ij} - \bar N)^2}
\end{equation}
is the fraction of the observed field's spatial variance the forecast explains, one for a perfect field, zero for the constant mean field $\bar N$, and negative for a worse one.

\section{Additional forecast examples}
\label{si:examples}

\begin{figure}[p]
  \centering
  \includegraphics[width=\textwidth]{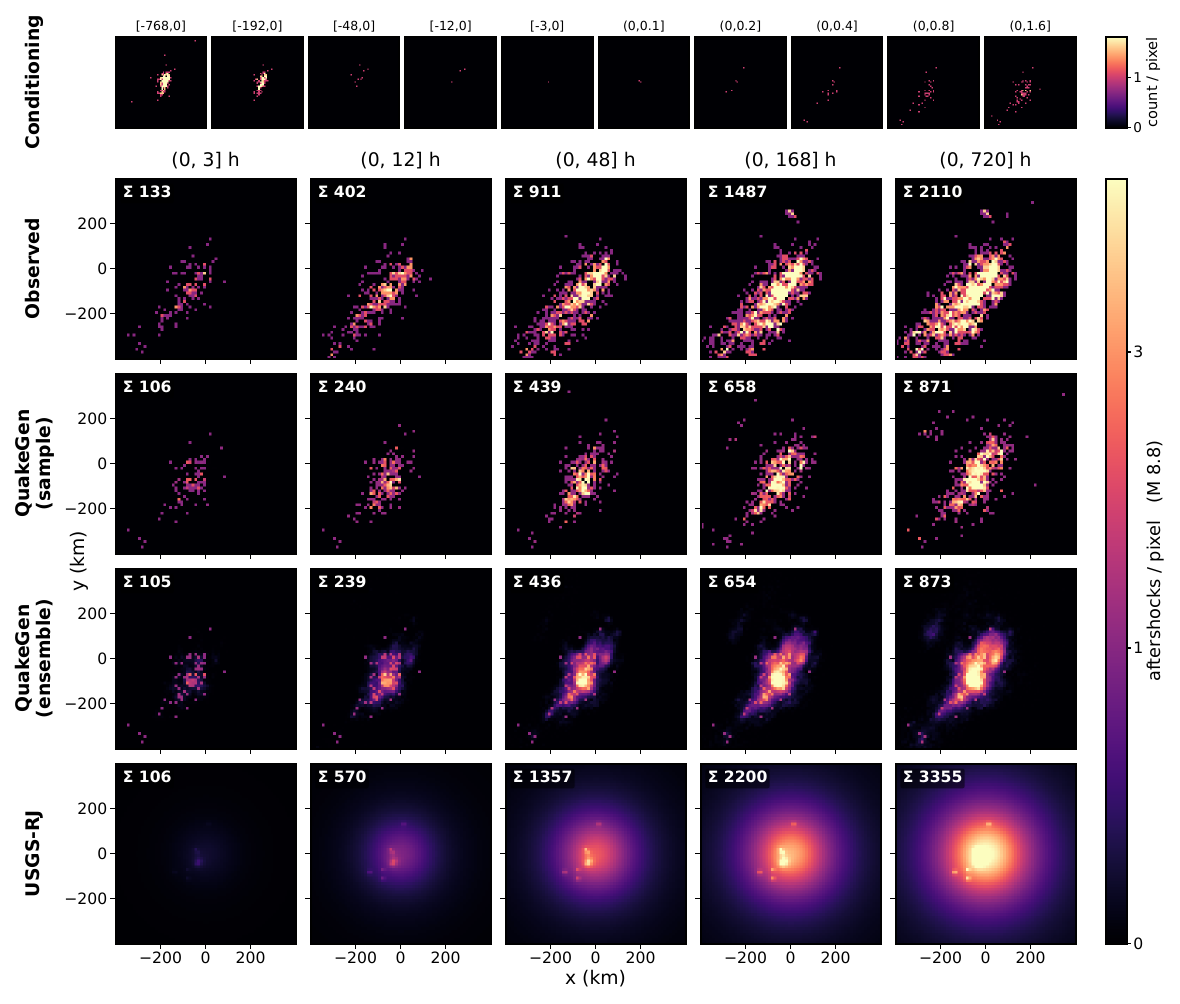}
  \caption{M~$8.8$ Kamchatka, Russia, 29 July 2025 in the layout of Fig.~\ref{fig:ex_noto}. The observed aftershocks form a band several hundred kilometers long along the megathrust. QuakeGen recovers the central along-strike elongation but not its full extent and undercounts the total.}
  \label{fig:si_freq_kamchatka}
\end{figure}

\begin{figure}[p]
  \centering
  \includegraphics[width=\textwidth]{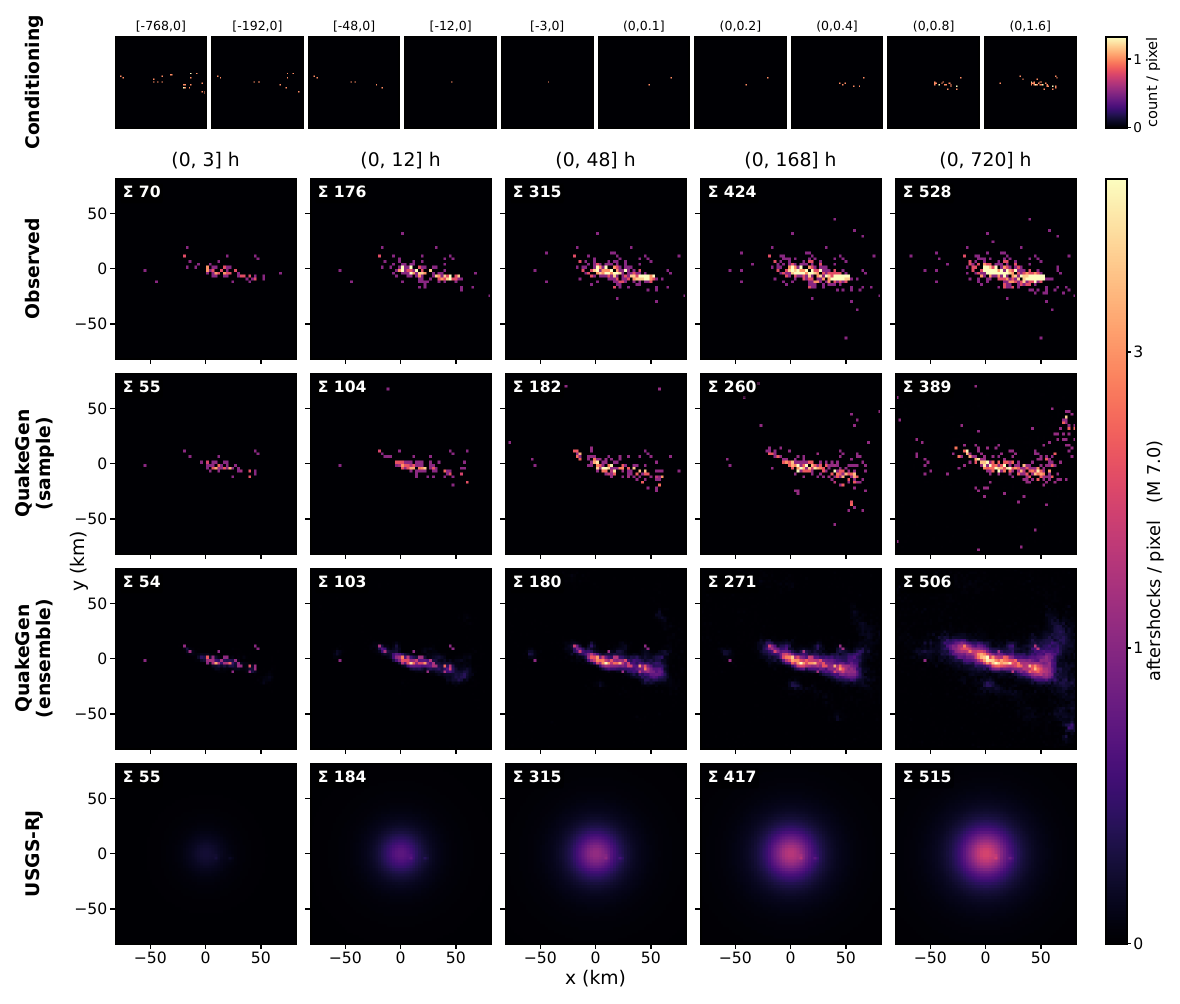}
  \caption{M~$7.0$ Cape Mendocino, California, 5 December 2024 in the layout of Fig.~\ref{fig:ex_noto}.}
  \label{fig:si_freq_norcal}
\end{figure}

\begin{figure}[p]
  \centering
  \includegraphics[width=\textwidth]{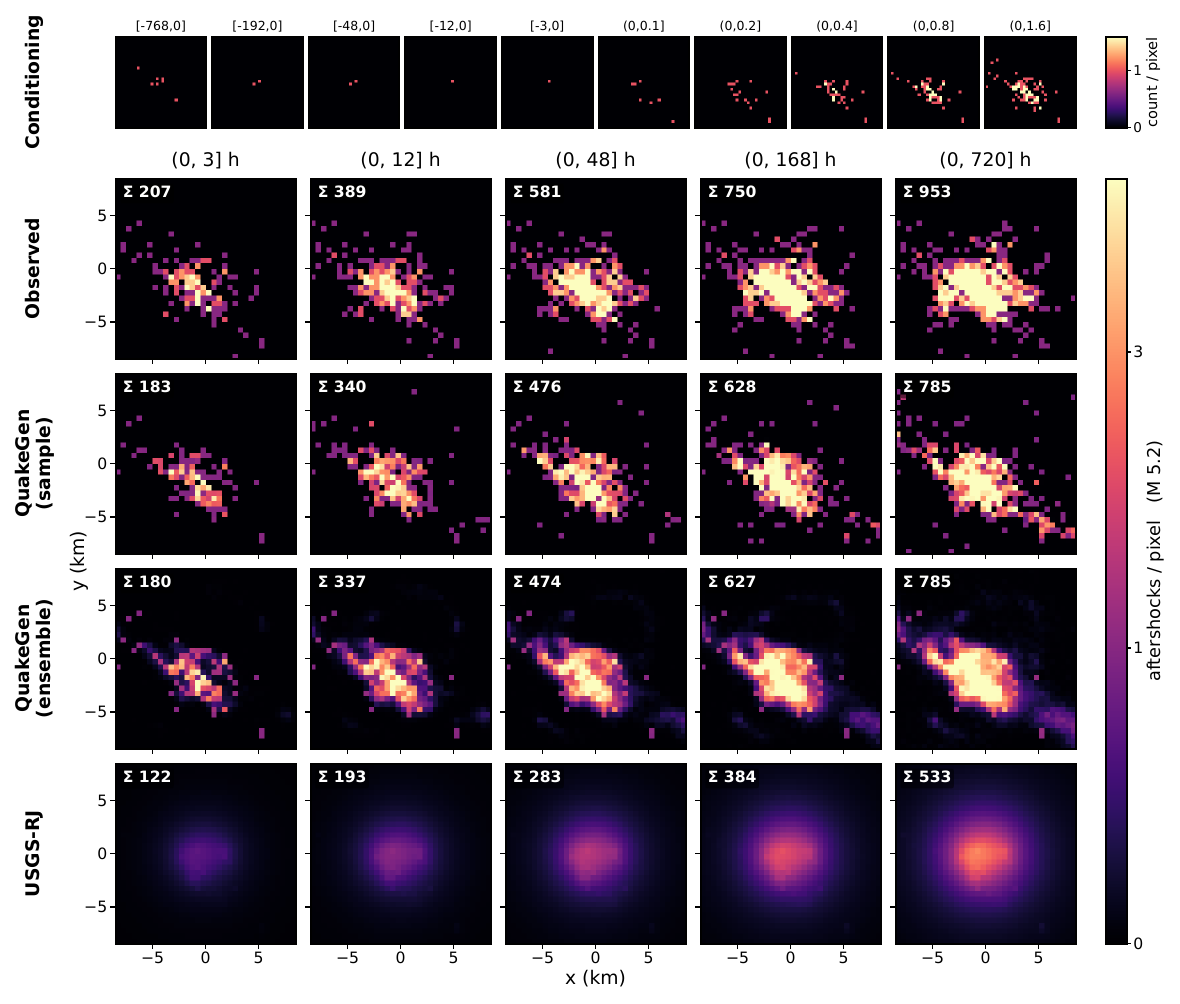}
  \caption{M~$5.2$ Bakersfield, California, 7 August 2024 in the layout of Fig.~\ref{fig:ex_noto}.}
  \label{fig:si_freq_socal}
\end{figure}

\begin{figure}[p]
  \centering
  \includegraphics[width=\textwidth]{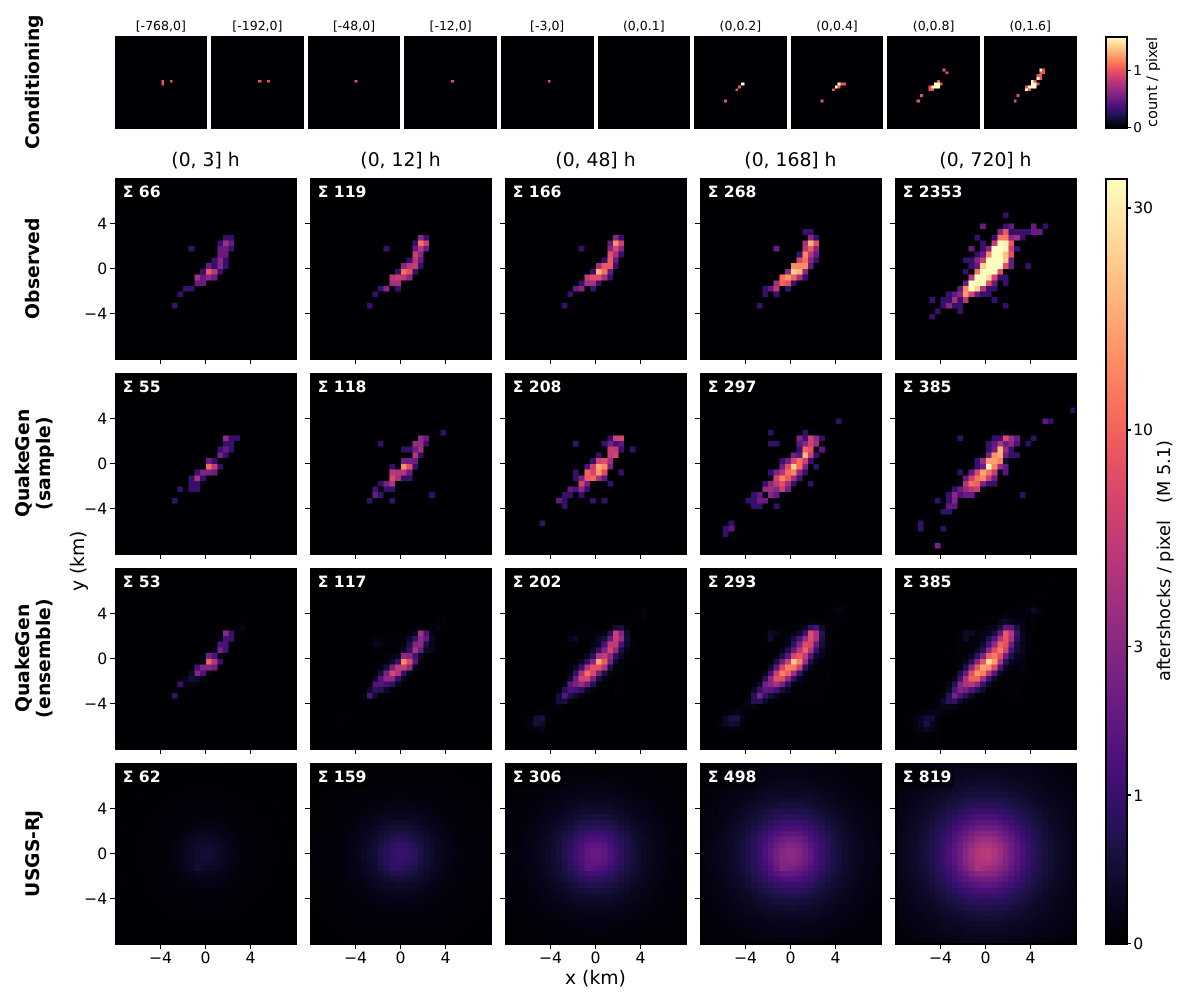}
  \caption{M~$5.1$ Prague, Oklahoma, 3 February 2024, an induced sequence in the layout of Fig.~\ref{fig:ex_noto}. QuakeGen traces the southwest--northeast fault trace, while a late secondary burst drives the observed total above both forecasts.}
  \label{fig:si_freq_oklahoma}
\end{figure}

\begin{figure}[p]
  \centering
  \includegraphics[width=\textwidth]{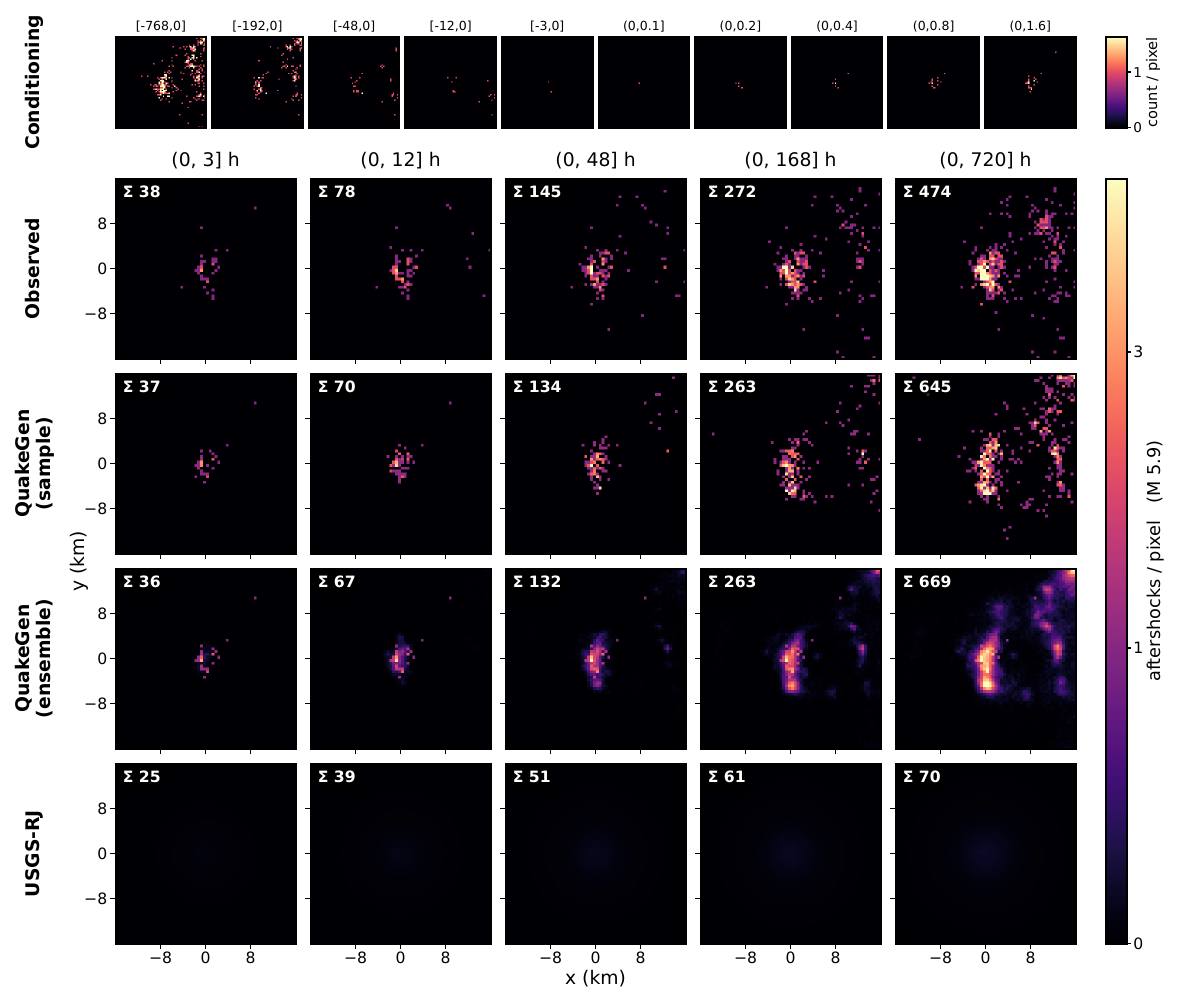}
  \caption{M~$5.9$ Pahala, Hawaii, 9 February 2024 in the layout of Fig.~\ref{fig:ex_noto}.}
  \label{fig:si_freq_hawaii}
\end{figure}

\begin{figure}[p]
  \centering
  \includegraphics[width=\textwidth]{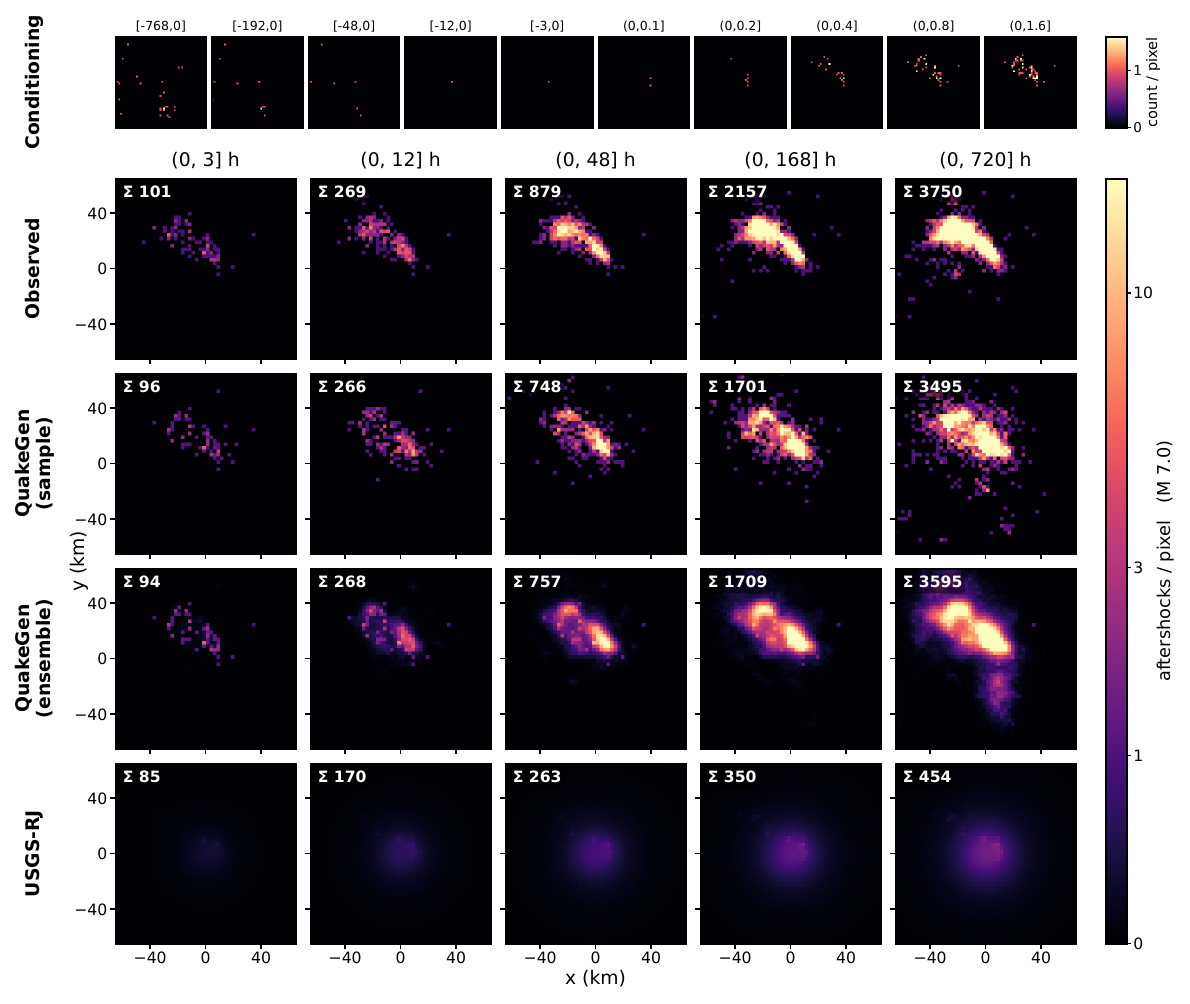}
  \caption{M~$7.0$ southeastern Alaska, 6 December 2025 in the layout of Fig.~\ref{fig:ex_noto}.}
  \label{fig:si_freq_alaska}
\end{figure}

\begin{figure}[p]
  \centering
  \includegraphics[width=\textwidth]{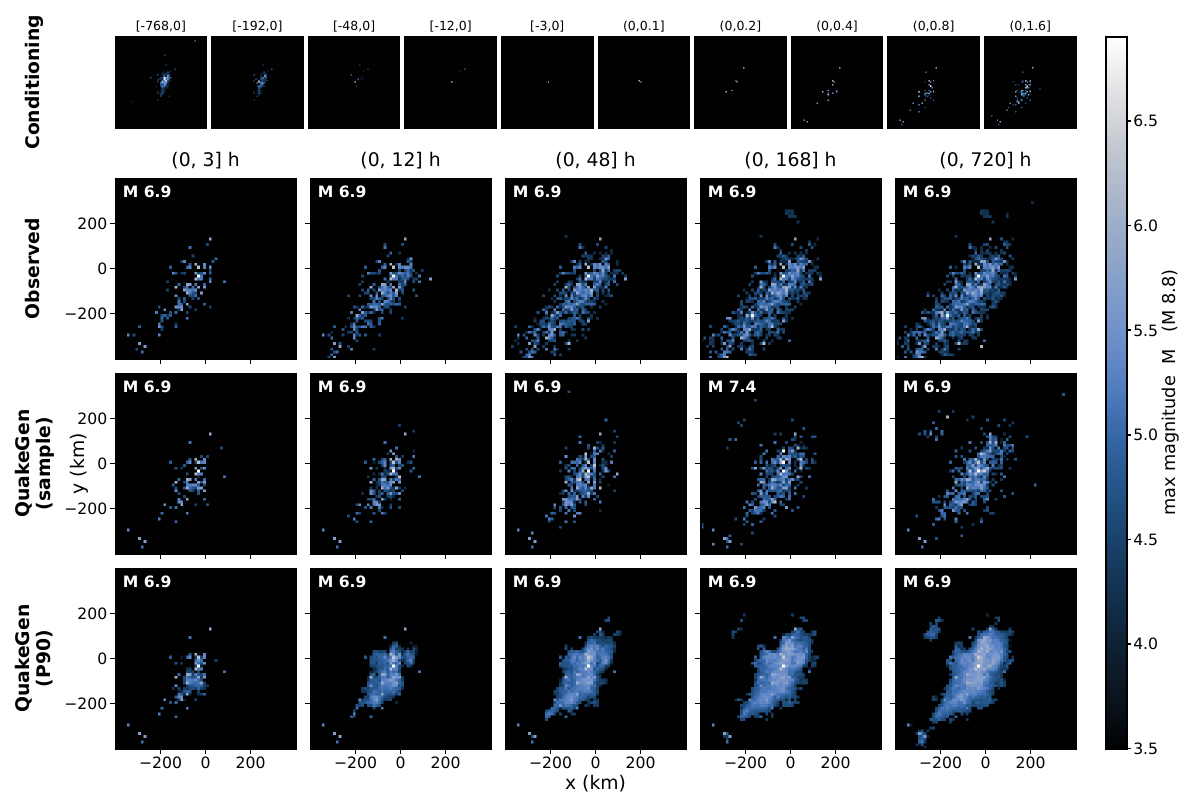}
  \caption{Maximum-magnitude field for the M~$8.8$ Kamchatka sequence in the layout of Fig.~\ref{fig:ex_mag}.}
  \label{fig:si_mag_kamchatka}
\end{figure}

\begin{figure}[p]
  \centering
  \includegraphics[width=\textwidth]{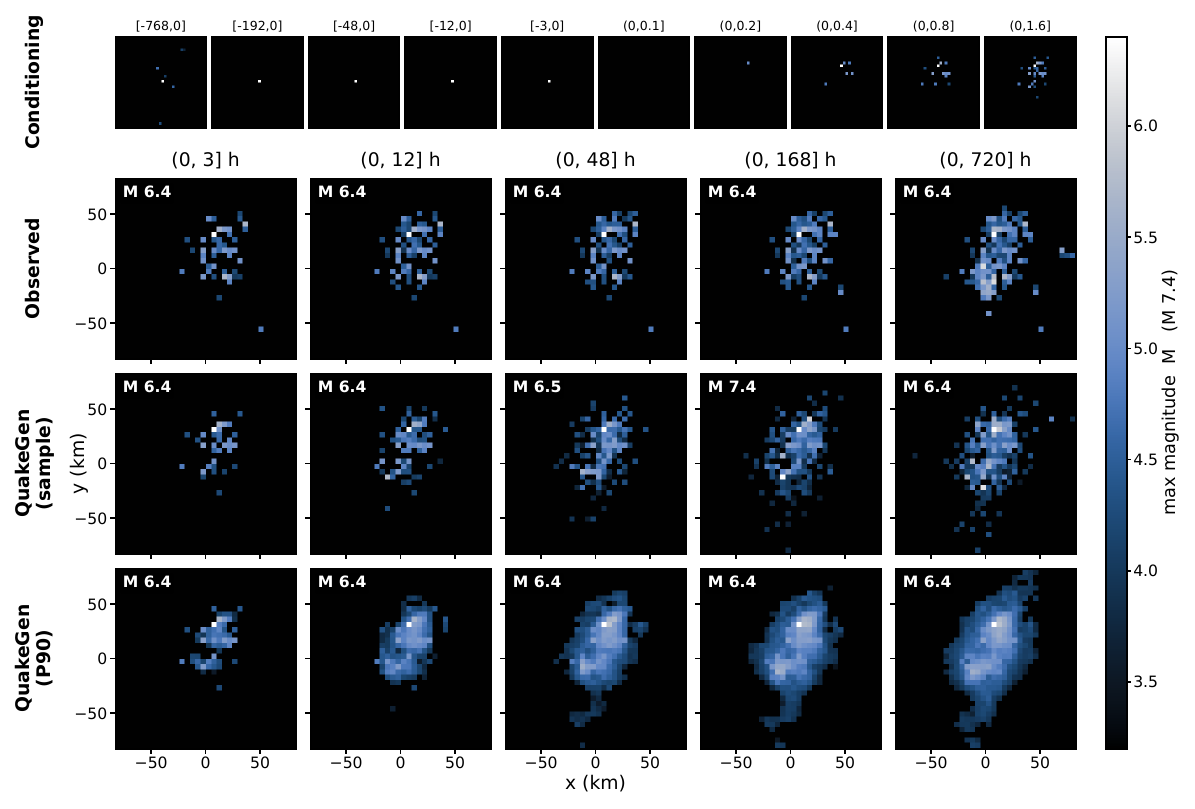}
  \caption{Maximum-magnitude field for the M~$7.4$ Hualien, Taiwan sequence in the layout of Fig.~\ref{fig:ex_mag}.}
  \label{fig:si_mag_taiwan}
\end{figure}

\begin{figure}[p]
  \centering
  \includegraphics[width=\textwidth]{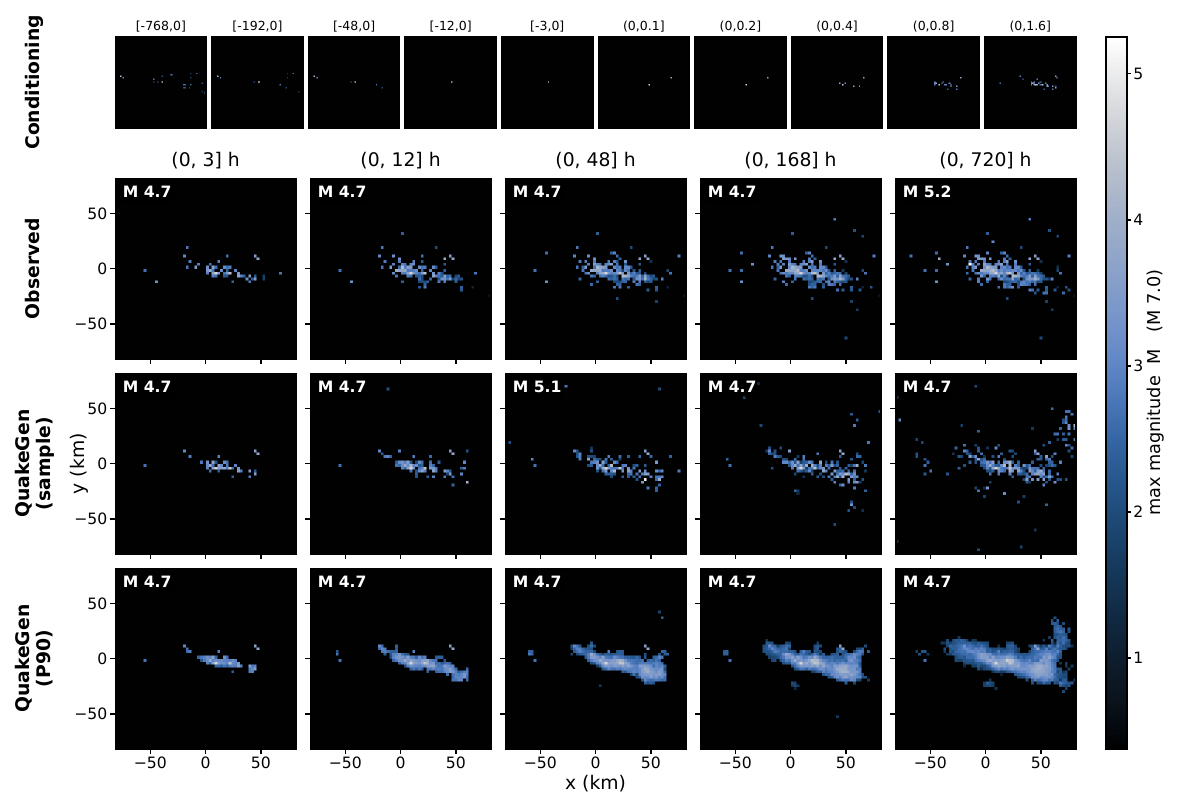}
  \caption{Maximum-magnitude field for the M~$7.0$ Cape Mendocino sequence in the layout of Fig.~\ref{fig:ex_mag}.}
  \label{fig:si_mag_norcal}
\end{figure}

\begin{figure}[p]
  \centering
  \includegraphics[width=\textwidth]{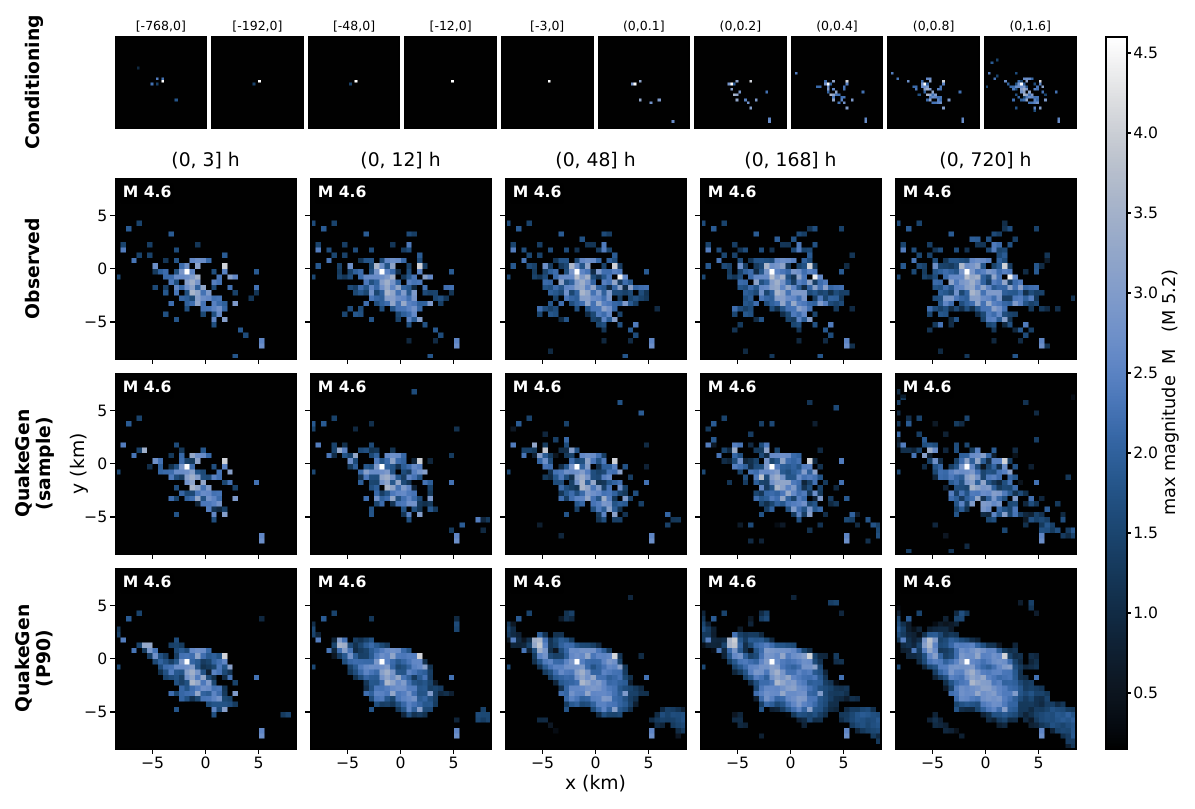}
  \caption{Maximum-magnitude field for the M~$5.2$ Bakersfield sequence in the layout of Fig.~\ref{fig:ex_mag}.}
  \label{fig:si_mag_socal}
\end{figure}

\begin{figure}[p]
  \centering
  \includegraphics[width=\textwidth]{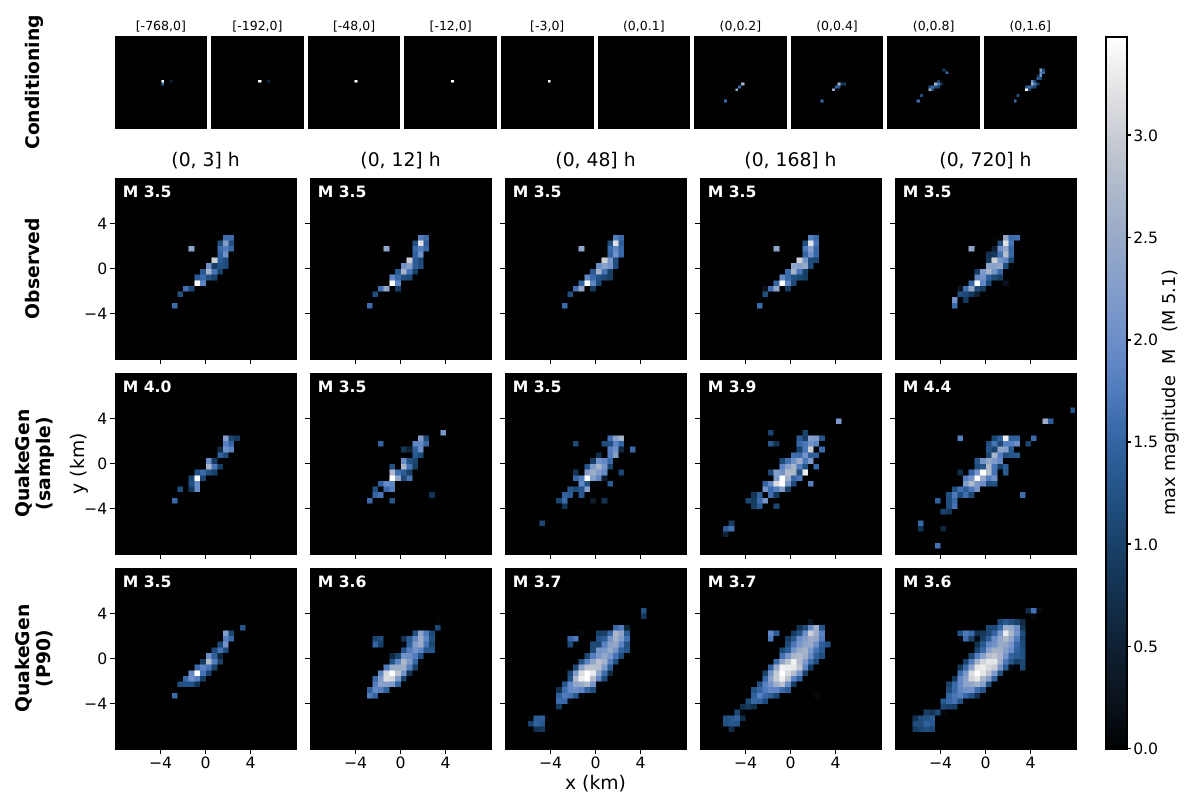}
  \caption{Maximum-magnitude field for the M~$5.1$ Prague, Oklahoma sequence in the layout of Fig.~\ref{fig:ex_mag}.}
  \label{fig:si_mag_oklahoma}
\end{figure}

\begin{figure}[p]
  \centering
  \includegraphics[width=\textwidth]{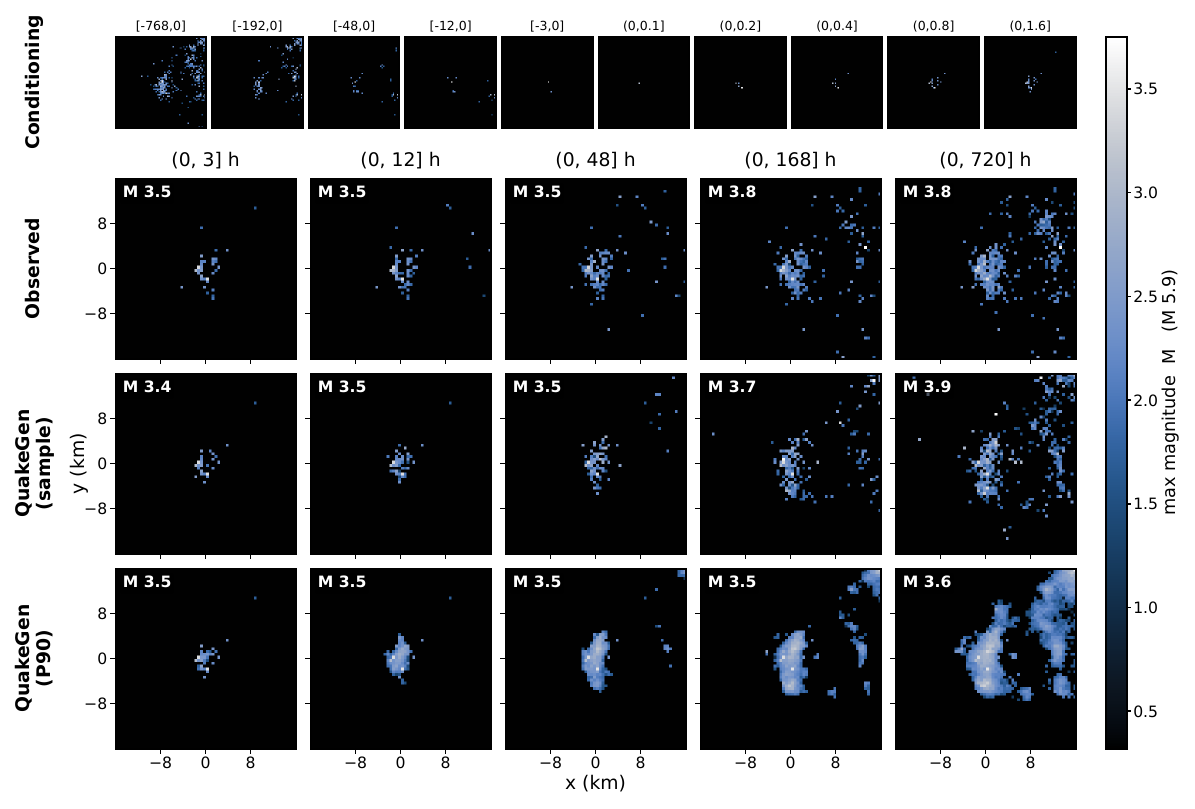}
  \caption{Maximum-magnitude field for the M~$5.9$ Pahala, Hawaii sequence in the layout of Fig.~\ref{fig:ex_mag}.}
  \label{fig:si_mag_hawaii}
\end{figure}

\begin{figure}[p]
  \centering
  \includegraphics[width=\textwidth]{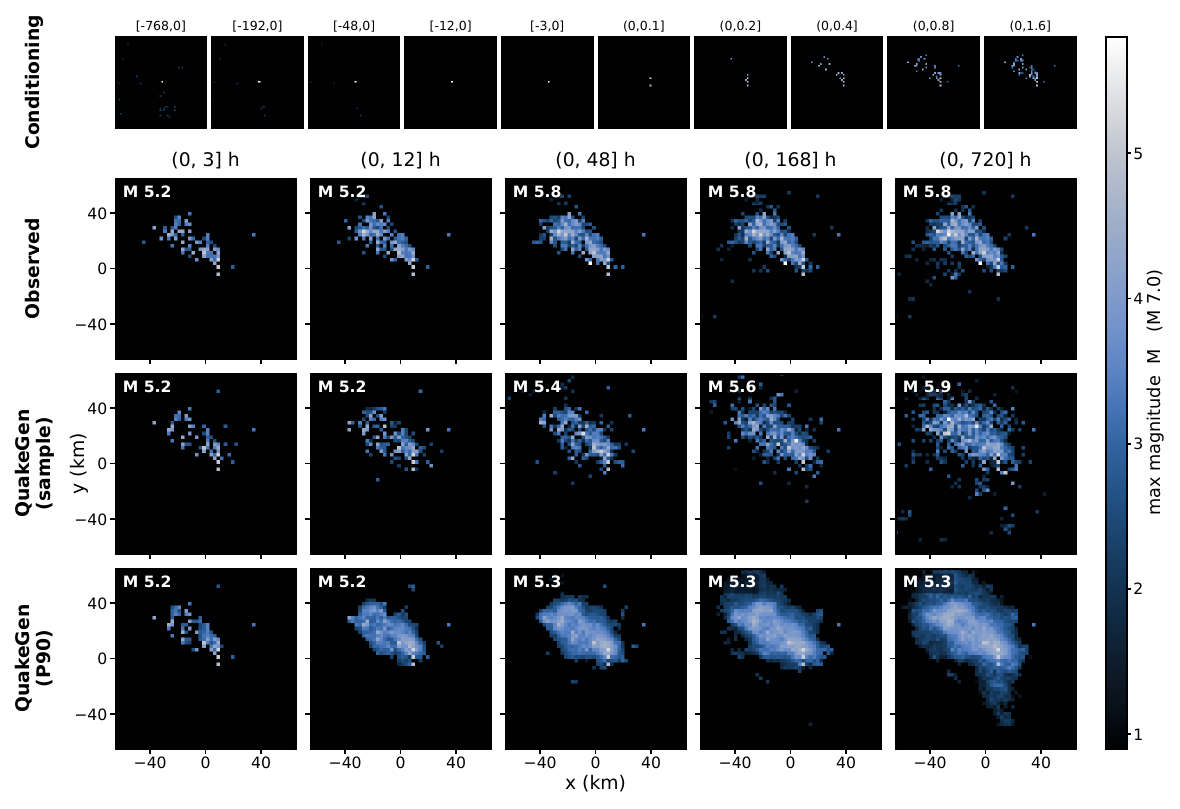}
  \caption{Maximum-magnitude field for the M~$7.0$ southeastern Alaska sequence in the layout of Fig.~\ref{fig:ex_mag}.}
  \label{fig:si_mag_alaska}
\end{figure}

Figures~\ref{fig:si_freq_kamchatka}--\ref{fig:si_freq_alaska} show the count-field forecasts for the six global test mainshocks not shown in the main text, in the layout of Fig.~\ref{fig:ex_noto}, and Figures~\ref{fig:si_mag_kamchatka}--\ref{fig:si_mag_alaska} the corresponding maximum-magnitude fields in the layout of Fig.~\ref{fig:ex_mag}. Across tectonic settings the QuakeGen sample reproduces the observed anisotropy, from tight linear ruptures to multi-lobed clusters, while the Reasenberg--Jones forecast spreads smooth halos about the mainshock and its early aftershocks.

\begin{figure}[p]
  \centering
  \includegraphics[width=\textwidth]{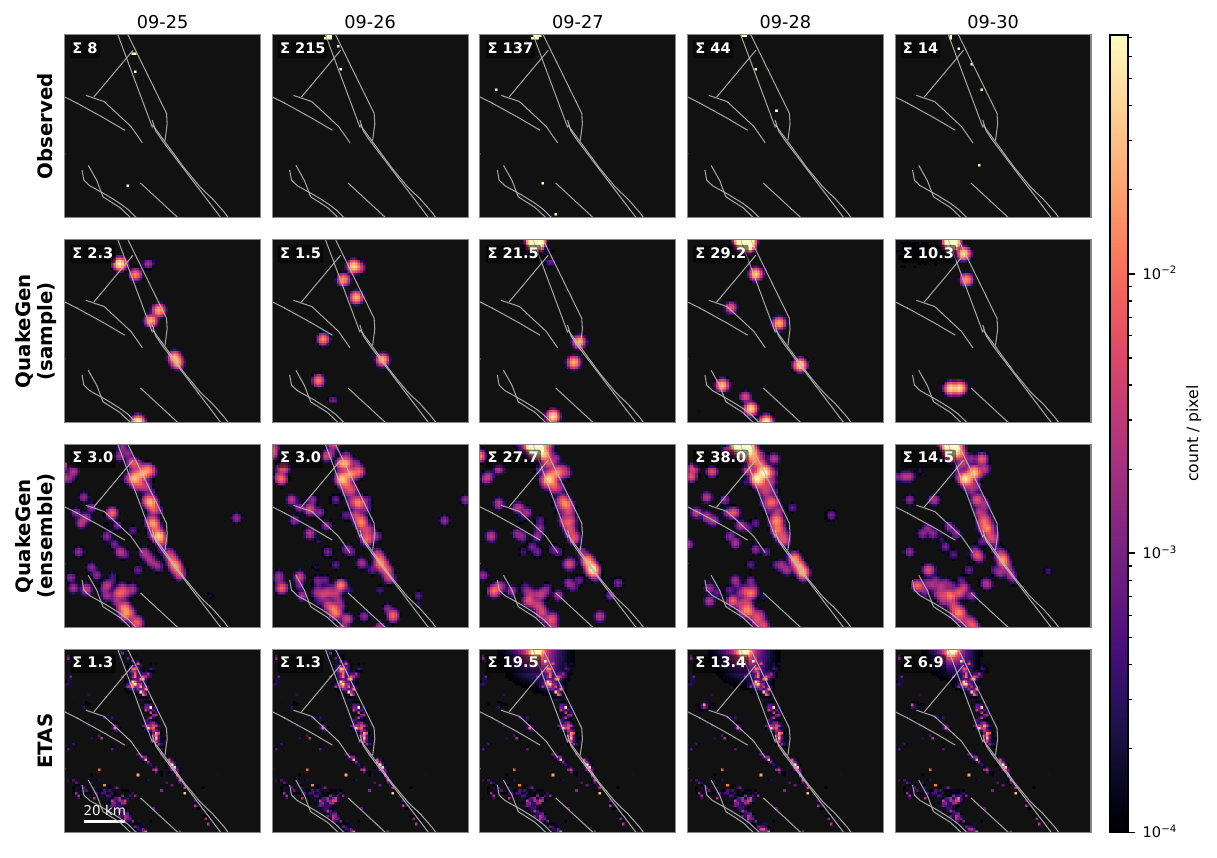}
  \caption{Daily 24-hour rate forecasts for the Salton Sea region during the September 2016 Salton Sea swarm. Rows: observed daily count, the median QuakeGen sample, the QuakeGen ensemble mean, and ETAS; mapped fault traces are overlaid in gray. Both the QuakeGen ensemble and ETAS concentrate the forecast rate on the active fault strands as the swarm builds. The single QuakeGen sample places discrete events only approximately on the cells where events later occur.}
  \label{fig:si_regional_rate}
\end{figure}

Figure~\ref{fig:si_regional_rate} shows the regional daily rate for the Salton Sea test region on selected days spanning a swarm. Both QuakeGen and ETAS trace the mapped fault network; QuakeGen reaches that fault-aligned pattern with no prescribed spatial kernel, while a single generated sample places its discrete events only approximately, reflecting the limited day-to-day predictability of background seismicity.

\begin{figure}[p]
  \centering
  \includegraphics[width=\textwidth]{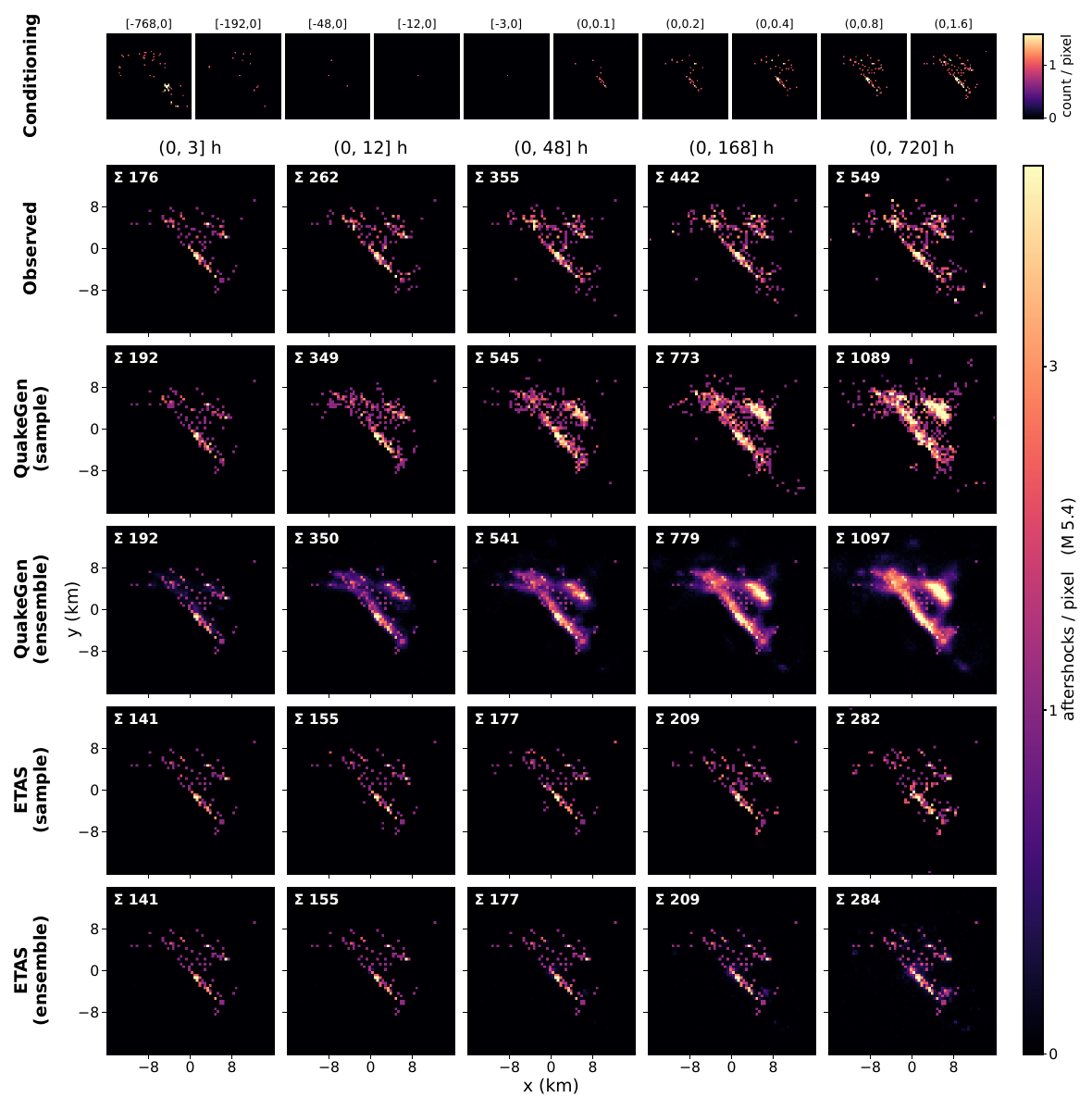}
  \caption{Global model applied to the M~$5.4$ San Jacinto sequence of 7 July 2010, in the layout of Fig.~\ref{fig:qtm_example}.}
  \label{fig:si_qtm_1}
\end{figure}

\begin{figure}[p]
  \centering
  \includegraphics[width=\textwidth]{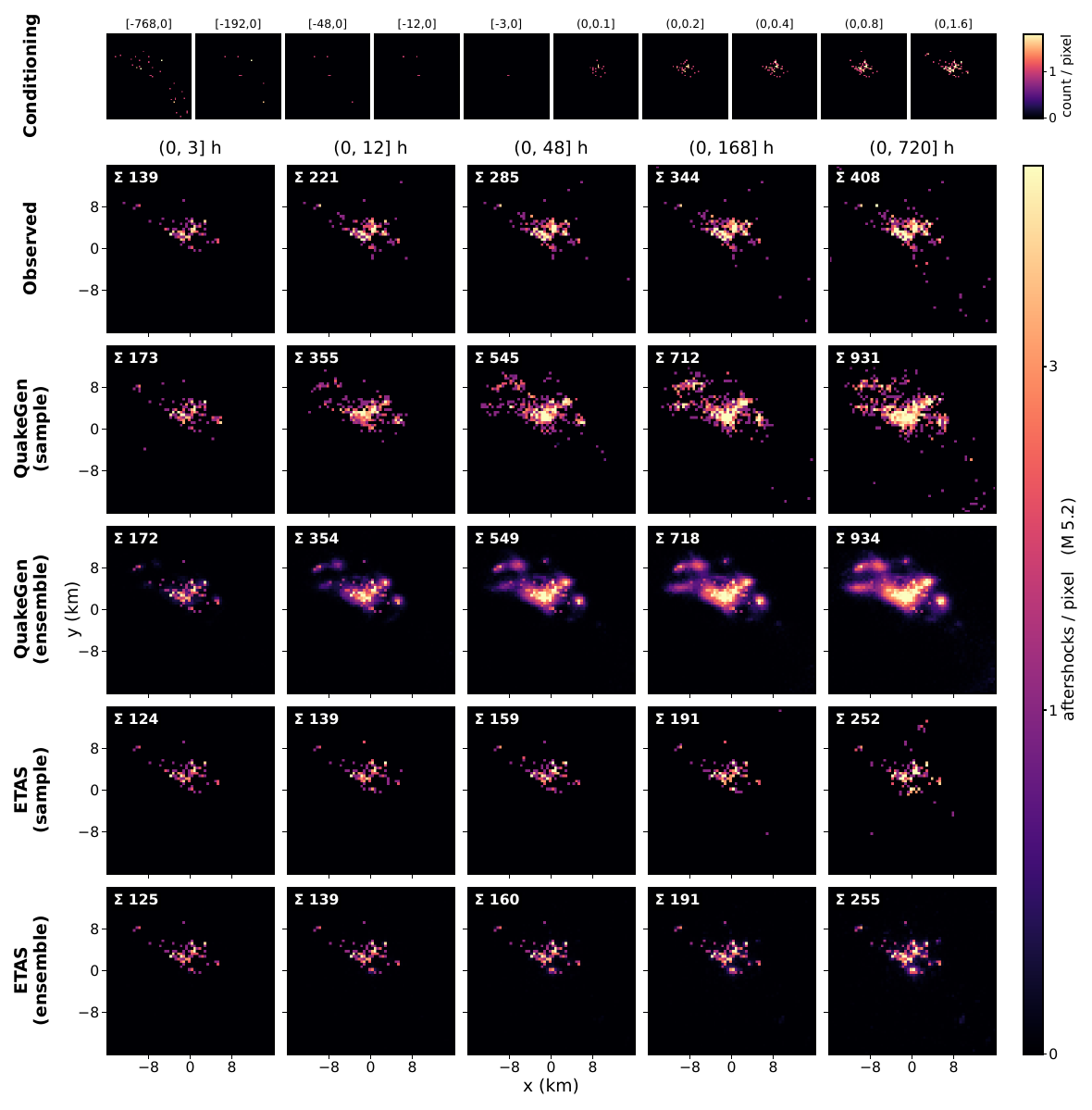}
  \caption{Global model applied to the M~$5.2$ Borrego Springs, San Jacinto sequence of 10 June 2016, in the layout of Fig.~\ref{fig:qtm_example}.}
  \label{fig:si_qtm_2}
\end{figure}

\begin{figure}[p]
  \centering
  \includegraphics[width=\textwidth]{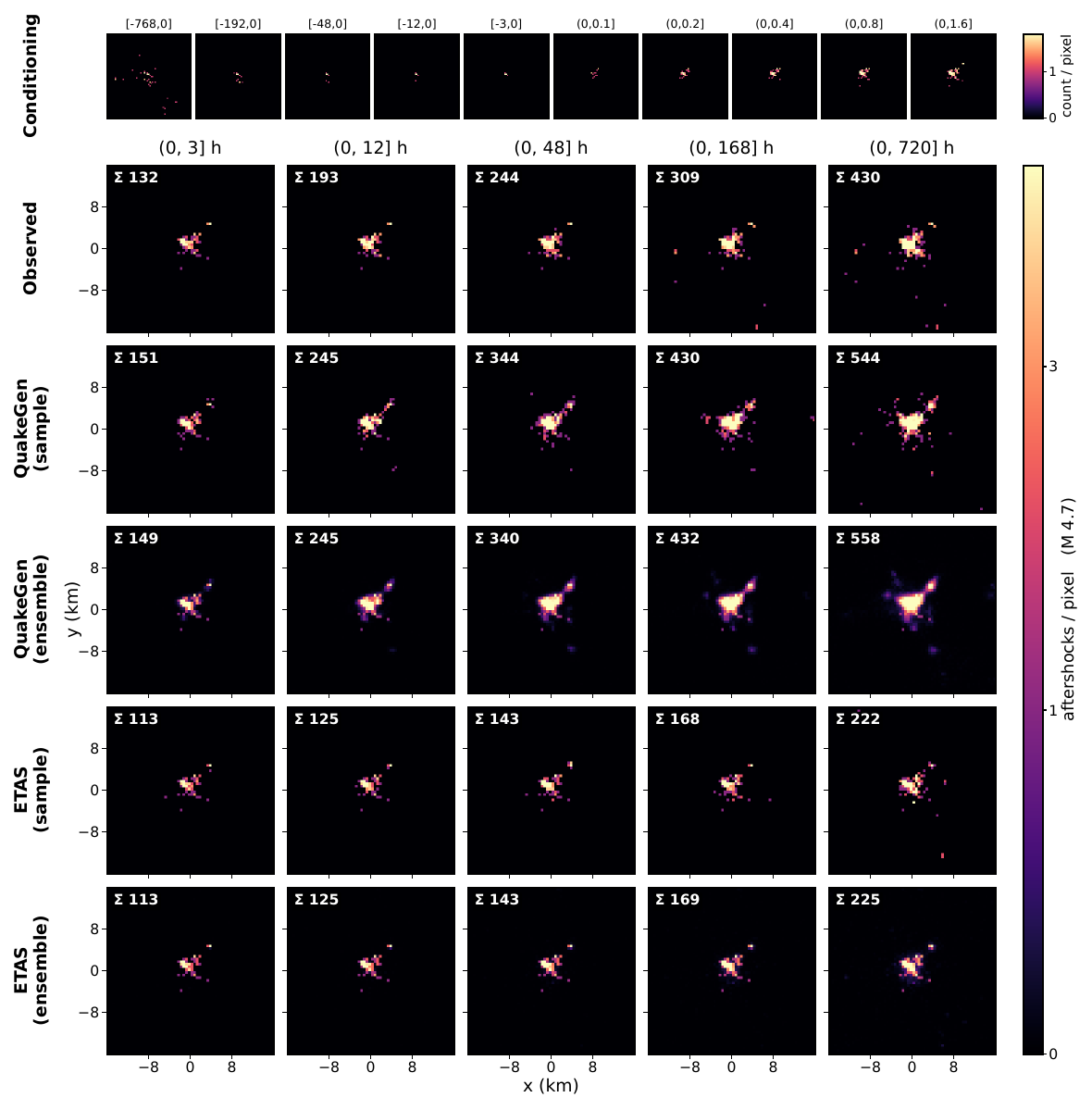}
  \caption{Global model applied to the M~$4.7$ San Jacinto sequence of 11 March 2013, in the layout of Fig.~\ref{fig:qtm_example}.}
  \label{fig:si_qtm_3}
\end{figure}

\begin{figure}[p]
  \centering
  \includegraphics[width=\textwidth]{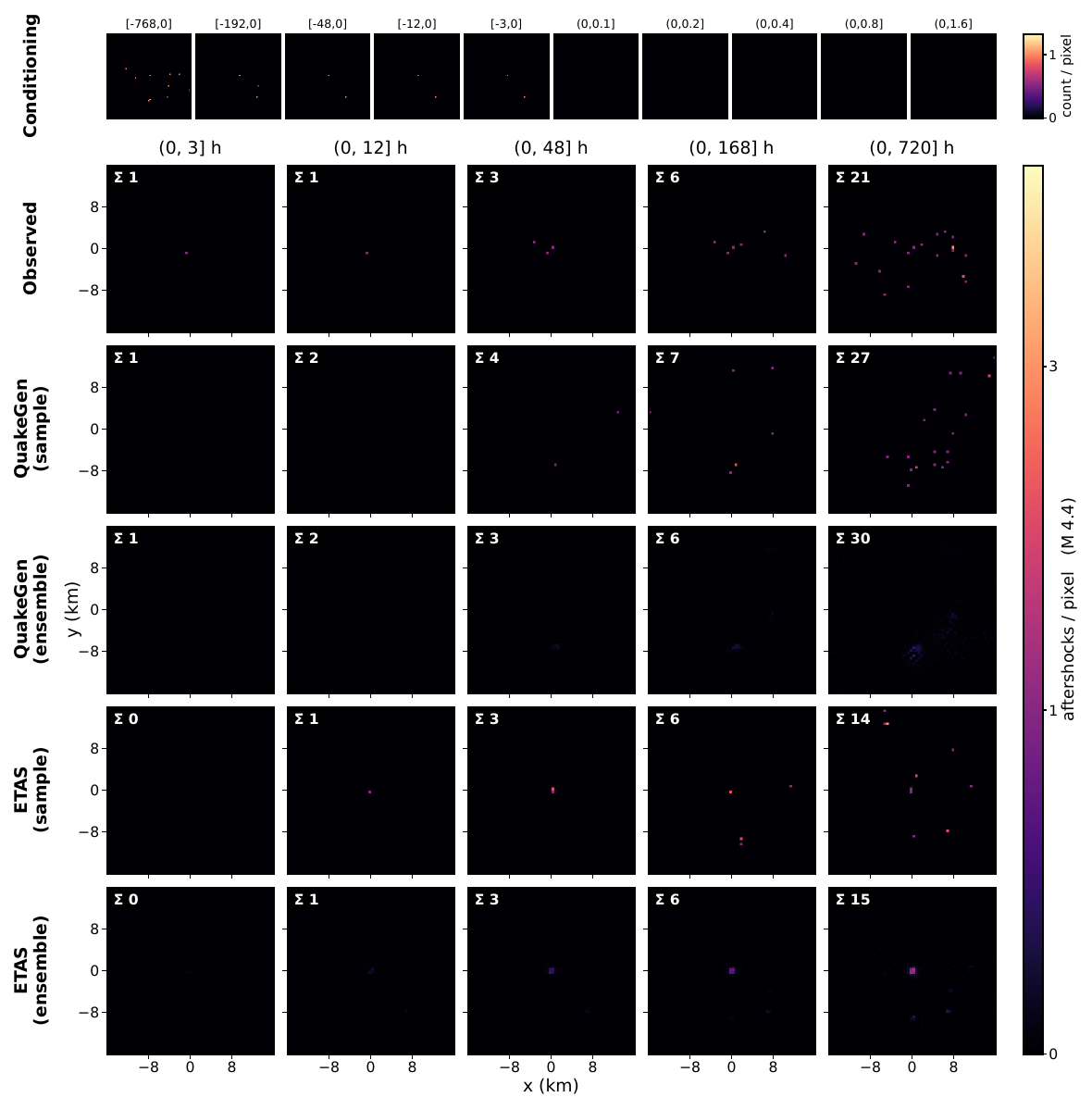}
  \caption{Global model applied to the M~$4.4$ San Jacinto sequence of 6 January 2016, in the layout of Fig.~\ref{fig:qtm_example}. This event is below the model's M~$4.5$ training range and has only 21 scored aftershocks.}
  \label{fig:si_qtm_4}
\end{figure}

\begin{figure}[p]
  \centering
  \includegraphics[width=\textwidth]{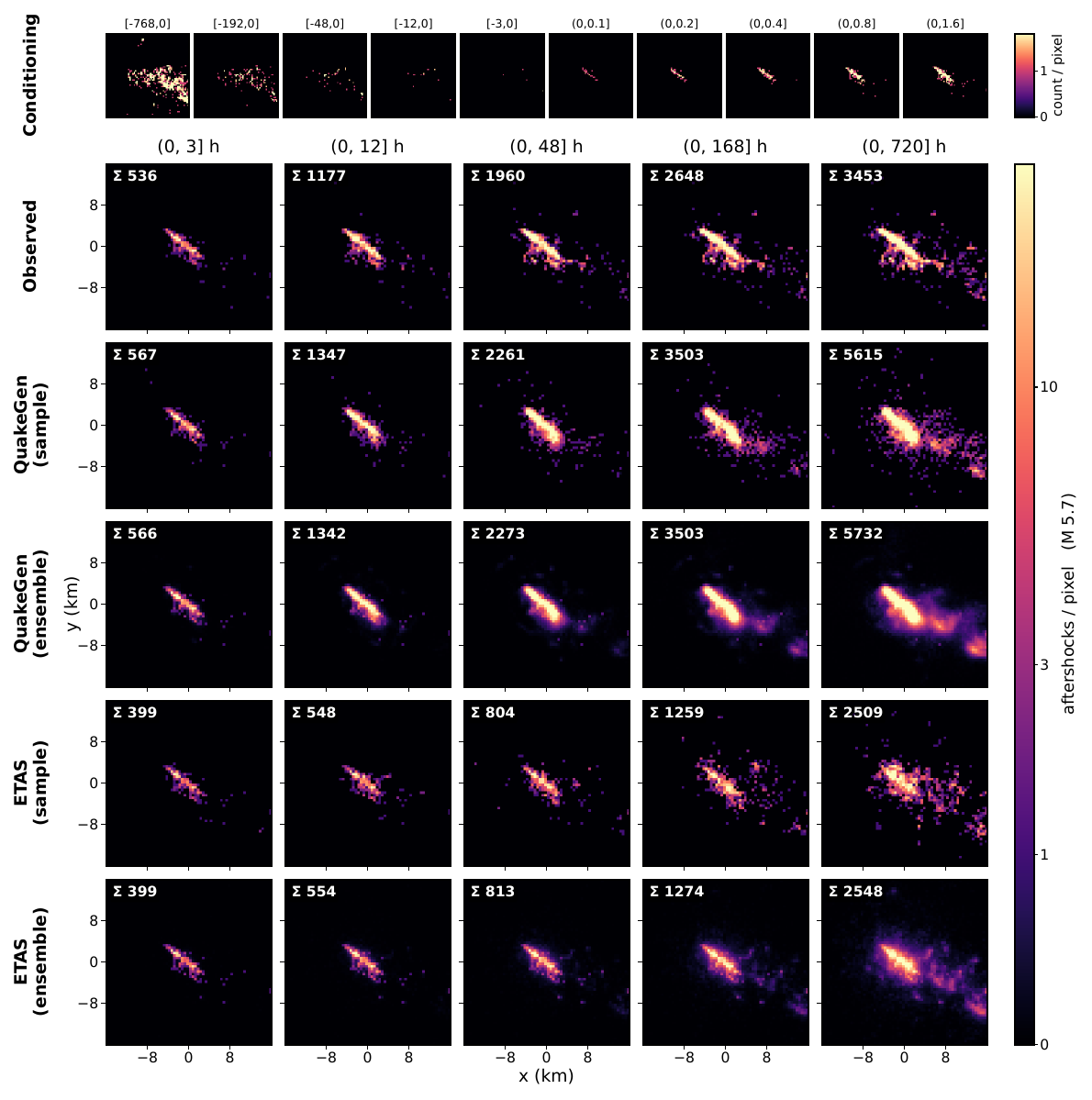}
  \caption{Global model applied to the M~$5.7$ Salton Sea sequence of 15 June 2010, in the layout of Fig.~\ref{fig:qtm_example}.}
  \label{fig:si_qtm_5}
\end{figure}

\begin{figure}[p]
  \centering
  \includegraphics[width=\textwidth]{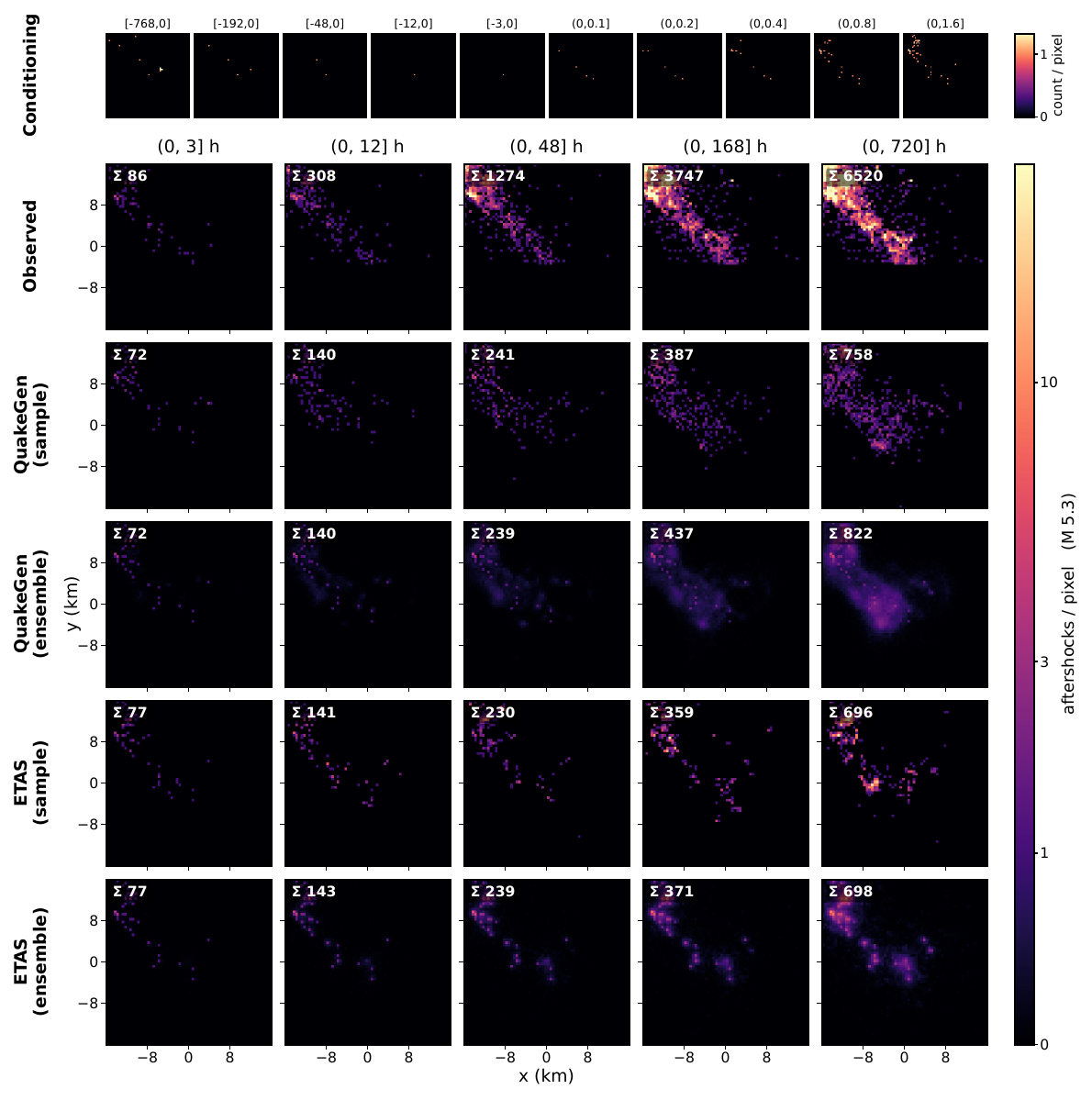}
  \caption{Global model applied to the M~$5.3$ Salton Sea sequence of 4 April 2010, in the layout of Fig.~\ref{fig:qtm_example}. This is the El Mayor--Cucapah sequence entering the box from outside the region, which both forecasts undershoot.}
  \label{fig:si_qtm_6}
\end{figure}

\begin{figure}[p]
  \centering
  \includegraphics[width=\textwidth]{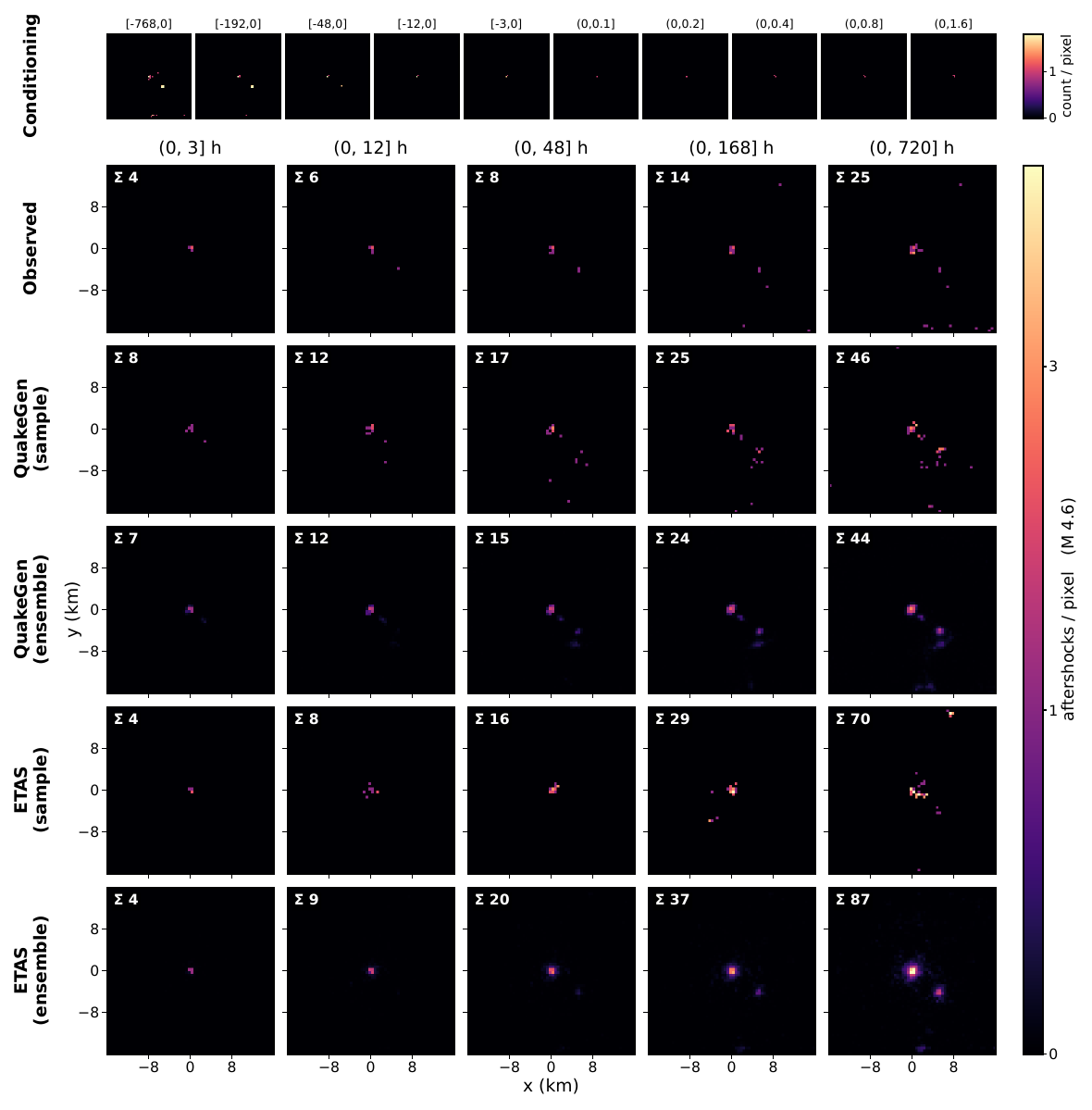}
  \caption{Global model applied to the M~$4.6$ Salton Sea sequence of 4 November 2010, in the layout of Fig.~\ref{fig:qtm_example}.}
  \label{fig:si_qtm_7}
\end{figure}

\begin{figure}[p]
  \centering
  \includegraphics[width=\textwidth]{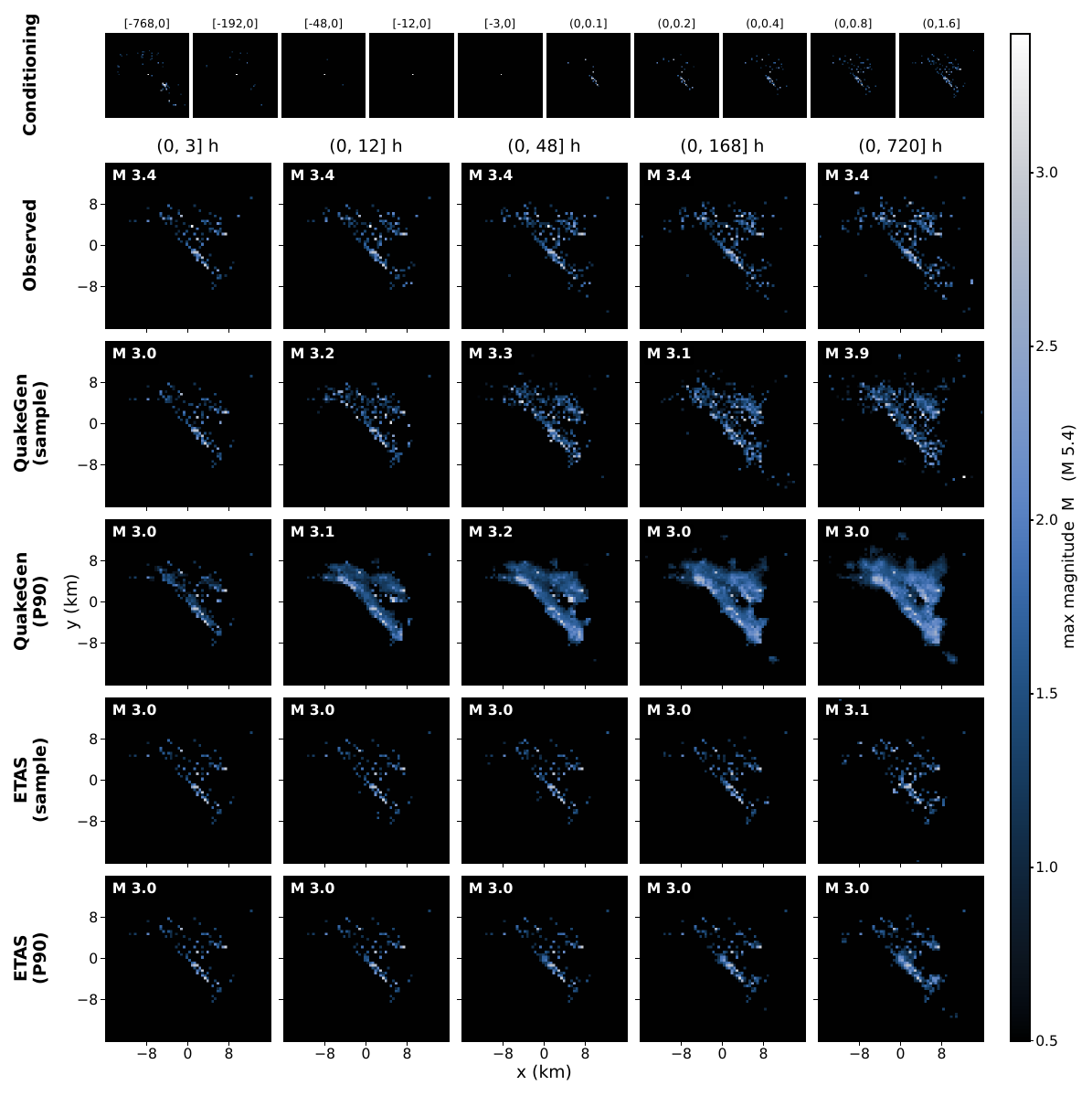}
  \caption{Maximum-magnitude field for the M~$5.4$ San Jacinto sequence of 7 July 2010.}
  \label{fig:si_qtm_mag_1}
\end{figure}

\begin{figure}[p]
  \centering
  \includegraphics[width=\textwidth]{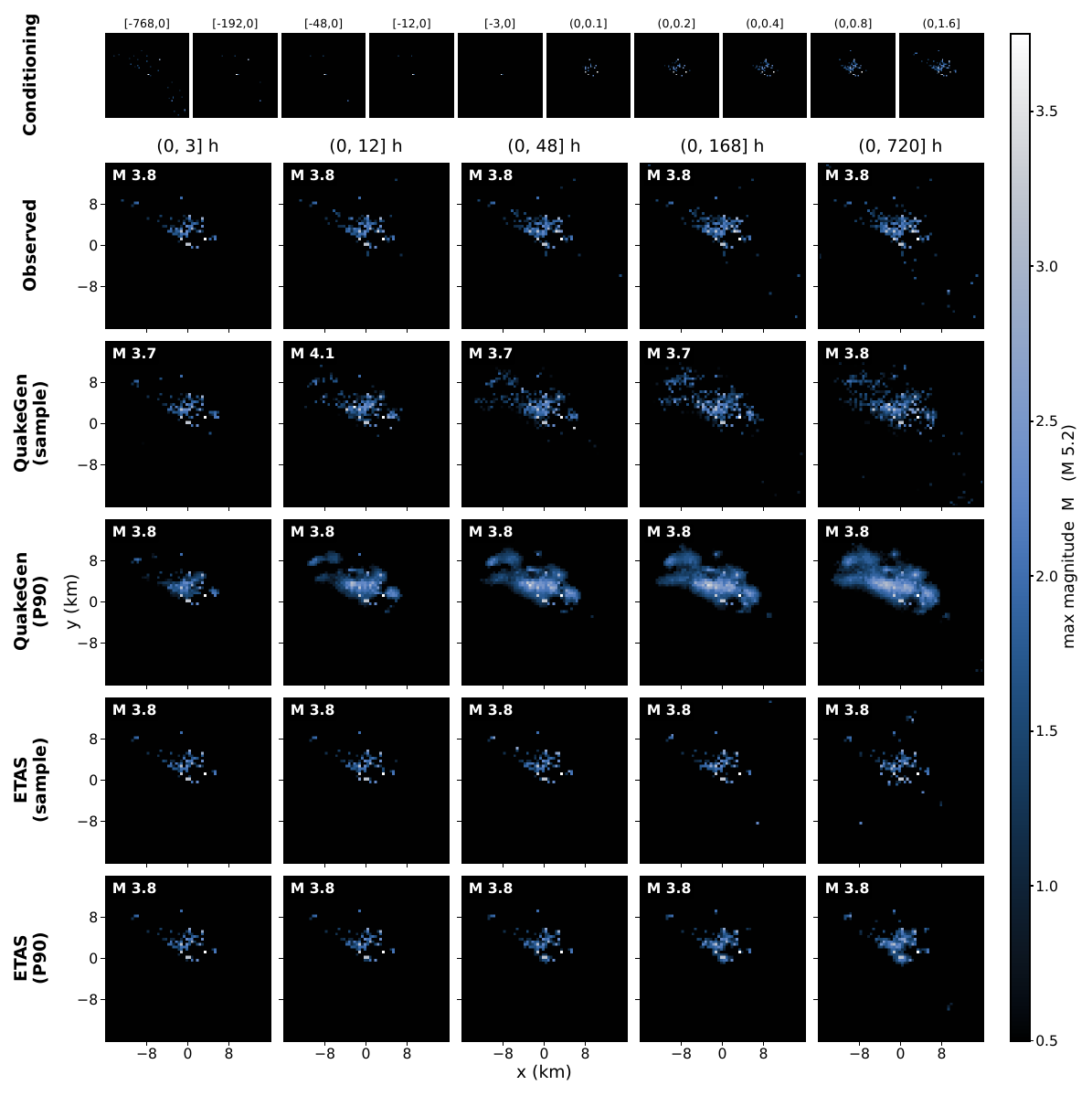}
  \caption{Maximum-magnitude field for the M~$5.2$ Borrego Springs, San Jacinto sequence of 10 June 2016.}
  \label{fig:si_qtm_mag_2}
\end{figure}

\begin{figure}[p]
  \centering
  \includegraphics[width=\textwidth]{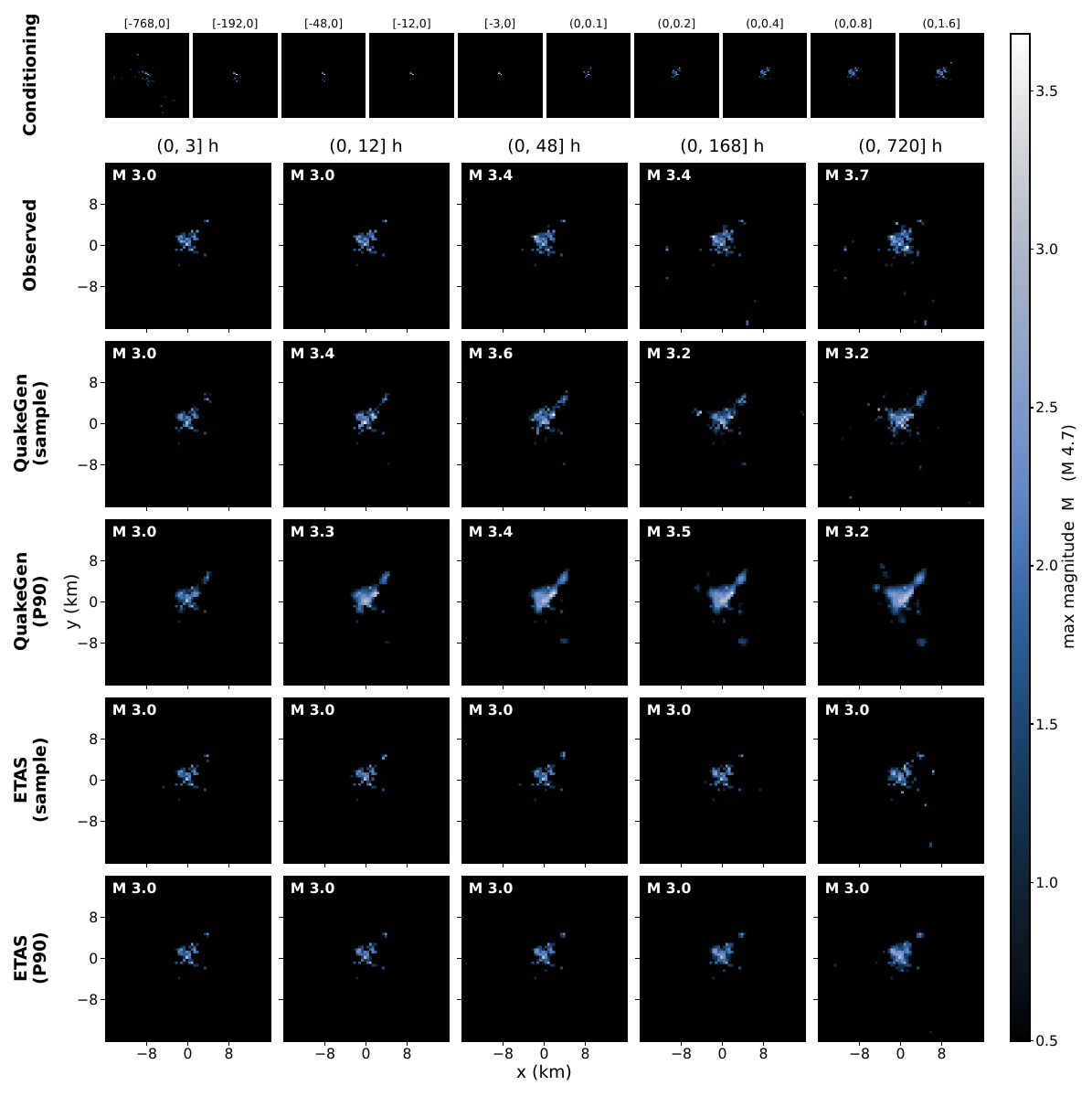}
  \caption{Maximum-magnitude field for the M~$4.7$ San Jacinto sequence of 11 March 2013.}
  \label{fig:si_qtm_mag_3}
\end{figure}

\begin{figure}[p]
  \centering
  \includegraphics[width=\textwidth]{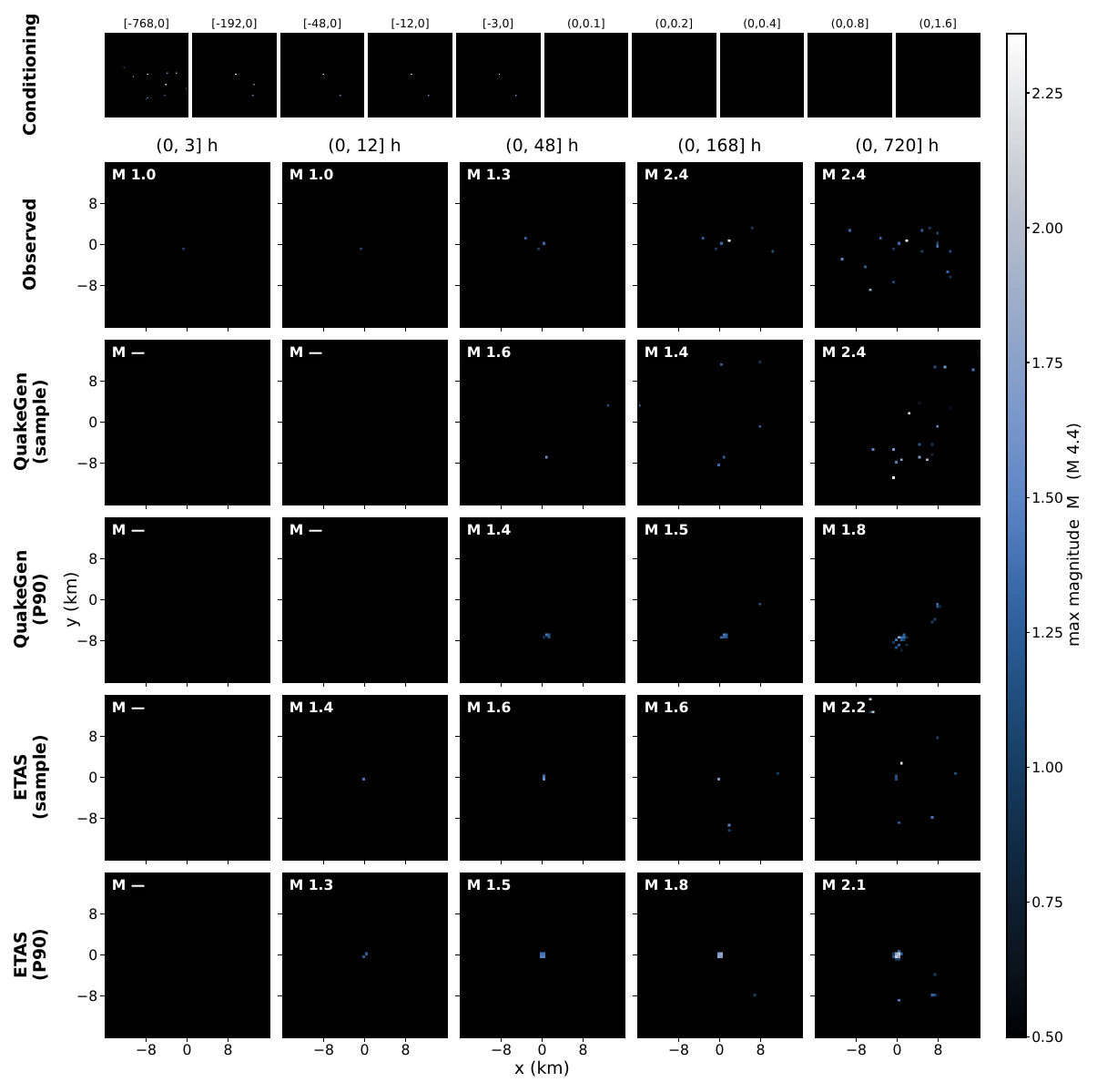}
  \caption{Maximum-magnitude field for the M~$4.4$ San Jacinto sequence of 6 January 2016.}
  \label{fig:si_qtm_mag_4}
\end{figure}

\begin{figure}[p]
  \centering
  \includegraphics[width=\textwidth]{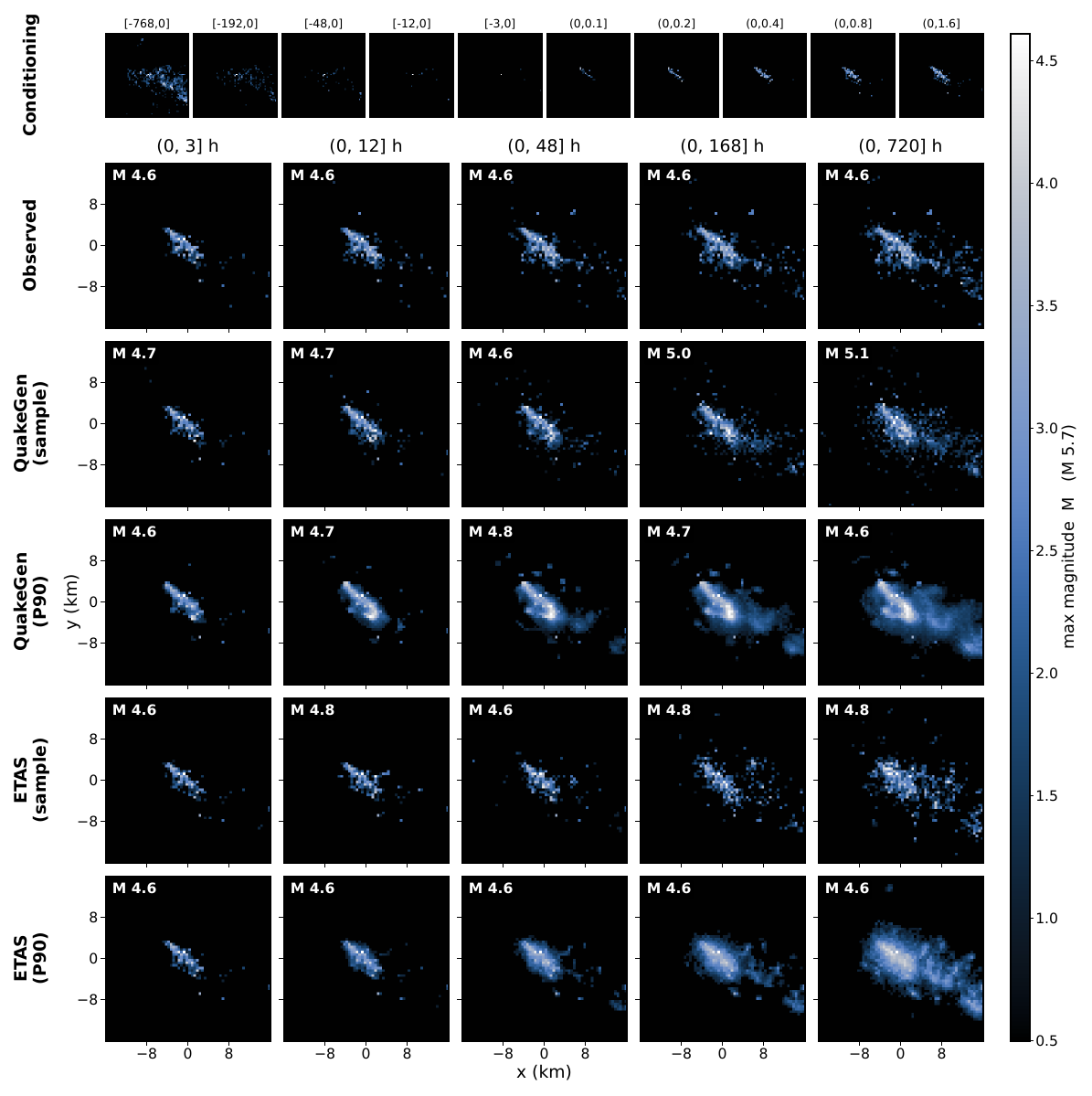}
  \caption{Maximum-magnitude field for the M~$5.7$ Salton Sea sequence of 15 June 2010.}
  \label{fig:si_qtm_mag_5}
\end{figure}

\begin{figure}[p]
  \centering
  \includegraphics[width=\textwidth]{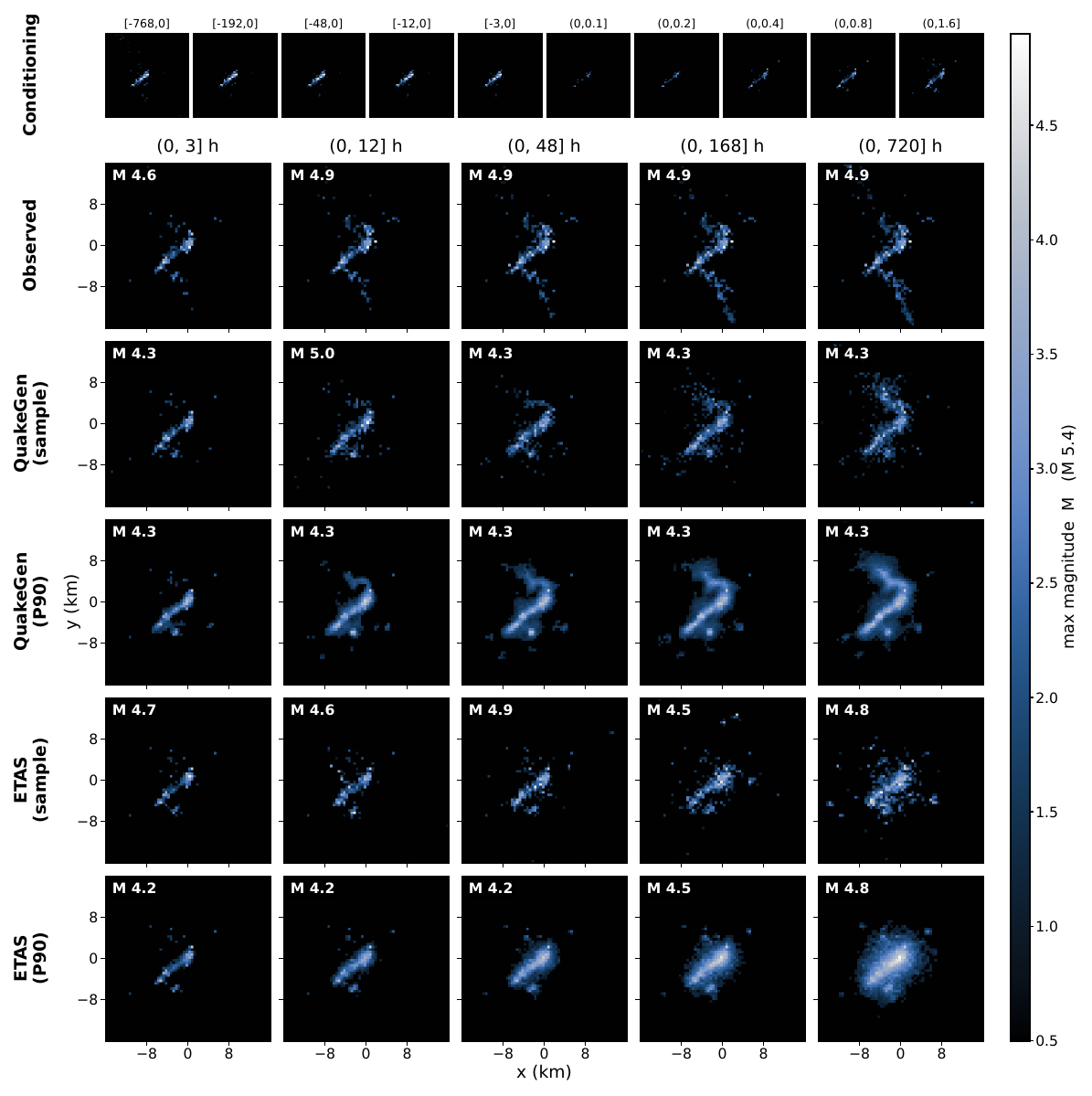}
  \caption{Maximum-magnitude field for the M~$5.4$ Salton Sea (Brawley) swarm of 26 August 2012.}
  \label{fig:si_qtm_mag_6}
\end{figure}

\begin{figure}[p]
  \centering
  \includegraphics[width=\textwidth]{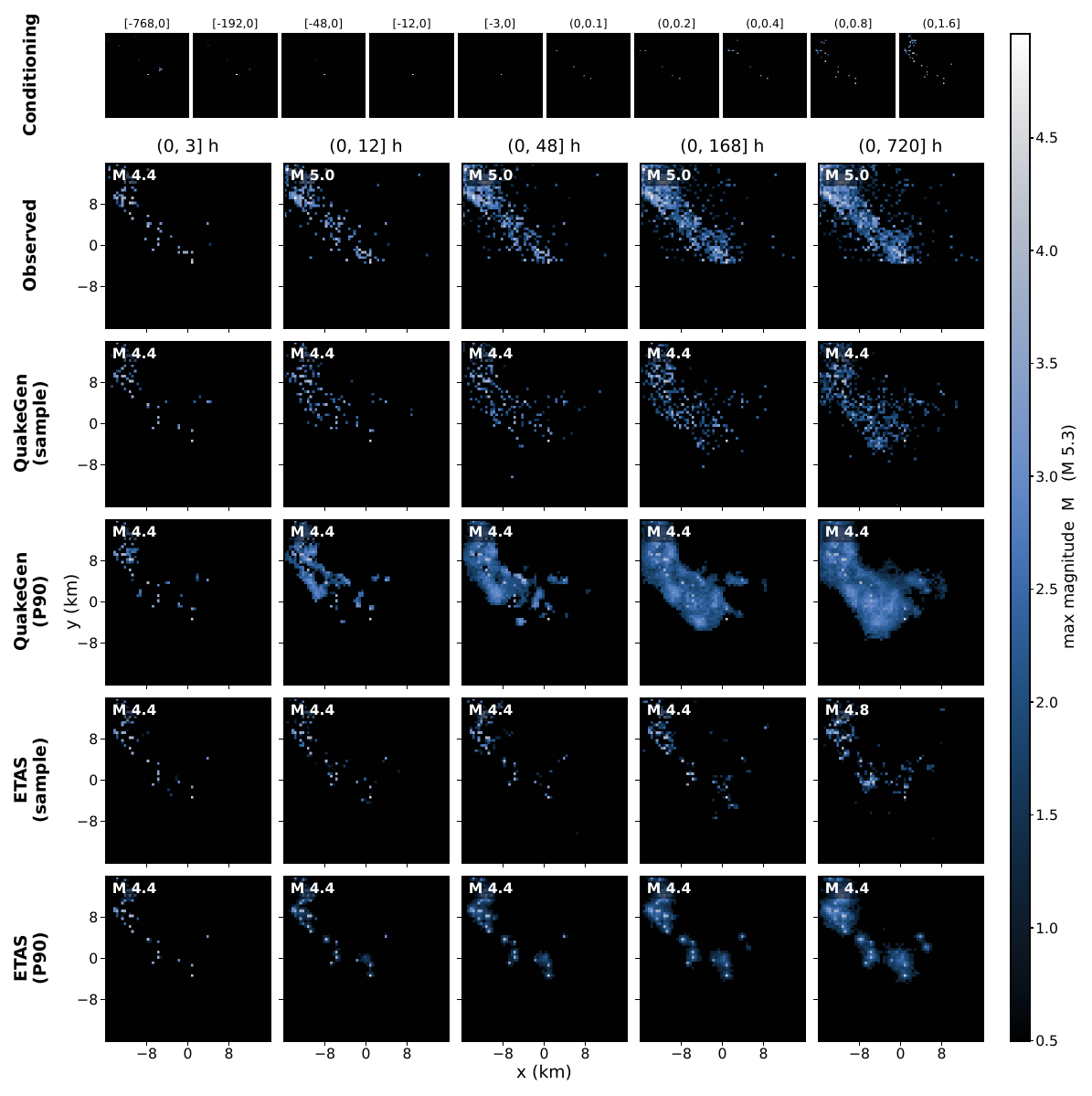}
  \caption{Maximum-magnitude field for the M~$5.3$ Salton Sea sequence of 4 April 2010, the El Mayor--Cucapah sequence entering the box from outside the region.}
  \label{fig:si_qtm_mag_7}
\end{figure}

\begin{figure}[p]
  \centering
  \includegraphics[width=\textwidth]{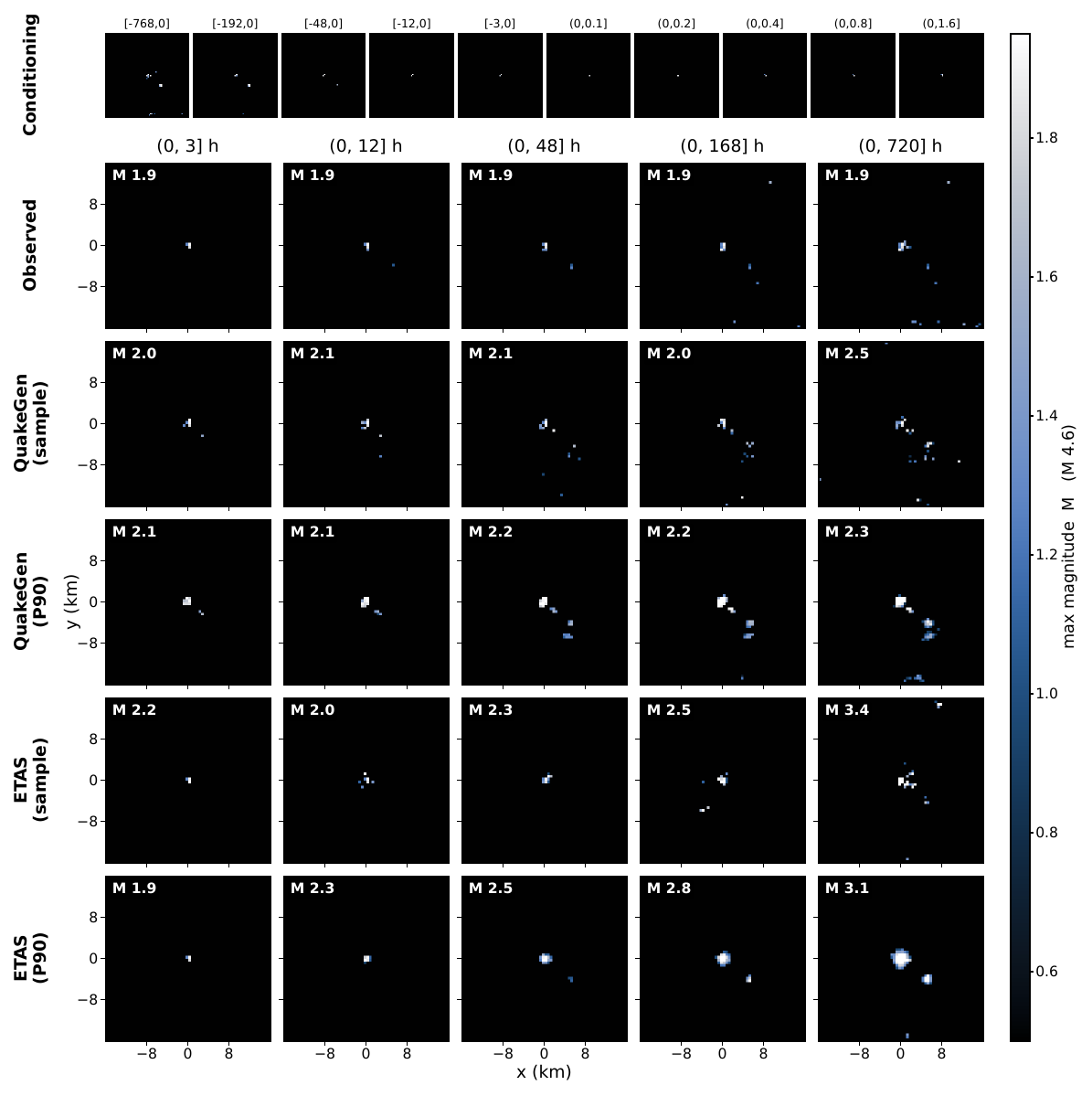}
  \caption{Maximum-magnitude field for the M~$4.6$ Salton Sea sequence of 4 November 2010.}
  \label{fig:si_qtm_mag_8}
\end{figure}

Figures~\ref{fig:si_qtm_1}--\ref{fig:si_qtm_7} show the count-field forecasts for the seven QTM mainshock sequences of Section~\ref{sec:generalization} not shown in the main text, in the layout of Fig.~\ref{fig:qtm_example}, with the branching ETAS simulation as the baseline. Figures~\ref{fig:si_qtm_mag_1}--\ref{fig:si_qtm_mag_8} show the maximum-magnitude fields for all eight sequences, in the same layout but with the QuakeGen and ETAS 90th percentiles (P90) in place of the ensemble means.

\clearpage
\bibliography{quakegen}

\end{document}